\documentclass[twocolumn]{aastex631}
\usepackage{amsmath}

\newcommand{\mgfe}[0]{[{\rm Mg/Fe}]}

\newcommand{\afe}[0]{[\alpha/{\rm Fe}]}
 
\newcommand{\xmg}{[{\rm X}/{\rm Mg}]} 
\newcommand{\mgh}{[{\rm Mg}/{\rm H}]}
\newcommand{\feh}[0]{[{\rm Fe/H}]}

\newcommand{\xfe}{[{\rm X}/{\rm Fe}]}

\newcommand{\logg}{\log(g)}
\newcommand{\teff}{T_{\rm eff}}
\newcommand{\kpc}{\rm \; kpc}
\newcommand{\kel}{\rm \; K}
\newcommand{\msun}{M_{\odot}}

\newcommand{\Msun}{M_{\odot}}

\newcommand{\angstrom}{\textup{\AA}}

\shorttitle{Low $Z$ Abund. Dispersion}
\shortauthors{Griffith et al.}

\begin{document}

\title{Untangling the Sources of Abundance Dispersion in Low-Metallicity Stars}

\correspondingauthor{Emily Griffith}
\email{Emily.Griffith-1@colorado.edu}

\author[0000-0001-9345-9977]{Emily J. Griffith}
\altaffiliation{NSF Astronomy and Astrophysics Postdoctoral Fellow}
\affiliation{Center for Astrophysics and Space Astronomy, Department of Astrophysical and Planetary Sciences, University of Colorado, 389 UCB, Boulder, CO 80309-0389, USA}
\affiliation{The Department of Astronomy and Center of Cosmology and AstroParticle Physics, The Ohio State University, Columbus, OH 43210, USA}

\author[0000-0001-7258-1834]{Jennifer A. Johnson}
\affiliation{The Department of Astronomy and Center of Cosmology and AstroParticle Physics, The Ohio State University, Columbus, OH 43210, USA}

\author[0000-0001-7775-7261]{David H. Weinberg}
\affiliation{The Department of Astronomy and Center of Cosmology and AstroParticle Physics, The Ohio State University, Columbus, OH 43210, USA}
\affiliation{The Institute for Advanced Study, Princeton, NJ, 08540, USA}

\author{Ilya Ilyin}
\affiliation{Leibniz-Institut for Astrophysics Potsdam (AIP), An der Sternwarte 16, D14482 Potsdam, Germany}

\author[0000-0002-6534-8783]{James W. Johnson}
\affiliation{The Department of Astronomy and Center of Cosmology and AstroParticle Physics, The Ohio State University, Columbus, OH 43210, USA}

\author[0000-0003-1445-9923]{Romy Rodriguez-Martinez}
\affiliation{The Department of Astronomy and Center of Cosmology and AstroParticle Physics, The Ohio State University, Columbus, OH 43210, USA}

\author[0000-0002-6192-6494]{Klaus G. Strassmeier}
\affiliation{Leibniz-Institut for Astrophysics Potsdam (AIP), An der Sternwarte 16, D14482 Potsdam, Germany}

\begin{abstract}

We measure abundances of 12 elements (Na, Mg, Si, Ca, Sc, Ti, V, Cr, Mn, Fe, Co, Ni) in a sample of 86 metal poor ($-2 \lesssim \feh \lesssim -1$) subgiant stars in the solar neighborhood. Abundances are derived from high-resolution spectra taken with the Potsdam Echelle Polarimetric and Spectroscopic Instrument on the Large Binocular Telescope, modeled using iSpec and MOOG.  By carefully quantifying the impact of photon-noise ($< 0.05$ dex for all elements) we robustly measure the {\it intrinsic} scatter of abundance ratios.  At fixed [Fe/H] the RMS intrinsic scatter in [X/Fe] ranges from 0.04 dex (Cr) to 0.16 dex (Na), with a median of 0.08 dex.  Scatter in [X/Mg] is similar, and accounting for [$\alpha$/Fe] only reduces the overall scatter moderately.  We consider several possible origins of the intrinsic scatter with particular attention to fluctuations in the relative enrichment by core-collapse supernovae (CCSN) and Type Ia supernovae (SNIa) and stochastic sampling of the CCSN progenitor mass distribution.  The stochastic sampling scenario provides a good quantitative explanation of our data if the effective number of CCSN contributing to the enrichment of a typical sample star is $N \sim 50$.  At the median metallicity of our sample, this interpretation implies that the CCSN ejecta are mixed over a gas mass $\sim 10^5 \msun$ before forming stars.  The scatter of elemental abundance ratios is a powerful diagnostic test for simulations of star formation, feedback, and gas mixing in the early phases of the Galaxy.

\end{abstract}

\keywords{Abundance ratios; High resolution spectroscopy, Chemical enrichment, Supernovae}

\section{Introduction} \label{sec:intro}

Over the past decade, large spectroscopic surveys have characterized the elemental abundance ratios of Galactic stellar populations in steadily increasing detail (e.g., \citealt{majewski2017,conroy2019,buder2021,gilmore2022}), yielding insight on the enrichment and assembly history of the Milky Way.  Many studies focus on the mean or median trends of [X/Fe] vs.\ [Fe/H] for elements X with different nucleosynthetic origins, with less attention to the star-to-star scatter about these trends.  Mean [X/Fe] trends provide diagnostics of star formation efficiency, gas accretion, outflows, and star formation history, and of relative contributions of prompt enrichment from core collapse supernovae (CCSN) vs.\ time-delayed sources such as Type Ia supernovae (SNIa) or asymptotic giant branch (AGB) stars (e.g., \citealt{tinsley1980,matteucci1986,mcwilliam1997,andrews2017,rybizki2017,weinberg2017,weinberg2019,griffith2019,spitoni2019}).  The existence of distinct low-Ia (high-$\alpha$) and high-Ia (low-$\alpha$) sequences in the Milky Way disk \citep{fuhrmann1998,bensby2003,hayden2015,vincenzo2021} suggests complexity in the disk's accretion and star formation history, modulated by the impact of stellar radial migration (e.g., \citealt{chiappini1997,schonrich2009,haywood2013,minchev2014,clarke2019,johnson2021,chen2022}).
Scatter about the mean trends can arise from star-to-star variation in the relative amounts of CCSN/SNIa/AGB enrichment or from stochastic sampling of the supernova and AGB populations themselves, providing insight on the number and diversity of enrichment events, the degree of mixing within the interstellar medium (ISM), and the mixing of stellar populations with distinct enrichment histories.  In this paper, we measure the intrinsic scatter of abundance ratios for twelve elements in a sample of 86 nearby sub-giant stars with $-2 < \feh < -1$.  We offer a tentative interpretation of our results in terms of stochastic sampling of the CCSN population.

The challenge in measuring intrinsic scatter arises from removing the effects of observational scatter, which can arise both from photon-noise in the stellar spectra and from systematic abundance determination errors that vary from star to star.  Contributions of the latter, differential systematics, can be mitigated by studying samples with a narrow range of physical properties (e.g., $\teff$, $\logg$).  Applying this approach to APOGEE abundances, \cite{vincenzo2021} find intrinsic scatter in $\afe$ at fixed [Fe/H] of $\sim 0.04$ dex for the disk low-Ia and high-Ia populations, in agreement with earlier analysis by \cite{bertran2016}.  This scatter can plausibly arise from variations in the ratio of SNIa/CCSN enrichment, since CCSN dominate the production of $\alpha$-elements while both classes of supernovae contribute to Fe.  At fixed [Fe/H] {\it and} $\afe$, which largely removes CCSN/SNIa variation as a source of scatter, \cite{ratcliffe2022} and \cite{ting2022} find smaller intrinsic scatter, $\sim 0.01-0.02$ dex for most $\alpha$ and Fe-peak elements, with larger values (up to $\sim 0.05$ dex) for some elements such as C, N, Na, and Ce that may have substantial AGB contributions. \cite{griffith2022} and \cite{weinberg2022} draw similar conclusions by fitting abundance patterns from the GALAH and APOGEE surveys with a 2-process model that accounts for IMF-averaged CCSN and SNIa contributions.

The above studies focus on the metallicity range of the Galactic disk, roughly $-0.7 \leq \feh \leq 0.4$.  At much lower metallicities, the intrinsic abundance scatter is clearly larger.  In a study of very metal poor stars ($-4.3 \lesssim \feh \lesssim -1.7$) \citet{li2022} presented [X/Fe] vs. [Fe/H] abundance trends and measured the dispersion about a linear fit. They report scatter of 0.05 to 0.1 dex in $\alpha$ and Fe-peak elements, and scatter near 0.2 dex for [Na/Fe]. \citet{lombardo2022} find similar levels of dispersion about the mean abundance of metal poor ($-3.6<\feh<-1.5$) giants.

Here we examine the intermediate metallicity range $-2 < \feh < -1$. Many studies in this regime have focused on distinguishing {\it in situ} and accreted components of the stellar halo (e.g., \citealt{nissen2010,nissen2011,naidu2020,belokurov2022,horta2022}) or identifying metal-poor stars in the bulge (e.g., \citealt{howes2016, lucey2019}).  Plots in these papers suggest $\sim 0.05-0.2$ dex scatter of [X/Fe] ratios for many elements within each population, but they do not disentangle observational and intrinsic contributions.  Searching the JINAbase, an abundance and parameter database for metal-poor stars \citep{abohalima2018}, we find $\sim 170$ stars with $-2 < \feh < -1$ from 18 different studies. The collection of data sets show scatter of $0.05-0.1$ dex about the mean [Mg/Fe] value. However, none of the works compiled from the JINAbase achieve abundance precision $< 0.05$ dex, and each work has a unique parameter and abundance determination scheme.  Reliably measuring intrinsic scatter of [X/Fe] ratios in this regime requires homogeneous analysis of a large ($\sim 100$ star) sample with observational errors that are small (less the the expected intrinsic error, i.e. $\lesssim 0.04$ dex; \citealp{vincenzo2021}) and well characterized.  We provide such an analysis in this paper.

Most Galactic stars in this [Fe/H] range lie close to the $\afe \approx 0.3$ plateau that is conventionally interpreted as representing nearly pure CCSN enrichment (though see \citealt{conroy2022} for an alternative interpretation).  Thus, one can reasonably expect that variations in the CCSN/SNIa/AGB enrichment fractions are smaller than those in the higher metallicity disk populations.  However, since the number of individual supernova events contributing to a given star's elemental abundances is likely smaller at low metallicity, stochastic sampling of the initial mass function (IMF) may be a more important source of scatter.  Because of differences in pre-supernova structure, explosive nucleosynthesis, and the boundary between ejected and fallback material, the element-by-element yields of CCSN depend strongly and often non-monotonically on progenitor mass (\citealt{sukhbold2016}; see Figure 12 of \citealt{griffith2021b}).  If the supernova-to-supernova variance of a given element ratio is order unity, then a population of stars enriched by $\sim N_{\rm SN}$ supernovae should exhibit $\sim N_{\rm SN}^{-1/2}$ fractional scatter in that ratio.  If stochastic IMF sampling can be demonstrated as the dominant source of abundance scatter for some elements, then measurements of scatter can constrain the typical mass scale of element mixing in the ISM and test supernova yield models.  Furthermore, the IMF-averaged yields and supernova-to-supernova variance depend on the landscape of black hole formation \citep{pejcha2015, ertl2016,sukhbold2016}, since, e.g., more compact progenitors produce large yields of Fe-peak elements if they explode but none if they collapse into black holes \citep{griffith2021b}.  With an appropriate suite of progenitor/supernova models (which does not currently exist), our measurements could provide insights on black hole formation in the $-2 < \feh < -1$ metallicity range.

Section~\ref{sec:data} describes our sample selection and details of our spectroscopic observations. Section~\ref{sec:pipeline} outlines our methods for stellar parameter and abundance determination. Section~\ref{sec:abunds} presents our stellar abundance trends, comparing trends to a higher metallicity GALAH sample for context.   Section~\ref{sec:scatter} presents our primary observational results, determinations of the intrinsic scatter of [X/Fe] and [X/Mg] ratios in our sample for $\alpha$, light odd-$Z$, and Fe-peak elements (with X = Mg, Si, Ca, Ti, Na, Sc, V, Cr, Mn, Co, Fe, Ni). In Section~\ref{sec:kin} we divide our sample into kinematically defined subsamples to compare scatter between \textit{in situ} and accreted stars. We discuss potential sources of scatter and estimate the scatter from a stochastically sampled IMF in Section~\ref{sec:discussion}. Finally, in Section~\ref{sec:summary}, we summarize our conclusions and describe our planned future work on this subject.

\section{Data} \label{sec:data}

In order to robustly measure the intrinsic abundance scatter in metal-poor stars, we must measure stellar abundances to a precision of $\leq 0.04$ dex. Such precision is needed in the differential scatter, but not in the zero-point uncertainties. To achieve differential scatter $\leq0.04$ dex, we require (1) spectra of high enough signal to noise (S/N) that the photon-noise contribution to the errors on abundance ratios will be smaller than 0.04 dex and (2) a sample population of sufficient homogeneity that the errors in modeling the spectra should not introduce star-to-star scatter as large as 0.04 dex. 

\subsection{Sample Selection} \label{subsec:select}

Stellar abundances often bear systematic uncertainties due to abundance correlations with stellar parameters such as $\teff$ and $\logg$ \citep[e.g.,][]{holtzman2018, griffith2021a}. Uncertainties in the derived stellar parameters translate to uncertainty in the derived abundances of 0.01 to 0.1 dex \citep[e.g.,][]{jonsell2005, jacobson2015, jofre2019}. To minimize the effect of systematic uncertainties, maximize number of targets observed, and achieve our desired abundance precision, we construct a uniform stellar sample of bright stars, targeting local subgiants. 

\begin{figure}[!htb]
    \centering
    \includegraphics[width=\columnwidth]{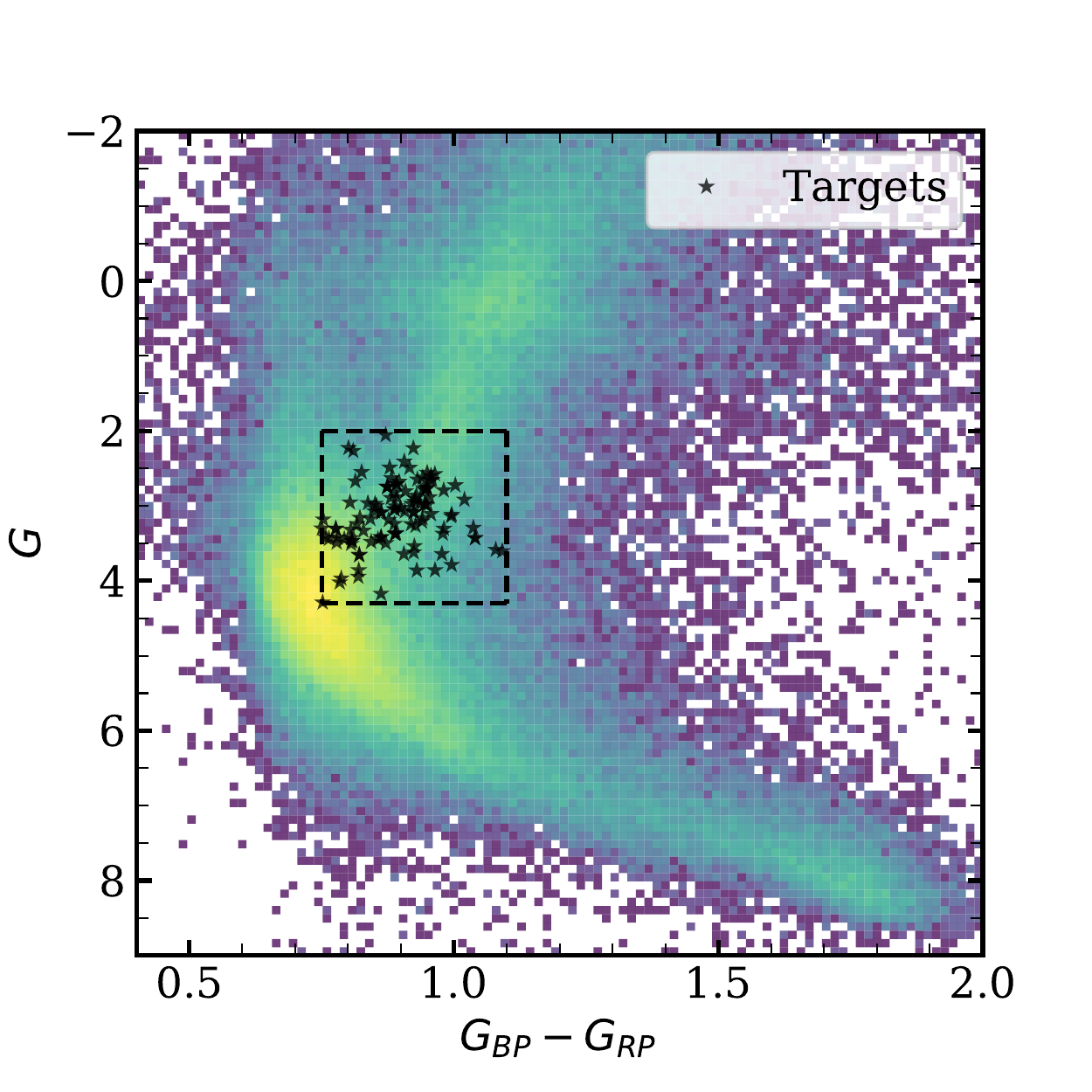}
    \caption{Gaia $G_{BP} - G_{RP}$ vs. $G$ HR diagram of metal-poor ($-2.1 \leq \feh \leq -1$) LAMOST stars (2D density plot) and our targets (black stars). The box formed by the black dashed lines denotes our magnitude and color cuts for the subgiant population.}
    \label{fig:subgiants}
\end{figure}

We select stellar candidates from LAMOST\footnote{Large sky Area Multi-Object fiber Spectroscopic Telescope} DR5 \citep{xiang2019}, a low-resolution spectroscopic survey. We cross-match the LAMOST catalog with Gaia eDR3 \citep{gaia2021} and leverage the astroNN Gaia tools\footnote{\url{https://astronn.readthedocs.io/en/latest/tools_gaia.html}} \citep{bovy2017} to calculate stellar distances and absolute V band magnitudes and thus place stars on an HR diagram (Figure~\ref{fig:subgiants}). From LAMOST, we select metal-poor, bright, and nearby subgiants that meet the following criteria: 
\begin{itemize}
    \item $-2.1 \leq \feh \leq -1.0$
    \item $V < 12.5$
    \item $d < 2500$ pc
    \item $ 3.3 < \logg < 4$
    \item $2 < G < 4.3$ 
    \item $0.75 < G_{BP} - G_{RP} < 1.1$
\end{itemize}
using $\feh$ and $\logg$ from LAMOST and $V$, $d$, $G$, and $G_{BP}-G_{RP}$ derived from Gaia data. The magnitude and color cuts are indicated by the dashed box in Figure~\ref{fig:subgiants}, and our targets are shown as the black stars. These cuts supply a sample of 102 observable targets. 

\subsection{Observations} \label{subsec:obs}

We obtained high resolution optical spectra of 98 stars from the Potsdam Echelle Polarimetric and Spectroscopic Instrument \citep[PEPSI;][]{strassmeier2015} on the Large Binocular Telescope (LBT) between April 30th, 2021 and February 12th, 2022. 
Observations were taken with the 300 $\mu$m fiber and cross-dispersers (CD) II and IV, to obtain $R = \lambda / \Delta \lambda = 50,000$ spectra in the wavelength ranges $4260 \angstrom - 4800\angstrom$ and $5440\angstrom-6270\angstrom$. 
We chose PEPSI's largest fiber diameter (and lowest resolution setting) to minimize the exposure time needed per target and maximize the number of targets observed. With this spectral configuration, we need exposure lengths of 300 to $1200$ s to reach S/N $\approx 100$ per pixel in CD II and S/N $> 100$ in CD IV. We achieve a median S/N of 125 in CD II and 236 in CD IV.

Details for each target, including the 2MASS ID, V band magnitude, observation date, exposure length, and S/N achieved in CD II and CD IV, are given in Table~\ref{tab:obs}.

\begin{deluxetable*}{lrrrrr}
\tablecaption{Observing Details 
\label{tab:obs}}
\tablehead{
\colhead{Object} &      \colhead{V} &        \colhead{Date} &      \colhead{Exp (s)} &  \colhead{SNR$_{\text{CD II}}$} &  \colhead{SNR$_{\text{CD IV}}$}
}
\startdata
2MASS J07295084+3251585 &  10.41 &  2021-04-30 &  600 &    174 &    340 \\
2MASS J09144031+5200589 &  10.68 &  2021-04-30 &  600 &    176 &    349 \\
2MASS J13135193+1619226 &  10.13 &  2021-04-30 &  600 &    226 &    432 \\
2MASS J11374200+3305528 &   9.42 &  2021-04-30 &  300 &    230 &    443 \\
2MASS J11571896-0102018 &  11.15 &  2021-04-30 &  600 &    130 &    267 \\
\enddata
\tablecomments{V magnitude, observing date, exposure time, and SNR per pixel in CD II and CD IV for all objects. Full table available online.}
\end{deluxetable*}

\section{Stellar Parameter and Abundance Determination} \label{sec:pipeline}

We determine stellar parameters and stellar abundances with the radiative transfer code MOOG \citep{sneden1973} implemented through the iSpec interface \citep{blanco2014,blanco2019}. 

\subsection{Line List} \label{subsec:linelist}

Referencing the line lists of \citet{fulbright2000}, GALAH DR3 \citep{buder2021}, and Gaia-ESO \citep{gilmore2012, heiter2021} as well as the solar spectrum \citep{moore1966}, we construct a line list that is optimized for the metallicity range of our sample. We adopt line parameters such as excitation potential and oscillator strength ($\log (gf)$) from version 6 of the Gaia-ESO line list \citep[included in iSpec;][]{gilmore2012, heiter2021} for all lines except those of Mg. To ensure we have the most reliable $\log (gf)$ values for our five \ion{Mg}{1} lines (4571, 4700, 4730, 5528, 5711$\angstrom$), we use Mg oscillator strengths from \citet{pehlivan2017} as recommended by \citet{bergemann2017}.  The adopted \ion{Mg}{1} oscillator strengths are within 0.04 dex of the Gaia-ESO values for the four reddest lines and $0.226$ dex lower than the Gaia-ESO value for the line at 4571$\angstrom$. After a preliminary line-by-line stellar abundance determination, we remove lines from our list for which the population's median line variation, the difference between the individual line's abundance and the star's median elemental abundance, exceeds 0.25 dex, similar to sigma clipping procedures discussed in \citet{jofre2019}. We do not remove lines from our list for elements with $<15$ lines in our spectral range. Our final line list includes species \ion{Na}{1}, \ion{Mg}{1}, \ion{Si}{1}, \ion{Ca}{1},  \ion{Sc}{2}, \ion{Ti}{1}, \ion{Ti}{2}, \ion{V}{1},  \ion{Cr}{1}, \ion{Cr}{2}, \ion{Mn}{1}, \ion{Fe}{1}, \ion{Fe}{2}, \ion{Co}{1}, \ion{Ni}{1}, \ion{Cu}{1}, \ion{Zn}{1}, \ion{Sr}{1}, \ion{Y}{2}, \ion{Zr}{1}, \ion{Zr}{2}, \ion{Ba}{2}, \ion{La}{2}, and \ion{Ce}{2}. In this paper we will only present abundances of elements lighter than Cu, leaving analysis of the heavy element scatter to future work. 

\subsection{Spectral Reduction}\label{subsec:spec_reduc}

The majority of the spectral reduction is done through the Spectroscopic Data Systems for PEPSI pipeline, as described in \citet{strassmeier2018}. In brief, this pipeline applies bias corrections, estimates photon-noise, flat field corrects images for CCD fixed pattern noise, subtracts scattered light contamination, defines the spectral orders of the \'{e}chelle, removes cosmic ray contamination, and optimizes the flux extraction. It wavelength calibrates the spectra against Th-Ar lines and shapes the spectra by correcting for the \'{e}chelle blaze function, vignetting effects, and fringing. Finally, the pipeline applies a global continuum normalization by fitting a 2D smoothing spline. 

The remaining spectral reduction is done within an integrated spectroscopic framework called iSpec \citep{blanco2014}. We use iSpec to correct spectra for radial velocity (RV) shifts by cross correlating with the NARVAL solar spectrum and mask spectral regions within $\pm 30\angstrom$ of strong telluric features. The cross correlation routine finds two RV peaks for star 2MASS J14042846+1330190, suggestive of a spectroscopic binary, so we remove this star from our sample. 

\begin{figure*}[!htb]
    \centering
    \includegraphics[width=\textwidth]{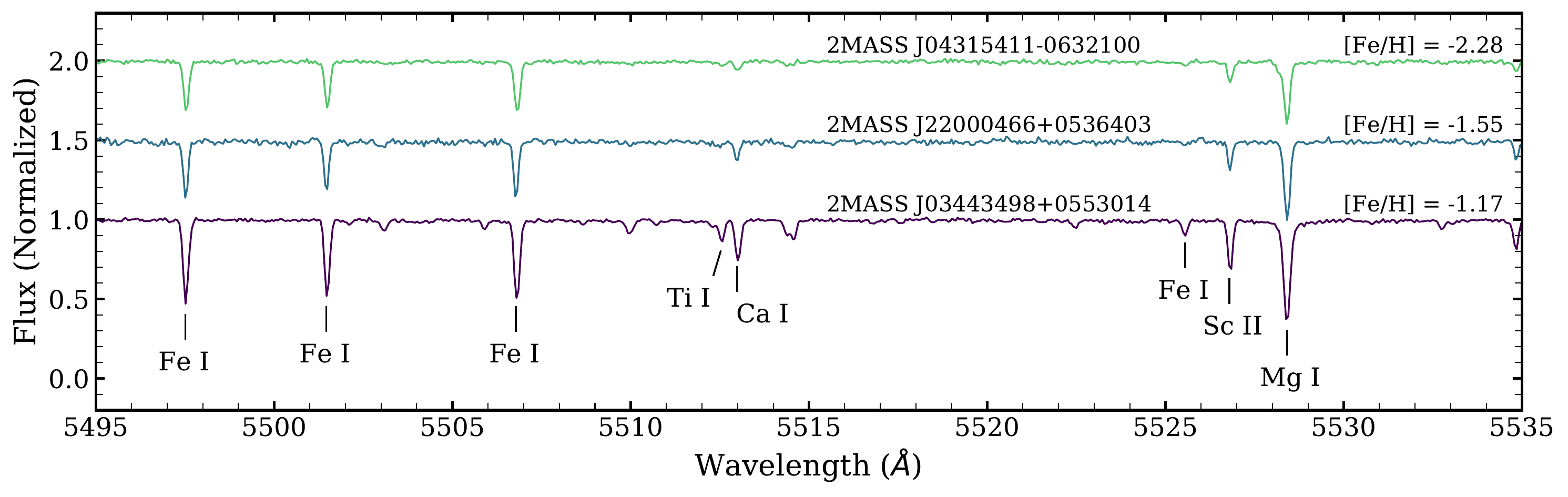}
    \caption{Spectral section from $5495-5535 \, \angstrom$ of three stars with $\feh$ near $-2$ (green), $-1.5$ (blue), and $-1.0$ (purple). Line features for \ion{Fe}{1}, \ion{Ti}{1}, \ion{Ca}{2}, \ion{Sc}{2}, and \ion{Mg}{1} are labeled. Fluxes are normalized, with the $\feh\approx-1.5$ spectra offset by 0.5 and the $\feh\approx-2$ spectra offset by 1.}
    \label{fig:spectra}
\end{figure*}

While the PEPSI pipeline applies a continuum normalization, we re-fit the continuum as a second degree spline with 50 knots, about one for every 50$\angstrom$, as recommended by \citet{blanco2014}, and normalize the spectrum. We plot a small region of three RV-corrected, continuum normalized spectra of different $\feh$ in Figure~\ref{fig:spectra} with labeled line features.

\subsection{Photometric Stellar Parameter Determination}\label{subsec:exo}

To derive preliminary physical properties of our targets, we use the IDL package \texttt{EXOFASTv2} \citep{eastman2013, eastman2017, eastman2019}, a planet modeling software that also provides stellar parameter determinations/estimates. To do this, we perform a spectral energy distribution (SED) fit of the stars from available broadband photometry using the \citet{kurucz1992} atmosphere models. We used the Gaia $G$, $G_{BP}$, and $G_{RP}$ magnitudes \citep{gaia2018, gaia2021}, the J, H, and K magnitudes from the 2MASS catalog \citep{cutri2003, zacharias2004,skrutskie2006}, and the W1, W2, W3, and W4 magnitudes from the WISE catalog \citep{cutri2014}. We combined these with the stars' parallaxes from Gaia DR3 and an interstellar visual extinction, $A_V$, from the \citet{schlegel1998} dust maps. In addition to the SED fit, we employ the MIST stellar evolutionary models to estimate the stellar properties $\teff$, $\logg$, $\feh$, and $M_*$.

\subsection{Equivalent Width Stellar Parameter Determination} \label{subsec:params}

With these reliable photometric parameters as initial guesses, we derive spectroscopic stellar parameters from Fe lines in the spectra. We use iSpec to fit \ion{Fe}{1} and \ion{Fe}{2} lines with Gaussian or Voigt profiles and measure their equivalent widths. Voigt profiles are typically selected for stronger lines, where Gaussian profiles do not capture the broadened wing features. Because our spectra have many Fe features, we remove the strongest and weakest lines from this analysis, requiring that the reduced equivalent width be between $-4.3$ and $-6.0$ and that the lines' lower energy state be between 0.5 and 5.0, as recommended by \citet{mucciarelli2013} and \citet{blanco2014, blanco2019}. We then run the equivalent width parameter determination, utilizing local thermodynamic equilibrium (LTE) radiative transfer code MOOG \citep{sneden1973}, MARCS model atmospheres \citep{gustafsson2008}, and solar abundances from \citet{asplund2009}, implemented and automated through iSpec. The program iterates through values of $\teff$, $\logg$, $v_{\text{micro}}$, [M/H], and [$\alpha$/Fe] until Fe lines of varying strengths and ionization states return the same [Fe/H] abundance. Once ionization and excitation equilibrium has been achieved, the stellar parameters are reported. 

\begin{figure}[!htb]
    \centering
    \includegraphics[width=\columnwidth]{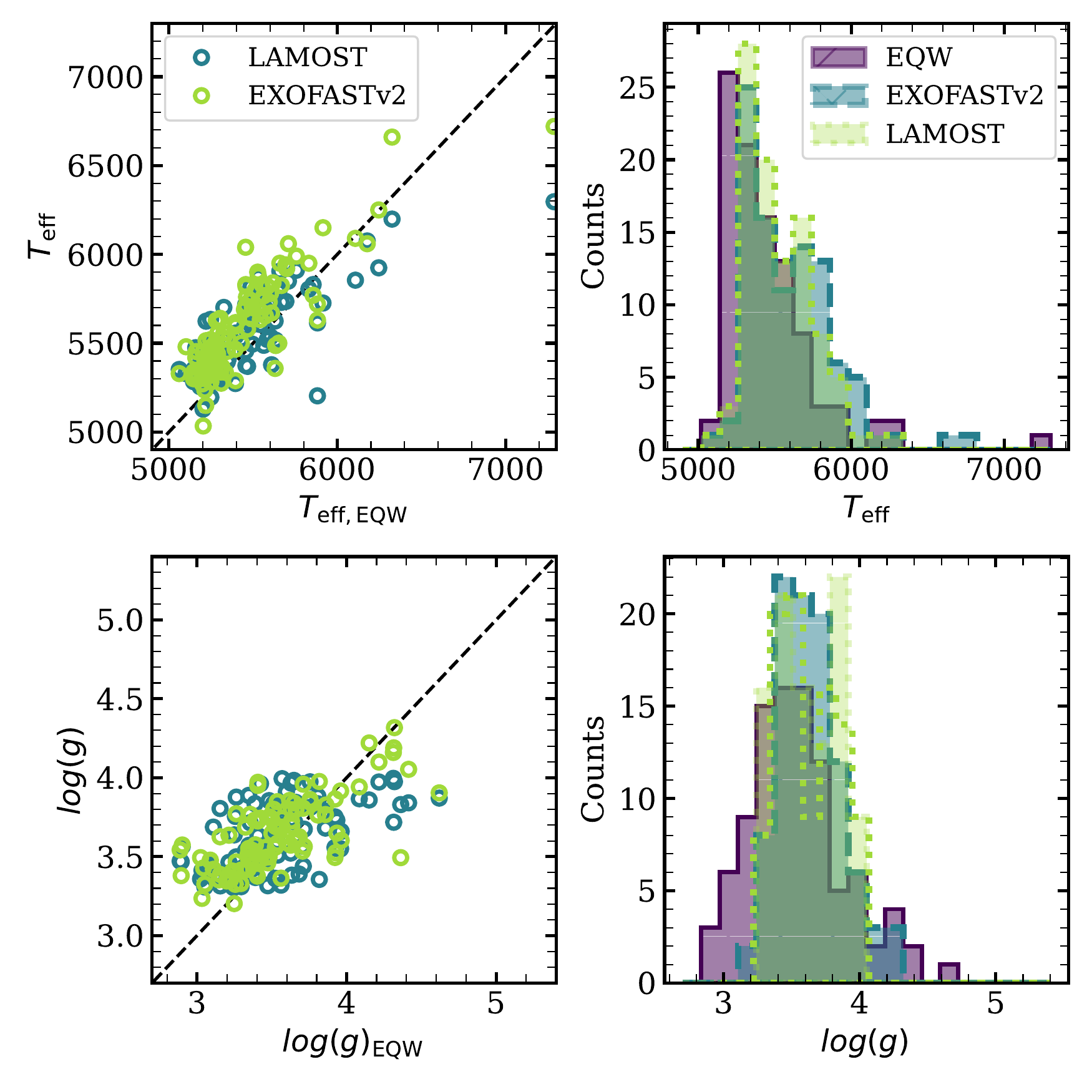}
    \caption{A comparison of stellar $\teff$ (top row) and $\logg$ (bottom row) values. Left: the $\teff$ and $\logg$ values derived from equivalent widths plotted against those from LAMOST (open blue circles) and \texttt{EXOFASTv2} (open green circles). A 1:1 relationship is plotted as the dashed line for reference. Right: Histograms of the sample's $\teff$ and $\logg$ distributions for parameters from iSpec equivalent widths (purple, this work), \texttt{EXOFASTv2} (blue), and LAMOST (green).}
    \label{fig:teff_logg}
\end{figure}

In Figure~\ref{fig:teff_logg} we plot a comparison of the spectroscopic and photometric stellar parameters. Overall, we see good agreement between the $\teff$ and $\logg$ values derived from LAMOST, \texttt{EXOFASTv2}, and iSpec. The $\teff$ derived from iSpec are slightly lower than both the LAMOST and \texttt{EXOFASTv2} values. The $\logg$ derived from iSpec span a slightly wider range of values than both the LAMOST and \texttt{EXOFASTv2} values. There are eight noticeable outliers:
\begin{enumerate}
    \item 2MASS J04163444+0812444 
    \item 2MASS J09404750+2956449
    \item 2MASS J13182135+0610410
    \item 2MASS J13515802+4002269
    \item 2MASS J03004383+0218124	
    \item 2MASS J04572322+1338535	
    \item 2MASS J16132003+0515209	
    \item 2MASS J23591304+4046438
\end{enumerate}
We derive spectroscopic $\logg>4.5$ for the first target, $\logg<3$ for $2-4$, and $\teff>6000 \kel$ for 1 and $5-8$. Because the spectroscopic $\logg$ of the first four stars push them out of our target subgiant parameter space, we exclude them from further analysis. We also exclude stars with $\teff > 6000 \kel$ to ensure a small range of stellar temperatures, but we report stellar parameters and abundances for all stars in our tables.

While the photometric stellar parameters have a more solid physical basis than the spectroscopic parameters, the adoption of these values pushes the \ion{Fe}{1} and \ion{Fe}{2} abundances further from ionization and excitation equilibrium. We proceed in our analysis employing the spectroscopic stellar parameters derived from equivalent widths, as our goal is to determine precise stellar abundances with minimal relative influence from systematic uncertainties, rather than absolute abundances. The spectroscopic $\logg$ and $\teff$ reported here should not be used to derive stellar masses or radii, but they are optimized for our abundance goals. Table~\ref{tab:params} lists the photometric and spectroscopic stellar parameters for our sample. 

\begin{deluxetable*}{cccc|cc}
\singlespace
\tablecaption{Spectroscopic and Photometric Stellar Parameters \label{tab:params}}
\tablehead{
\multicolumn{1}{c}{} & \multicolumn{3}{c|}{Spectroscopic} & \multicolumn{2}{c}{Photometric} \\
\colhead{Object} & \colhead{$\teff$ (K)} & \colhead{$\logg$} & \multicolumn{1}{c|}{$v_{\text{micro}}$ (km/s)} & \colhead{$\teff$ (K)} & \colhead{$\logg$} 
}
\startdata
2MASS J00021423+3228190 &	5482.0 &	3.66 &	1.20 &	5740 &	3.80\\
2MASS J00091409+1728209	&   5255.3 &	3.35 &	1.09 &	5480 &	3.49\\
2MASS J00383315+3433115 &	5218.7 & 	3.41 &	1.18 &	5150 &	3.97\\
2MASS J01051165+3103568 & 	5430.4 &	3.72 &	1.43 &	5500 &	3.56\\
2MASS J01591792+0212080 &	5249.2 &	3.68 &	1.07 &	5282 &	3.63\\
\enddata
\tablecomments{Spectroscopic (iSpec) and photometric (\texttt{EXOFASTv2}) stellar parameters for all stars except 2MASS J14042846+1330190 (no RV solution). Full table available online.}
\end{deluxetable*}

\subsection{Spectral Synthesis} \label{subsec:synth}

We calculate elemental abundances with an automated stellar synthesis pipeline, implemented through iSpec. We use MARCS model atmospheres \citep{gustafsson2008}, \citet{asplund2009} solar abundances, and the MOOG synthesizer \citep{sneden1973}. Synthetic abundance determination is better able to fit a range of line profiles, including blended features, than the equivalent width method. It calculates more reliable abundances for elements with few and/or weak spectral features.

In addition to the stellar parameters determined in Section~\ref{subsec:params}, the synthesis routine requires values for the macroturbulence ($v_{\text{macro}}$), rotational velocity ($v\sin(i)$), limb-darkening coefficient, and resolution ($R$). As suggested by \citet{blanco2014}, we set the limb-darkening coefficient to 0.6 and fix the resolution to that of the instrument ($R=50 000$). At this resolution, the line broadening terms $v_{\rm macro}$ and $v\sin(i)$ are degenerate.

We choose $v\sin(i)$ to be the broadening term, as done in \citet{buder2021}, and we fit spectral windows around \ion{Fe}{1} and \ion{Fe}{2} lines to synthetically determine its value. In these fits we set $v_{\text{macro}}=0$ and fix the $\teff$, $\logg$, $v_{\text{micro}}$, [M/H], and [$\alpha$/Fe] to the values fit in Section~\ref{subsec:params}. iSpec iteratively computes synthetic stellar spectra with such parameters and finds the value of $v\sin(i)$ that minimizes the residual between the observed and synthetic spectra. We report best fit $v\sin(i)$ values in Table~\ref{tab:params}.

With all stellar parameters in hand, we then determine the stellar elemental abundances with our synthesis pipeline. For each element we create spectral windows with a width of $5 \angstrom$ around each line feature. The synthesizer computes synthetic spectra for these spectral regions using all lines in the Gaia-ESO line list (with updated Mg line oscillator strengths, as noted in Section~\ref{subsec:linelist}), simultaneously fitting all line profiles for an individual element. As with the broadening velocity, iSpec iteratively calculates synthetic spectra, attempting to minimize the residual. We set the maximum number of iterations to six, as recommended by \citet{blanco2014}. The pipeline returns the best fit abundance for each element for each star.

\section{Stellar Abundances} \label{sec:abunds}

In this paper, we present the LTE abundances of Mg, Si, Ca, Ti, Na, Sc, V, Cr, Mn, Fe, Co, and Ni. Abundances of heavier elements will be presented in later work, and the impact of non-LTE (NLTE) effects on abundance scatter will be discussed in Section~\ref{subsec:nlte}. 

\subsection{Line-by-Line Mg Abundances} \label{subsec:mg}

\begin{figure}[!htb]
    \centering
    \includegraphics[width=\columnwidth]{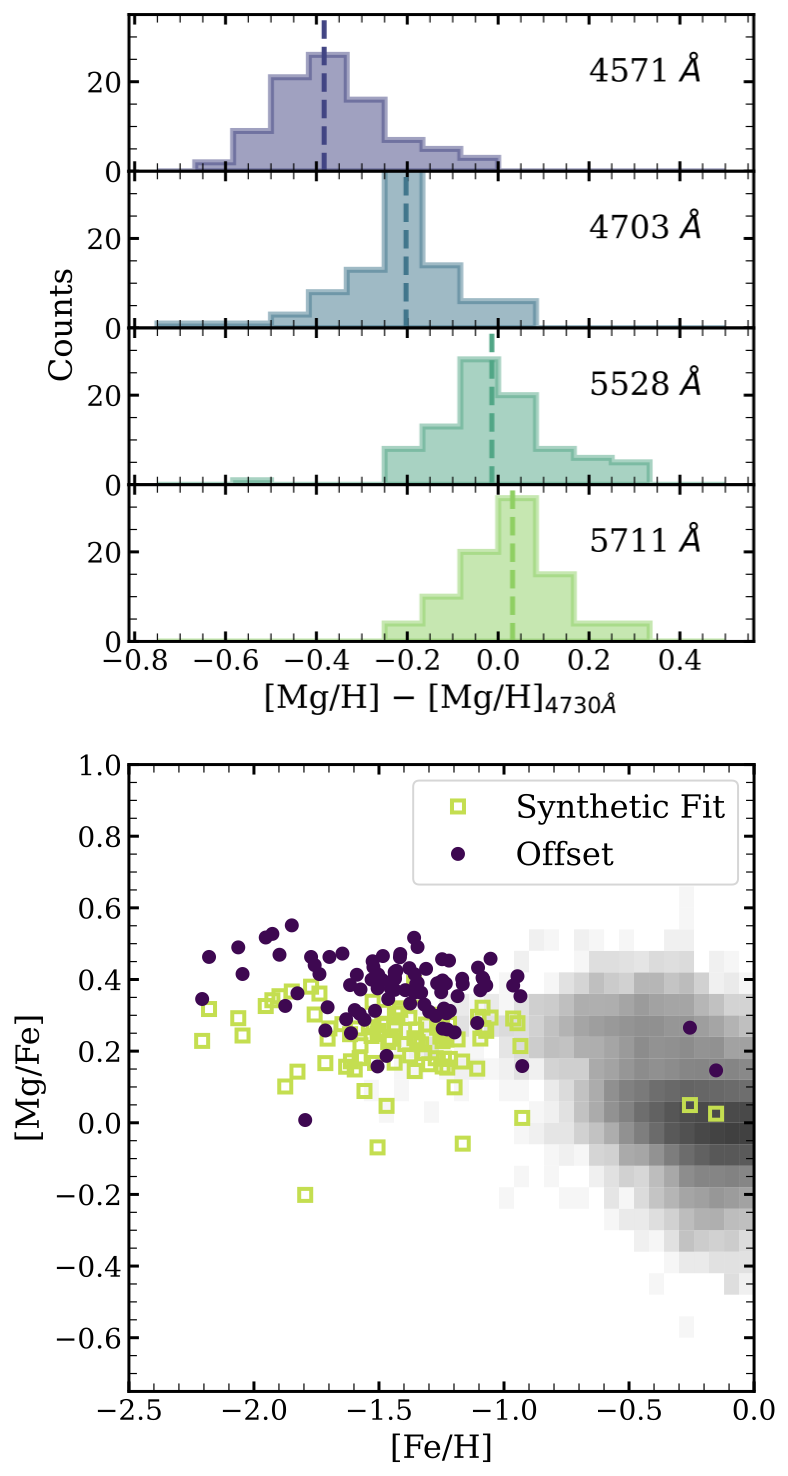}
    \caption{Top: Histograms of the line-by-line abundance deviations from the [Mg/H] value derived from the $4730\angstrom$ line. The medians of each distribution (and the applied offset) are shown as the dashed lines. Bottom: Comparison of [Mg/Fe] vs. [Fe/H] abundances using the [Mg/H] value synthetically fit to five Mg lines (open green squares) and after incorporating the median [Mg/H] of offset line-by-line abundances (dark purple filled circles). The GALAH DR3 dwarf sample is shown in greyscale. The purple circles represent the abundances used in our analysis.}
    \label{fig:mg_offset}
\end{figure}

As in prior work by our group \citep[e.g.,]{weinberg2019, griffith2019, weinberg2022}, we are interested in stellar abundance ratios relative to both Fe and Mg. While Fe is produced by both CCSN and SNIa, Mg is of pure CCSN origin \citep{andrews2017}, making it a clearer tracer of nucleosynthesis. From [X/Mg] vs. [Mg/H] abundance trends we can discern the relative contributions of prompt and delayed sources to elemental production as well as the metallicity dependence of these production mechanisms \citep{weinberg2019}. Oxygen could serve a similar purpose, as it is also purely produced by CCSN, but it is inaccessible in the wavelength ranges of our spectra.

While we would like to use Mg as our reference element, the best fit [Mg/H] values lie below the $\mgfe=0.3$ plateau at $\feh\sim-1$ that is seen in GALAH \citep[][Figure~\ref{fig:mg_offset}]{buder2021}, APOGEE \citep{jonsson2020}, and smaller abundance studies \citep[e.g.,][]{bensby2014}. To investigate what may be causing this discrepancy, we synthetically fit each Mg line individually, using the methods described in Section~\ref{subsec:synth} to determine line-by-line abundances for each star. We find that all lines do not return the same [Mg/H] for a given star and are instead systematically offset above ($4730\angstrom$, $5528\angstrom$, $5711 \angstrom$) and below ($4571\angstrom$, $4703\angstrom$) the median [Mg/H] value. The variation in line-by-line abundances could artificially inflate the abundance scatter if weak lines are undetected in a subset of stars. 

To alleviate the previously observed discrepancy and align our [Mg/Fe] values with the abundance plateau, we apply line-by-line offset relative to the $4730\angstrom$ Mg line and adopt the median [Mg/H] abundance of the five updated line values. We offset to the $4730\angstrom$ line as it is well fit by the synthesis routine and returns [Mg/H] abundance values that place $\feh\approx -1$ stars at $\mgh\approx0.3$. In Figure~\ref{fig:mg_offset} we show the distribution of deviations between the line-by-line [Mg/H] values and the [Mg/H] fit to the $4730\angstrom$ line. The median deviation is shown in each panel as the dashed line. The $4571\angstrom$ and $4703\angstrom$ lines deviate most strongly, with median offsets from the $4730\angstrom$ line of -0.38 dex and -0.20 dex, respectively. We apply global offsets to each line's [Mg/H] abundance to shift the medians of these distributions to zero. Offsetting in such a way provides uniform [Mg/H] abundances for each star without artificially reducing the population scatter. Figure~\ref{fig:mg_offset} shows the [Mg/Fe] values fit by the synthesis routine compared to the adopted offset values. The new values line up with the plateau observed in the GALAH DR3 dwarf sample.

Although this offsetting procedure is somewhat unsatisfying, we consider it the best way to mitigate differential scatter in Mg abundances that would arise if the lines have different effective weights in stars of our sample (e.g., because of different $\teff$). The $\log(gf)$ values are uncertain, and one can view our procedure as an astrophysical relative calibration of these values, though the abundance differences might arise from some other effect. The overall normalization of [Mg/Fe] is set by our choice to match GALAH at $\feh\approx-1$ and should not be considered an independent measure of the plateau level.

We note that \citet{bergemann2017} also find that the LTE and NLTE Mg abundances derived for the $4571\angstrom$ line in metal-poor dwarf HD 84937 (and other stars) are 0.15 to 0.22 dex lower than the values derived from other lines, including those at $5528\angstrom$ and $5711 \angstrom$. They are unsure if line weakness or atmospheric inhomogeneities cause the inconsistency. Recent work by \citet{lind2022} finds similar discrepancies in the optical Mg lines of HD 84937, with the LTE and NLTE abundances of the $4571\angstrom$ line systematically 0.2 to 0.4 dex lower than those from the other Mg lines used in our work.

We performed a similar check on other elements to evaluate if discrepant lines skew the elemental abundance. Other elements with few lines (Na, Sc, and Co) do not exhibit such large systematic line offsets. For elements with $>15$ lines, we remove discrepant lines from our list, as mentioned in Section~\ref{subsec:linelist}.

\subsection{Abundance Trends \& Comparison with GALAH} \label{subsec:ab_trends}

\begin{figure*}[!htb]
    \centering
    \includegraphics[width=\textwidth]{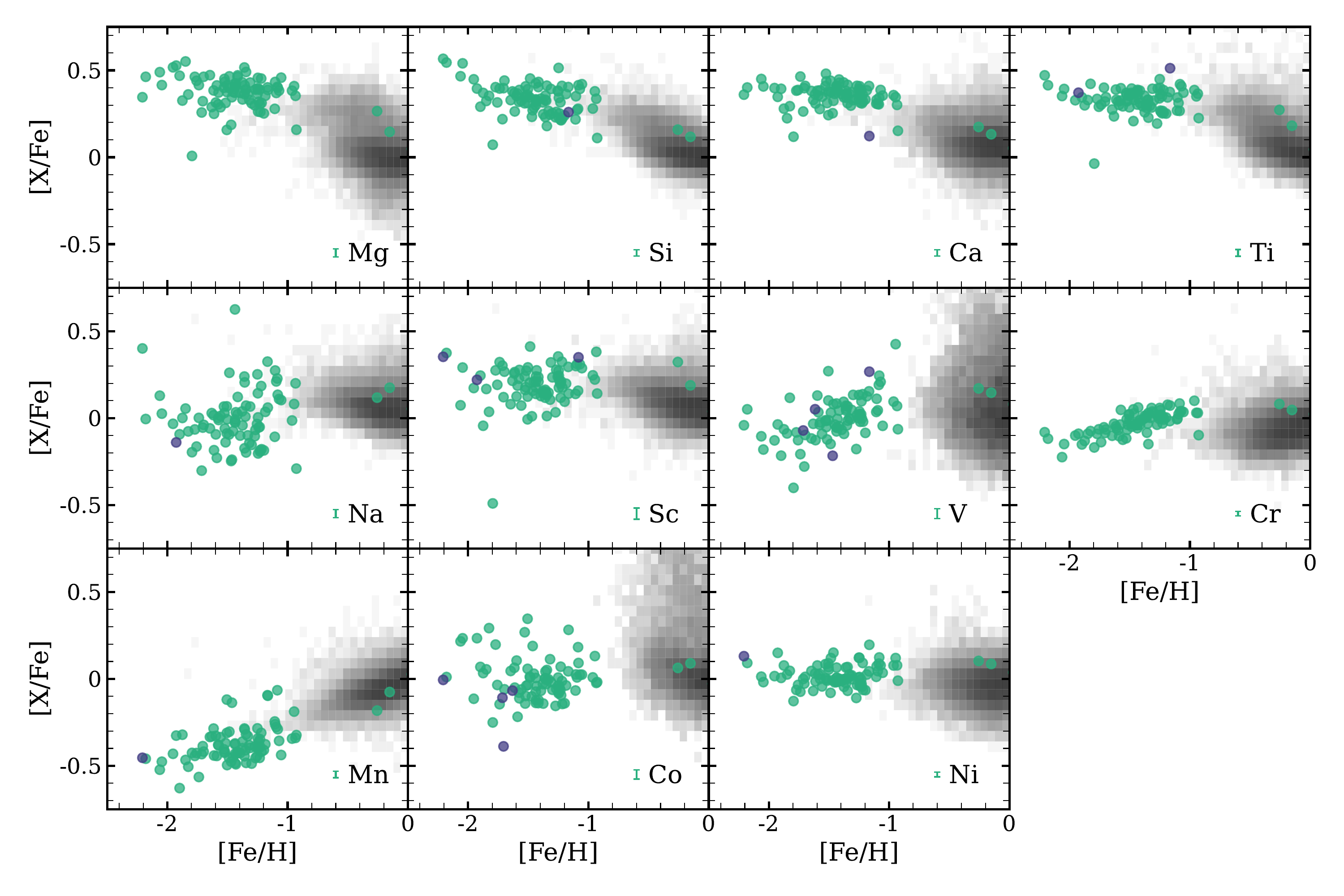}
    \caption{[X/Fe] vs. [Fe/H] abundances derived in this work. The median [X/Fe] observational error ($\sigma_{\rm phot, \, med}$) is shown in each panel as the error bar to the left the of element symbol. Stars with $\sigma_{\rm phot}$ less than 3 times $\sigma_{\rm phot, \, med}$ are shown in teal. Those with $\sigma_{\rm phot}$ greater than 3 times $\sigma_{\rm phot, \, med}$ are shown in purple. A 2D density plot of GALAH DR3 dwarfs is shown in greyscale for comparison. }
    \label{fig:x_fe}
\end{figure*}

\begin{figure*}[!htb]
    \centering
    \includegraphics[width=\textwidth]{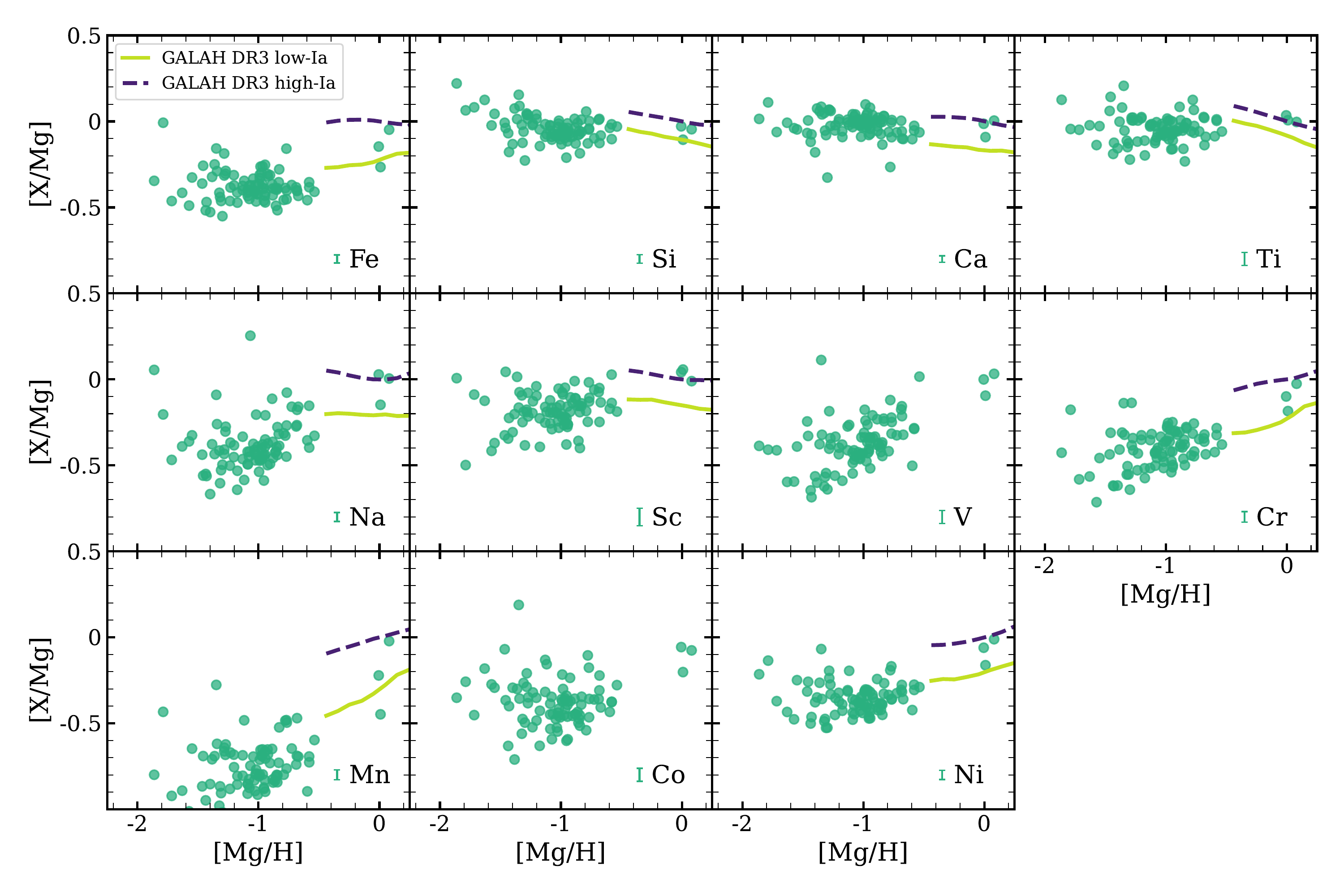}
    \caption{[X/Mg] vs. [Mg/H] abundances derived in this work (green circles). The typical [X/Mg] observational error is shown in each panel as the error bar to the left of the element symbol. The median high-Ia (purple dashed line) and low-Ia (lime green solid line) trends from \citep[][GALAH DR3]{griffith2022} are included for comparison.}
    \label{fig:x_mg}
\end{figure*}

We report the derived stellar abundances in Table~\ref{tab:abunds} and show the [X/Fe] vs. [Fe/H] as well as [X/Mg] vs. [Mg/H] abundance trends in Figures~\ref{fig:x_fe} and~\ref{fig:x_mg}, respectively. In the former plot, we include 2D density histograms of the GALAH DR3 population with unflagged stellar parameters, $4200<\teff<6700$, and $3.3<\logg<4$. While we include GALAH DR3 abundance trends for all elements, \citet{buder2021} caution against the use of V and Co as their line features are frequently blended. In Figure~\ref{fig:x_mg} we include median high-Ia and low-Ia [X/Mg] vs. [Mg/H] abundance trends from \citep[][using GALAH DR3 data]{griffith2022}. We note that these trends have offsets of up to 0.05 dex applied to ensure that the median high-Ia sequence passes through $\xmg=0$ at $\mgh=0$. We do not apply such offsets to our data.

\begin{deluxetable*}{cccccccccccccc}
\singlespace
\tablecaption{Stellar Abundances \label{tab:abunds}}
\tablehead{
\colhead{Object} & \colhead{[Fe/H]} & \colhead{[Mg/H]} & \colhead{[Si/H]} & \colhead{[Ca/H]} & \colhead{[Ti/H]} & \colhead{[Na/H]} & \colhead{[Sc/H]} & \colhead{[V/H]} & \colhead{[Cr/H]} & \colhead{[Mn/H]} & \colhead{[Co/H]} & \colhead{[Ni/H]}
}
\startdata
2MASS J00021423+3228190	& -1.48	& -1.02	& -1.03	& -1.11	& -1.09	& -1.22	& -1.07	& -1.45	& -1.53	& -1.79	& -1.48	& -1.36 \\
2MASS J00091409+1728209	& -2.05	& -1.62	& -1.51	& -1.64	& -1.65	& -2.02	& -1.75	& -2.23	& -2.20	& -2.52	& -1.81	& -2.06 \\
2MASS J00383315+3433115	& -1.42	& -0.95	& -1.01	& -0.97	& -1.03	& -1.39	& -1.18	& -1.37	& -1.43	& -1.84	& -1.56	& -1.43 \\
2MASS J01051165+3103568	& -1.52	& -1.20	& -1.16	& -1.20	& -1.19	& -1.66	& -1.24	& -1.64	& -1.55	& -1.89	& -1.56	& -1.53 \\
2MASS J01591792+0212080	& -1.07	& -0.68	& -0.68	& -0.68	& -0.66	& -0.96	& -0.76	& -0.86	& -1.03	& -1.43	& -1.05	& -0.99 \\
\enddata
\tablecomments{Full table available online}
\end{deluxetable*}

As discussed in Section~\ref{subsec:select}, we select target stars based on metallicity with selection criteria of $-2.1 \leq\feh\leq -1.0$. While LAMOST metallicities placed all targets within range, we find [Fe/H] $> -0.9$ for three targets: 
\begin{enumerate}
    \item 2MASS J08452357+1758115 ($\feh = 0.03$)
    \item 2MASS J12172774-0154034 ($\feh -0.15$)
    \item UCAC4 473-046951 ($\feh = -0.26$).
\end{enumerate}
While we report the abundances for these stars, we drop them from further analysis as they fall outside of our desired metallicity range. We proceed with 86 metal-poor subgiant stars. Overall, the abundance trends of our low-metallicity stars appear as an extension of the GALAH DR3 sample. We see smaller scatter in the [X/Fe] trends for Mg, Si, Ca, Ti, Cr, Mn, and Ni, and larger scatter for Na, V, and Co.

Amongst the other $\alpha$-elements, dominantly produced by CCSN \citep{andrews2017}, we see that the [Mg/Fe] values align with the low-Ia plateau at [Fe/H]$=-1$ due to our applied offsets (Section~\ref{subsec:mg}). We find that as the metallicity decreases, the [Mg/Fe] values continue to increase, approaching $\sim0.5$ dex at $\feh=-2.0$, in agreement with results from the H3 survey \citep{conroy2022} and dwarf satellites \citep{kirby2011}. The [Si/Mg] and [Ca/Mg] abundance trends also align with the low-Ia median trends from GALAH DR3. The [Si/Mg] and [Si/Fe] abundances decrease with increasing metallicity, while the [Ca/Mg] and [Ca/Fe] trends show a weaker metallicity dependence and plateau near [Ca/Mg]$=0$ and [Ca/Fe]$=0.35$. The Ti trend appears most similar to Ca. The [Ti/Fe] abundances are metallicity independent, with [Ti/Fe]$\approx 0.4$. The [Ti/Mg] abundances are also metallicity independent---in contrast to the negative metallicity dependence of the [Ti/Mg] trends in GALAH DR3 \citep{griffith2022}.

Light odd-$Z$-elements Na and Sc show more scatter than the $\alpha$-elements. Na abundances fall below an extrapolation of the GALAH DR3 trends and display a positive metallicity gradient in [Na/Mg]. While the [Sc/Fe] abundances appear to scatter around a constant value, we observe a weak positive metallicity dependence in [Sc/Mg], different from the weak negative trend at higher metallicity in the GALAH DR3 low-Ia median. 

V and Co display the largest scatter of the Fe-peak elements, with scatter noticeably increasing at lower metallicity. The [Co/Fe] and [V/Fe] trends appear consistent with the GALAH DR3 abundance distribution, though these should be interpreted cautiously. The abundances for Fe-peak elements included in \citep{griffith2022} (Fe, Cr, Mn, Ni) appear as continuations of the low-Ia [X/Mg] abundance trends from GALAH DR3. We observe a positive metallicity dependence in [Cr/Fe] and [Mn/Fe], as well as a near constant [Ni/Fe] abundance of $\sim0$. The abundance scatter in [X/Mg] for these three elements is noticeably larger than the scatter in [X/Fe].

\section{Abundance Scatter} \label{sec:scatter}

Our primary observational goal is to measure the intrinsic abundance scatter. This requires assessing the observational noise carefully. We have constructed our sample and analysis methods to minimize differential systematic scatter and hereafter assume it is negligible, though we cannot conclusively demonstrate this. Our measured abundances do, however, include photon-noise scatter. We discuss photon-noise scatter in Section~\ref{subsec:photon}, the potential impact of NLTE corrections in Section~\ref{subsec:nlte}, and estimations of the intrinsic scatter in Section~\ref{subsec:intrin}. We explore the impact of kinematic subsamples in Section~\ref{sec:kin}, and turn to possible interpretations in Section~\ref{sec:discussion}. Although Mg is a physically simpler reference element than Fe, our data provide significantly better Fe measurements, and Mg is subject to systematic uncertainties as discussed in Section~\ref{subsec:mg}. We therefore focus primarily on [X/Fe] scatter while presenting some results for [X/Mg].

\subsection{Photon-Noise} \label{subsec:photon}

Photon-noise is one extrinsic source of scatter in our observed abundances. For our choices of resolution and targeted S/N, we expect the photon-noise contributions to the errors to be $<0.04$ dex. We must determine the magnitude of the photon-noise in order to constrain the intrinsic scatter. In their review article, \citet{jofre2019} discuss multiple methods to evaluate random and systematic uncertainties used in the literature. To quantify abundance uncertainties, some use repeat observations of the same star to assess the variation in parameters \citep[e.g.][]{hayes2022}, observe a reference star with well known abundances, or observe homogeneous cluster members. Variations in the determined stellar abundances indicate the level of random error from photon-noise and data reduction. While we do not have repeat observations of any of our stars, nor do we have spectra of reference stars or cluster members taken with PEPSI, we simulate multiple observations. Using the error spectrum calculated in the PEPSI reduction pipeline, we create 10 realizations of each star's spectrum. At each wavelength, we vary the flux around the observed value by adding fluctuations drawn from a Gaussian distribution with standard deviation equal to the error value at that wavelength. We repeat RV correction, continuum normalization, parameter determination, and spectral synthesis as described in Sections~\ref{subsec:spec_reduc},~\ref{subsec:params}, and~\ref{subsec:synth}. We recalculate the [Mg/H] abundance for each star by applying the line-by-line offsets from Section~\ref{subsec:mg} and adopting the median value of the five offset lines.

This ``wiggling'' procedure produces 10 abundances for each element for each star. We calculate the standard deviation of the stellar [X/Fe] and [X/Mg] abundances and adopt this value as the photon-noise contributed error for each star ($\sigma_{\rm phot}$). Tables~\ref{tab:scat_xfe} and~\ref{tab:scat_xmg} list the $\sigma_{\rm phot, \, med}$ in [X/Fe] and [X/Mg], respectively. We also show $\sigma_{\rm phot, \, med}$ as error bars in Figures~\ref{fig:x_fe} and~\ref{fig:x_mg}. We find that $\sigma_{\rm phot, \, med} <0.04$ dex for all elements in [X/Fe], consistent with our observational design. The $\sigma_{\rm phot, \, med}$ is $<0.05$ dex for all elements but Sc (median $\sigma_{\rm phot}=0.052$) in [X/Mg]. [Ni/Fe] and [Cr/Fe] have the smallest photon-noise scatter, with median $\sigma_{\rm phot}\approx 0.013$, followed by [Ca/Mg], [Ca/Fe], and [Si/Fe] with median $\sigma_{\rm phot}\approx 0.018$. In Figure~\ref{fig:x_fe}, we color individual stars purple if $\sigma_{\rm phot, \, *}$ is greater than $3\times \sigma_{\rm phot, \, med}$. In most cases, the abundances of these stars align with rest of the population, and the same stars have high photon-noise errors across all panels. We inspect the spectra of each star and do not observe any anomalies. Stars with higher photon-noise tend to have a lower SNR.

The advantage of this procedure over simpler error estimates is that it accounts for all of the ways that photon-noise impacts a given abundance measurement, including the impact of $\teff$ and $\logg$ uncertainties and blended features---achieved by some through further error analysis such as evaluating the impact of varying the stellar parameters on the calculated abundances \citep[e.g.,][]{roederer2014, jofre2019}.

\subsection{NLTE Effects} \label{subsec:nlte}

In addition to photon-noise and systematic uncertainties in analysis, NLTE effects may be another source of abundance uncertainty without careful selection of lines that are less sensitive to these effects \citep[e.g.][]{jofre2019}. It is well understood that NLTE effects can impact the slope and zero point of abundance trends. We are interested in their impact on the overall abundance scatter. While iSpec does not allow for the integration of NLTE grids, the Max Planck Institute for Astronomy (MPIA) has published look up tables\footnote{\url{https://nlte.mpia.de/gui-siuAC_secE.php}} of NLTE abundance corrections for \ion{Mg}{1}, \ion{Si}{1}, \ion{Ca}{1}, \ion{Ti}{1}, \ion{Ti}{2}, \ion{Cr}{1}, \ion{Mn}{2}, and \ion{Co}{1} \citep{mashonkina2007, bergemann2008, bergemann2010a, bergemann2010b, bergemann2011, bergemann2013,  bergemann2017}.

To evaluate the potential impact of NLTE effects on our abundance trends and scatter, we calculate corrections using plane parallel MAFAGS-OS atmospheric models \citep{grupp2004a, grupp2004b} and the MPIA datasets for all available lines in our line list. For some lines, particularly in low metallicity stars, corrections are not available as the line is too weak. For each element for each star, we calculate the median NLTE correction of all lines. To evaluate the impact of NLTE corrections on the scatter, we offset each star's abundance by the median NLTE line correction. We find significant changes to the abundances (median $\Delta \xfe$ and $\Delta \xmg > 0.1$ dex) of Co, Mn, and Cr. The [Co/Fe] and [Co/Mg] abundances increase by $\sim 0.3$ dex at all metallicities, with the [Co/Mg] trend gaining a slight negative metallicity dependence.  The metallicity dependence of the [Mn/Fe] and [Mn/Mg] trends is unaffected by the NLTE offsets, but the stellar abundances increase by $\sim 0.1$ dex. Finally the NLTE corrections to [Cr/Fe] and [Cr/Mg] have a strong correlation with metallicity and cause the observed positive metallicity dependence to flatten, with the [Cr/Fe] plateauing at $-0.2$ dex and the [Cr/Mg] plateauing at $0.2$ dex. We note that the NLTE corrections are only for \ion{Cr}{1}, while iSpec determines the Cr abundance from a mix of \ion{Cr}{1} (29) and \ion{Cr}{2} (5) windows.

We calculate the RMS scatter about a linear model (see Section~\ref{subsec:intrin}) in [X/Fe] vs. [Fe/H] and [X/Mg] vs [Mg/H] for abundances with and without NLTE corrections. Although the NLTE corrections can change the shape of the abundance pattern as described above, we find that they \textit{do not} cause significant changes to the abundance scatter. We find that the RMS scatter between the LTE and NLTE abundance changes by $<0.02$ dex for all elements with available NLTE corrections. Further, the variation is less than 0.01 dex for [Mg/Fe], [Si/Fe], [Ca/Fe], [Ti/Fe], [Mn/Fe], [Cr/Fe], [Fe/Mg], [Si/Mg], [Ca/Mg], [Ti/Mg], and [Mn/Mg]. As the NLTE corrections cause scatter consistent with or smaller than random photon-noise for all elements, we continue in our analysis without their inclusion.

\subsection{Total and Intrinsic Scatter} \label{subsec:intrin}

As a measurement of the observed abundance scatter, we fit the [X/Fe] vs. [Fe/H] and [X/Mg] vs. [Mg/H] trends with a multi-parameter model and calculate the RMS deviations from the predictions. Compared to the minimal approach of computing scatter in bins of [Fe/H] or [Mg/H], our method removes the impact of trends within a bin and allows us to use the entire sample to calculate the scatter. We start with a two-parameter predictive model for abundance trends relative to both Fe and Mg: 
\begin{equation}
    \text{[X/Fe]}_{\text{pred}} = \text{[X/Fe]}_{\text{cen}} + Q_{\rm Fe}^{\rm X}(\feh - \text{[Fe/H]}_{\rm cen}),
\end{equation}
where $\text{[Fe/H]}_{\rm cen}$ is the sample median value of [Fe/H] and $Q_{\rm Fe}^X$ and $\text{[X/Fe]}_{\rm cen}$ are free, and 
\begin{equation}
    \text{[X/Mg]}_{\text{pred}} = \text{[X/Mg]}_{\text{cen}} + Q_{\rm Mg}^{\rm X}(\mgh - \text{[Mg/H]}_{\rm cen}),
\end{equation}
where $\text{[Mg/H]}_{\rm cen}$ is the sample median value of [Mg/H] and $Q_{\rm Mg}^X$ and $\text{[X/Mg]}_{\rm cen}$ are free. For each model, we fit for the abundance trends for both parameters simultaneously by minimizing the $\chi^2$ value with the Nelder-Mead minimization technique. These models fit a metallicity dependent linear trend to the data. We include the [Fe/H]$_{\rm cen}$ and [Mg/H]$_{\rm cen}$ to allow an overall offset relative to solar abundances, so that the slopes ($Q_{\rm Fe}^{\rm X}$ and $Q_{\rm Mg}^{\rm X}$) are the slopes within our sample. We calculate the RMS scatter between the observed and modeled abundances, taking this to be a measure of the abundance dispersion including contributions from observational error. 

To convert to intrinsic scatter ($\sigma_{\rm intrin}$), for each star we subtract an estimate of the photon-noise ($\sigma_{\rm phot}$) from the RMS scatter in quadrature such that, 
\begin{equation}
    \sigma_{\rm intrin} = \sqrt{\rm RMS^2 - \sigma_{\rm{phot}}^2}.
\end{equation}
Rather than use the median photon-noise contributions to scatter in [X/Fe] or [X/Mg] as $\sigma_{\rm phot}$, we estimate the photon-noise about the two-parameter model. For this estimate, we place each star at its model predicted value, add a random error drawn from a Gaussian distribution with standard deviation equal to the standard deviation in [X/Fe] or [X/Mg] and [Fe/H] or [Mg/H] for that star (Section~\ref{subsec:photon}), then calculate the RMS deviation between the model abundances and model abundances with the photon-noise scatter. We refer to this quantity as $\sigma_{\rm phot, \, 2-param}$. It accounts for the fact that stars have a range of photon-noise errors and that noise in [Fe/H] and [$\alpha$/Fe] also affects the RMS scatter about predicted model values. The values of $\sigma_{\rm phot, \, 2-param}$ are consistently larger than $\sigma_{\rm phot, \, med}$ (see Tables~\ref{tab:scat_xfe} and~\ref{tab:scat_xmg}), primarily because stars with larger photon-noise contribute more to the RMS scatter.

\begin{figure*}[!htb]
    \centering
    \includegraphics[width=\textwidth]{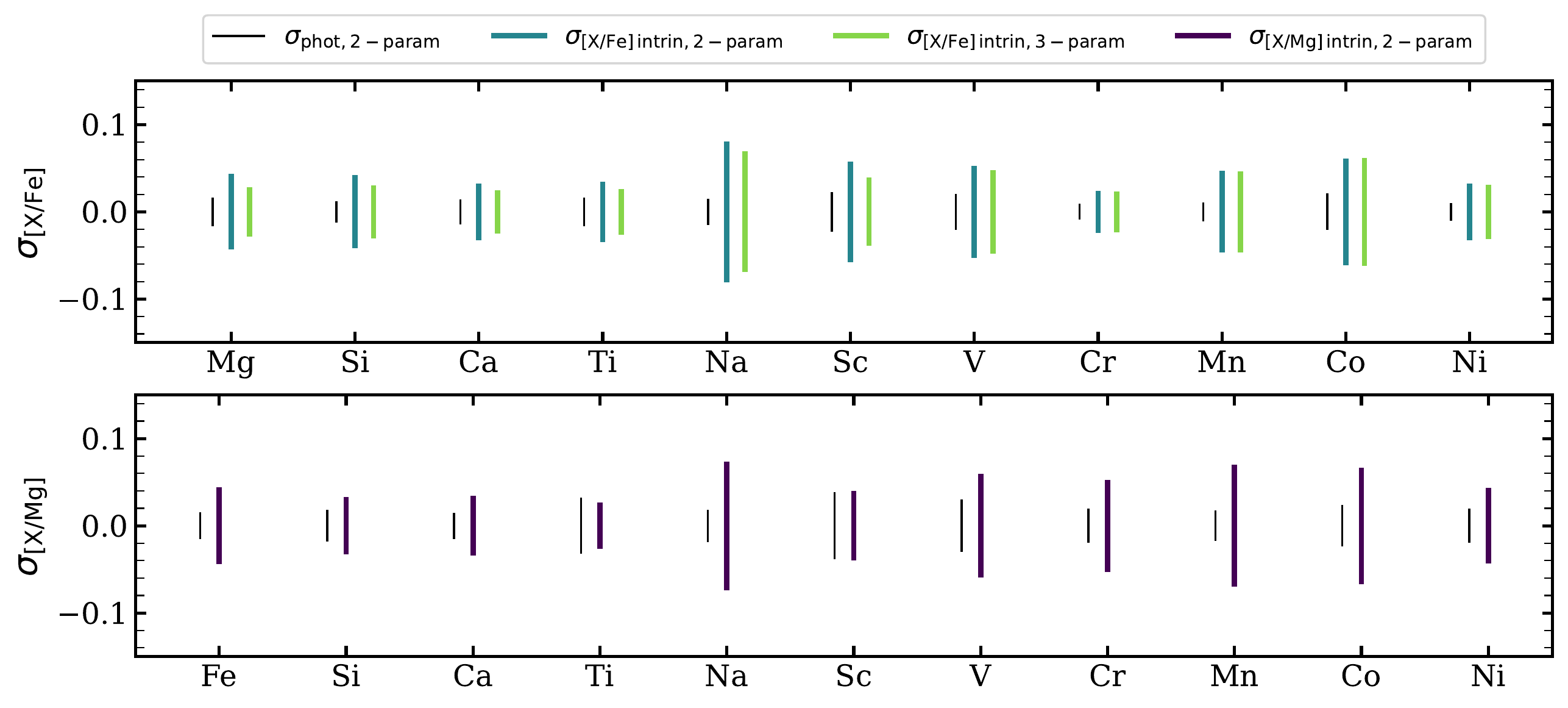}
    \caption{Top: photon-noise contributed scatter (black thin line) and intrinsic scatter about the two-parameter model (blue), and the three-parameter model (light green) in [X/Fe] for each element. Bottom: photon-noise contributed scatter (black thin line) and intrinsic scatter about the two-parameter model (dark purple) in [X/Mg]. In both panels the length of the line corresponds to the magnitude of the scatter.}
    \label{fig:rms_2v3}
\end{figure*}

\begin{figure*}[!htb]
    \centering
    \includegraphics[width=\textwidth]{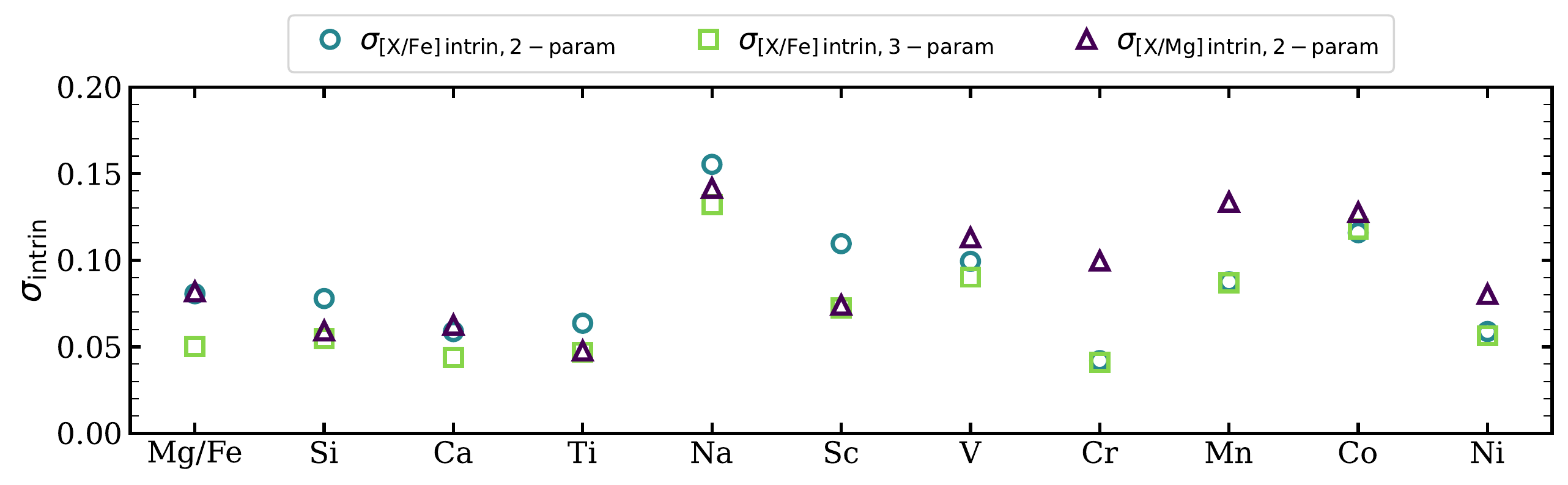}
    \caption{Intrinsic scatter about the two and three-parameter models, as shown in Figure~\ref{fig:rms_2v3}, but now comparing scatter in [X/Fe] and [X/Mg]. We show $\sigma_{\rm [X/Fe] \, intrin, \, 2-param}$ (blue circles), $\sigma_{\rm [X/Fe] \, intrin, \, 3-param}$ (light green squares), and  $\sigma_{\rm [X/Mg] \, intrin, \, 2-param}$ (dark purple triangles).}
    \label{fig:rms_2v3_scatter}
\end{figure*}

We find intrinsic scatter of $0.04-0.15$ dex in [X/Fe] and [X/Mg] for all elements. The magnitude of the intrinsic scatter is greater than $\sigma_{\rm phot, \, 2-param}$ for all elements, usually by a factor of two or more, an indication that we are robustly measuring true abundance variations within our population. The magnitude of two-parameter photon-noise is comparable to the intrinsic scatter for [Ti/Mg] and [Sc/Mg]. We plot the magnitude of the intrinsic scatter about the [X/Fe] (top) and [X/Mg] (bottom) two-parameter models for each element in Figure~\ref{fig:rms_2v3} as the blue lines. A direct comparison of [X/Fe] and [X/Mg] intrinsic scatter can be found in Figure~\ref{fig:rms_2v3_scatter}.

\begin{deluxetable*}{lrrrrrrrr}
\tablecaption{Scatter in metal-poor subgiants in [X/Fe]. \label{tab:scat_xfe}}
\tablehead{
\colhead{} & \colhead{$\sigma_{\rm phot, \,med}$} & \colhead{$\sigma_{\rm phot, \,2-param}$} & \colhead{$\sigma_{\rm phot, \,3-param}$} & \colhead{$\sigma_{\rm intrin, \,2-param}$} & \colhead{$\sigma_{\rm intrin, \,3-param}$} & \colhead{$\sigma_{N=10}$} & \colhead{$\sigma_{N=50}$} & \colhead{$\sigma_{N=100}$}  
}
\startdata
Mg & 0.024 & 0.030 & 0.035 & 0.081 & 0.050 & 0.237 & 0.077 & 0.051 \\
Si & 0.018 & 0.022 & 0.024 & 0.078 & 0.055 & 0.253 & 0.096 & 0.066 \\
Ca & 0.019 & 0.026 & 0.021 & 0.059 & 0.044 & 0.236 & 0.107 & 0.079 \\
Ti & 0.020 & 0.030 & 0.032 & 0.064 & 0.047 & 0.169 & 0.034 & 0.023 \\
Na & 0.023 & 0.028 & 0.028 & 0.155 & 0.132 & 0.364 & 0.133 & 0.094 \\
Sc & 0.034 & 0.043 & 0.056 & 0.110 & 0.072 & 0.411 & 0.180 & 0.129 \\
V  & 0.027 & 0.039 & 0.045 & 0.099 & 0.090 & 0.225 & 0.100 & 0.057 \\
Cr & 0.013 & 0.016 & 0.016 & 0.042 & 0.041 & 0.166 & 0.044 & 0.028 \\
Mn & 0.019 & 0.019 & 0.022 & 0.088 & 0.087 & 0.292 & 0.181 & 0.124 \\
Co & 0.027 & 0.039 & 0.031 & 0.116 & 0.118 & 0.145 & 0.049 & 0.034 \\
Ni & 0.013 & 0.018 & 0.013 & 0.059 & 0.056 & 0.058 & 0.018 & 0.013 \\
\enddata
\tablecomments{Median photon-noise scatter ($\sigma_{\rm phot, \, med}$), two-parameter model photon-noise scatter ($\sigma_{\rm phot, \,2-param}$), three-parameter model photon-noise scatter ($\sigma_{\rm phot, \,3-param}$), intrinsic scatter relative to the two-parameter model ($\sigma_{\rm intrin, \, 2-param}$), intrinsic scatter relative to the three-parameter model ($\sigma_{\rm intrin, \, 3-param}$), and standard deviation of 1000 independent draws of CCSN yields for $N_{\rm CCSN}=10$ ($\sigma_{N=10}$), $N_{\rm CCSN}=50$ ($\sigma_{N=50}$), and $N_{\rm CCSN}=100$ ($\sigma_{N=100}$) in [X/Fe].}
\end{deluxetable*}

\begin{deluxetable*}{lrrrrrr}
\tablecaption{Scatter in metal-poor subgiants in [X/Mg]. \label{tab:scat_xmg}}
\tablehead{
\colhead{} & \colhead{$\sigma_{\rm phot, \, med}$} & \colhead{$\sigma_{\rm phot, \,2-param}$} & \colhead{$\sigma_{\rm intrin, \, 2-param}$} & \colhead{$\sigma_{N=10}$} & \colhead{$\sigma_{N=50}$} & \colhead{$\sigma_{N=100}$}  
}
\startdata
Fe & 0.024 & 0.028 & 0.082 & 0.237 & 0.077 & 0.051   \\
Si & 0.025 & 0.034 & 0.059 & 0.233 & 0.094 & 0.063  \\
Ca & 0.019 & 0.028 & 0.063 & 0.210 & 0.097 & 0.070  \\
Ti & 0.038 & 0.062 & 0.048 & 0.239 & 0.071 & 0.048  \\
Na & 0.026 & 0.035 & 0.142 & 0.294 & 0.110 & 0.078   \\
Sc & 0.052 & 0.075 & 0.074 & 0.327 & 0.155 & 0.112   \\
V  & 0.038 & 0.057 & 0.113 & 0.254 & 0.108 & 0.063   \\
Cr & 0.030 & 0.037 & 0.100 & 0.254 & 0.079 & 0.053   \\
Mn & 0.030 & 0.033 & 0.134 & 0.305 & 0.177 & 0.120   \\
Co & 0.038 & 0.045 & 0.128 & 0.210 & 0.064 & 0.043   \\
Ni & 0.027 & 0.036 & 0.080 & 0.226 & 0.069 & 0.045   \\
\enddata
\tablecomments{Median photon-noise scatter ($\sigma_{\rm phot, \, med}$), two-parameter model photon-noise scatter ($\sigma_{\rm phot, \,2-param}$), intrinsic scatter relative to the two-parameter model ($\sigma_{\rm intrin, \, 2-param}$), and standard deviation of 1000 independent draws of CCSN yields for $N_{\rm CCSN}=10$ ($\sigma_{N=10}$), $N_{\rm CCSN}=50$ ($\sigma_{N=50}$), and $N_{\rm CCSN}=100$ ($\sigma_{N=100}$) in [X/Mg].}
\end{deluxetable*}

For most elements, we find similar levels of intrinsic scatter about the two-parameter model in [X/Fe] and [X/Mg]. The two clear exceptions are Cr and Mn, for which the [X/Fe] scatter is smaller by factors of 2.4 and 1.4, respectively. For other elements the differences are a factor of $\sim1.3$ (Si, Sc, Ni) or smaller. The elements exhibiting $\geq0.1$ dex intrinsic scatter in [X/Fe] are Na, Sc, V, and Co. For [X/Mg], Mn and Cr also exhibit scatter $\geq 0.1$ dex.

Figure~\ref{fig:x_fe} shows a few noticeable outliers from the abundance trends of the full population. The outlier stars do not show obvious signs of erroneous abundance measurements for some elements. To gauge their impact on our results, we recalculate the intrinsic scatter after clipping stars where [X/Fe] differs from the median value by $>3\sigma$. The $3\sigma$ clipping decreases the intrinsic scatter by a factor $1.1-1.3$ for Mg, Ca, Ti, Na, Sc, and V. Because these changes are fairly small and the measurements do not appear to be inaccurate, we proceed without the outlier clipping for our analysis.

As many elements studied here have contributions from delayed sources such as SNIa and AGB stars \citep[especially at higher metallicities, e.g.,][]{andrews2017, griffith2019, griffith2022}, we introduce a more complex three-parameter model to account for delayed contributions: 
\begin{equation} \label{eq:xfe_pred}
    \text{[X/Fe]}_{\text{pred}} = \text{[X/Fe]}_{\text{cen}} + Q_{\rm Fe}^{\rm X}(\feh - \text{[Fe/H]}_{\rm cen}) + Q_{\alpha}^{\rm X}[\alpha/\text{Fe}],
\end{equation}
where $\text{[Fe/H]}_{\rm cen}$ is the sample median value of [Fe/H], [$\alpha$/Fe] is the mean of the star's [Mg/Fe], [Ca/Fe], and [Si/Fe], and $Q_{\rm Fe}^X$, $\text{[X/Fe]}_{\rm cen}$, and $Q_{\alpha}^{\rm X}$ are free. When fitting Mg, Ca, or Si, we utilize the remaining two $\alpha$-elements in our calculating of [$\alpha$/Fe] so that we do not artificially reduce the scatter by fitting to it. Relative to the two-parameter model, the three-parameter model allows an additional dependence on [$\alpha$/Fe]; note that the typical value of [X/Fe] near the center of the sample is [X/Fe]$_{\rm cen}+Q_{\alpha}^{\rm X}[\alpha/\text{Fe}]_{\rm cen}$. For a pure $\alpha$ element, we expect $\xfe = [\alpha/\text{Fe}]$ and thus $Q_{\alpha}^{\rm X}=1$ and $Q_{\rm Fe}^{\rm X} = \xfe_{\rm cen} = 0$. For an element with the same relative CCSN and SNIa yields as Fe, we expect $\xfe=0$ and thus all three parameters equal zero. For an element with intermediate SNIa contribution, we expect $0<Q_{\alpha}^{\rm X}<1$.

We fit each abundance distribution with the three-parameter model to derive the values of [X/Fe]$_{\rm cen}$, $Q_{\rm Fe}^{\rm X}$, and $Q_{\alpha}^{\rm X}$ for each element, again minimizing the $\chi^2$ value with the Nelder-Mead method. We then calculate the intrinsic scatter about the three parameter model, as described above for the two-parameter case. We show the intrinsic scatter in [X/Fe] about the three-parameter model as the green line in the top panel of Figure~\ref{fig:rms_2v3}. Our procedure is analogous to the ``two-process model'' fitting of \citet{weinberg2019,weinberg2022} and \citet{griffith2019, griffith2022}, but the mathematically simpler model of Equation~\ref{eq:xfe_pred} is better suited to our relatively small sample. To the extent that scatter is driven by star-to-star variation in the CCSN/SNIa ratio at fixed [Fe/H], we would expect RMS variations to be smaller for the three-parameter model than for the two-parameter model. To more directly compare the intrinsic scatter in [X/Mg] and [X/Fe], we re-plot $\sigma_{\rm [X/Fe] \, intrin, \, 2-param}$, $\sigma_{\rm [X/Fe] \, intrin, \, 3-param}$, and  $\sigma_{\rm [X/Mg] \, intrin, \, 2-param}$ in Figure~\ref{fig:rms_2v3_scatter}.

For Mg, Ca, and Si the estimated intrinsic scatter drops by a factor of $1.3-1.6$ when using the three-parameter model rather than the two-parameter model. Ti shows a factor of 1.4 reduction in scatter, so in this sense it resembles an $\alpha$-element. Conversely, for V, Cr, Mn, Co, and Ni, the scatter reduction is a factor of $\lesssim 1.1$. This behavior demonstrates that the deviations in [X/Fe] among $\alpha$-elements are correlated, which is consistent with some of this scatter arising from variations in the SNIa/CCSN ratio (i.e., in the denominator of [X/Fe] rather than the numerator). However, it is difficult to distinguish this explanation from a scenario in which a fraction of CCSN are more efficient at producing Fe-peak elements (see Section~\ref{sec:discussion}). The clearest takeaway from this experiment is that accounting for [$\alpha$/Fe] removes only a rather small fraction of the intrinsic scatter for this metal-poor population. This contrasts with the Milky Way disk regime, where [$\alpha$/Fe] variations explain much of the overall [X/Fe] scatter even within the low-Ia and high-Ia population, and certainly between them \citep{ratcliffe2022, ting2022, weinberg2022}.

\begin{figure*}[!htb]
    \centering
    \includegraphics[width=\textwidth]{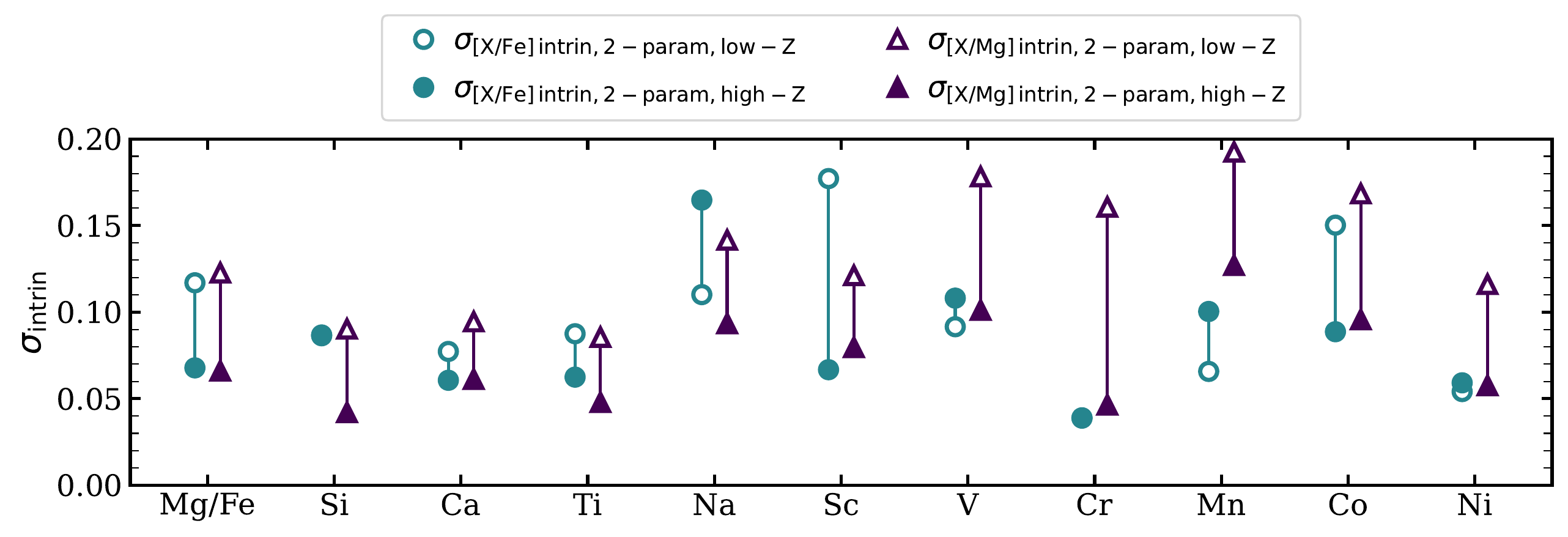}
    \caption{Intrinsic scatter in [X/Fe] (blue) and in [X/Mg] (dark purple) about the two-parameter model for subsamples above (solid points) and below (open points) the median [Fe/H] and [Mg/H] values, respectively. Where only solid points appear, the high and low-Z intrinsic scatter is nearly identical.}
    \label{fig:intrin}
\end{figure*}

The measures of scatter discussed above are calculated across the full metallicity range of our sample. However, stochastic enrichment effects should be larger at lower metallicity, and Figures~\ref{fig:x_fe} and~\ref{fig:x_mg} suggest that the scatter increases with decreasing metallicity for many elements. In Figure~\ref{fig:intrin}, we compare the intrinsic scatter in [X/Fe] and [X/Mg] above and below the median values of $\feh=-1.44$ and $\mgh=-1.01$, respectively. To do so, we re-fit the [X/Fe] vs. [Fe/H] and [X/Mg] vs. [Mg/H] abundances with the two-parameter models, once to the low-metallicity sample and once to the high-metallicity sample. Here, $\feh_{\rm cen}$ ($\mgh_{\rm cen}$) is -1.62 (-1.27) and -1.25 (-0.87) for the low and high-metallicity samples, respectively.  After determining the best-fit parameters, we calculate the predicted abundance and the intrinsic scatter about the predictions. 

The intrinsic scatter in [X/Mg] about the two-parameter model is always larger in the low-metallicity sample, by factors of 1.5 to 1.8 for all elements except Si, Cr, and Ni, for which the scatter in the low-metallicity sample is greater by a factor of $>2$. For [X/Fe], the trend is less consistent. Relative to the two-parameter model, Mg, Ti, Sc, and Co show substantially larger [X/Fe] scatter in the low-metallicity sample (by factors of 1.4 to 2.6), but for Si, Ca, Ti, V, Cr, Ni the scatter is similar (within a factor of 1.2) between low and high-metallicity. For Na and Mn, the high-metallicity sample shows larger scatter, by a factor of $\sim1.5$. Clipping the $3\sigma$ outliers does not affect the scatter at high or low-metallicity in [X/Mg], but it does significantly impact [Mg/Fe], [Ti/Fe], and [Sc/Fe]. For all three elements the scatter in the low-metallicity sample decreases with the removal of one strong outlier, leading to similar intrinsic scatter in the high and low-metallicity samples for Mg and Sc and larger scatter in the high-metallicity [Ti/Fe] than the low-metallicity [Ti/Fe].

We report the intrinsic scatter of the full population in [X/Fe] and [X/Mg] in Tables~\ref{tab:scat_xfe} and ~\ref{tab:scat_xmg}, respectively, relative to the two-parameter model (Mg and Fe) and three-parameter model (Fe only). We report $\sigma_{\rm phot, \, 2-param}$ and $\sigma_{\rm phot, \, 3-param}$ for the full population in these tables as well. We discuss interpretations of these results in Section~\ref{sec:discussion} below.
   
\section{Kinematic Subsamples}\label{sec:kin}

Within the metallicity range probed by our sample ($-2 \lesssim \feh \lesssim -1$), we expect a mix of stars with \textit{in situ} and accreted origin. The major accreted component within the solar neighborhood is the Gaia-Enceleadus Sausage \citep[GES;][]{helmi2018, belokurov2018, mackereth2019}, a disrupted massive satellite that merged with the Milky Way $10-11$ Gyrs ago \citep{helmi2018, chaplin2020}. Accreted stars have distinct kinematics but may also have distinct abundances and different sources/magnitudes of intrinsic abundance scatter than \textit{in situ} stars. Work by \citet{nissen2010, nissen2011} and \citet{hayes2018} identifies two distinct stellar populations at low metallicity (high-$\alpha$/\textit{in situ} and low-$\alpha$/accreted) and compares their abundance trends. Both works find a clear separation in [Mg/Fe] vs. [Fe/H] trends of the two stellar groups, as well as in the [X/Fe] vs. [Fe/H] trends of other $\alpha$, light odd-$Z$-elements, suggestive of distinct enrichment histories. Recent work by \citet{belokurov2022} finds that the scatter in the \textit{in situ} component of the APOGEE DR17 sample (separated chemo-dynamically) increases at $\feh < -1$, and that this scatter is larger than that of the accreted component.

Our target selection does not include any kinematic cuts, so our population likely includes both \textit{in situ} and accreted stars. Past works have used both kinematics and chemistry to separate stars with different origins. In particular, the [Mg/Mn] vs. [Al/Fe]  abundance space isolates accreted stars, as they tend to have larger [Mg/Mn] and smaller [Al/Fe] than \textit{in situ} population \citep{hawkins2015, das2020, horta2021}. Unfortunately, we cannot use this chemical diagnostic, as we do not have access to Al lines in our spectra. We instead use the energy-angular momentum plane to assess the origin of our stars. Using Gaia eDR3 proper motions and radial velocities, along with distances, calculated in Section~\ref{subsec:select} from Gaia parallax, we calculate orbital parameters with \texttt{galpy}\footnote{\url{https://github.com/jobovy/galpy}} \citep{bovy2015, mackereth2018} and a \citet{mcmillan2017} potential. Gaia radial velocities are not available for nine stars. In these cases we use radial velocities calculated by iSpec.

\begin{figure}[!htb]
    \centering
    \includegraphics[width=\columnwidth]{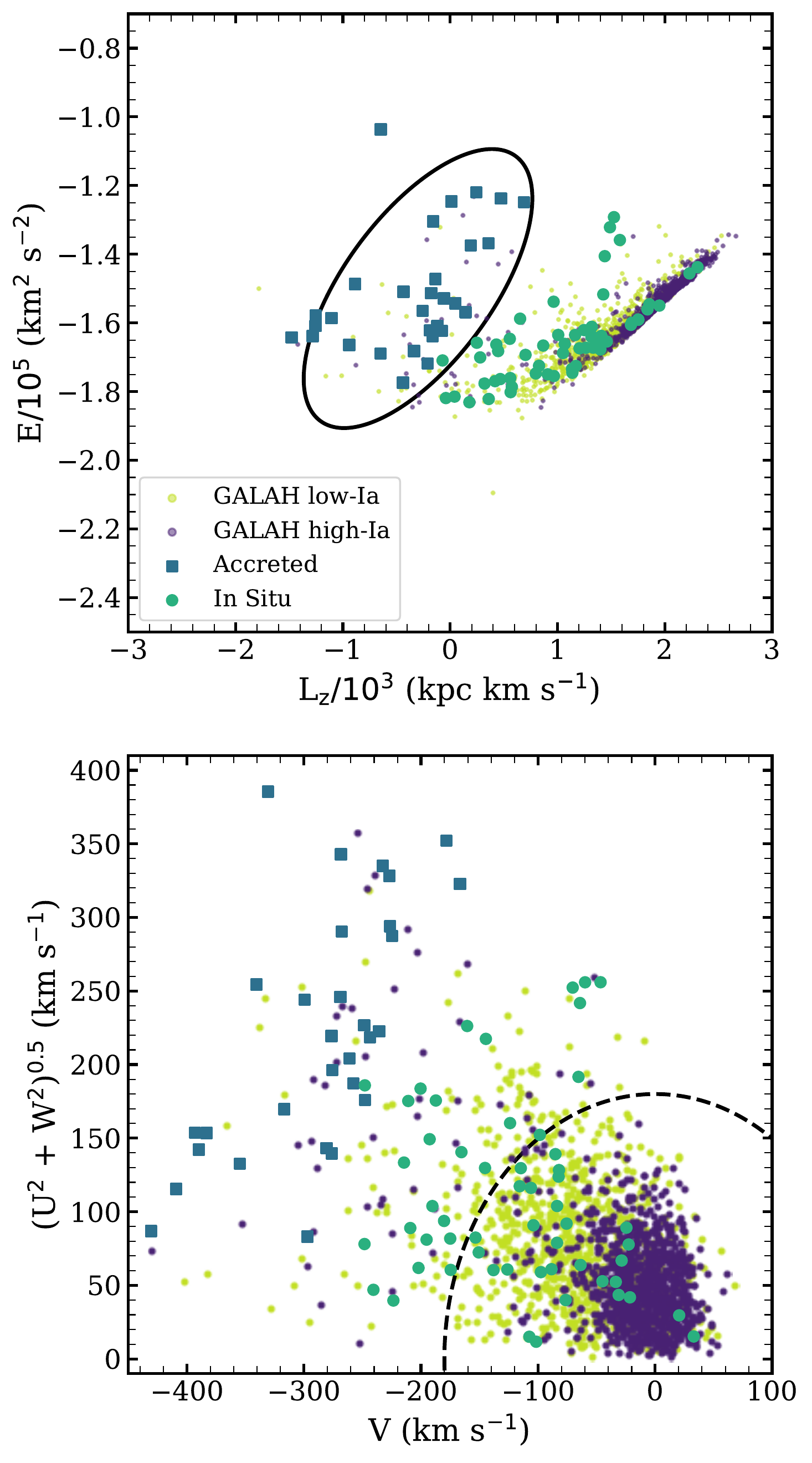}
    \caption{Top: $E/10^5$ (km$^2$s$^{-2}$) vs. $L_{\rm z}$ (kpc km s$^{-1}$) for our sample and a low-metallicity GALAH DR3 low-Ia (lime green) and high-Ia (purple) sample. The black ellipse indicates the energy-angular momentum space occupied by GES stars in \citet{horta2021}. Stars in our sample are color coded by their apparent kinematic origin, with accreted stars in dark blue (squares) and \textit{in situ} stars in teal (circles). Bottom: Toomre diagram for our sample and GALAH DR3 stars. The long dashed line indicates $V_{\rm total}=180$ km s$^{-1}$.}
    \label{fig:kin}
\end{figure}

We plot the stellar energy vs. angular momentum in the top panel of Figure~\ref{fig:kin} alongside a low-metallicity ($\feh < -0.5$) sample of high-Ia and low-Ia GALAH DR3 stars, with low-Ia population defined by Equation 1 in \citet{griffith2021a}. Most GALAH DR3 stars have $\feh > -1$, and formed within the Milky Way. The high-Ia and low-Ia GALAH DR3 samples form a tight sequence with positive $L_{\rm z}$ and $E/10^5$ extending from -1.8 to -1.4 km$^2$s$^{-2}$. A subset of our sample overlaps with the GALAH DR3 locus and has kinematics consistent with \textit{in situ} formation. Another subset of our sample appears clumped near $E/10^5\approx-1.5$ km$^2$s$^{-2}$ and $L_{\rm z}\approx0$ kpc km s$^{-1}$, overlapping with the kinematic space occupied by the GES system \citep[e.g.,][indicated with the black ellipse]{horta2021}. In Figure~\ref{fig:kin}, we color code stars by their \textit{in situ}-like (teal) and GES-like (dark blue) kinematics. We will refer to these subsamples as \textit{in situ} and accreted.

As an additional check of stars' kinematic origin, we plot their location on a Toomre diagram in the bottom panel of Figure~\ref{fig:kin}. We find that the stars whose energy and angular momentum are consistent with the \textit{in situ} disk have Galactic velocities closer to zero and overlapping with the GALAH low-Ia population. Stars whose energy and angular momentum overlap with the GES parameter space have larger velocities and are separated from the GALAH disk and \textit{in situ} stars. These two kinematic diagnostics confirm that stars of both \textit{in situ} and accreted origin reside in our stellar sample. We classify 30 stars as accreted, and 56 as \textit{in situ}. 

\begin{figure*}[!htb]
    \centering
    \includegraphics[width=\textwidth]{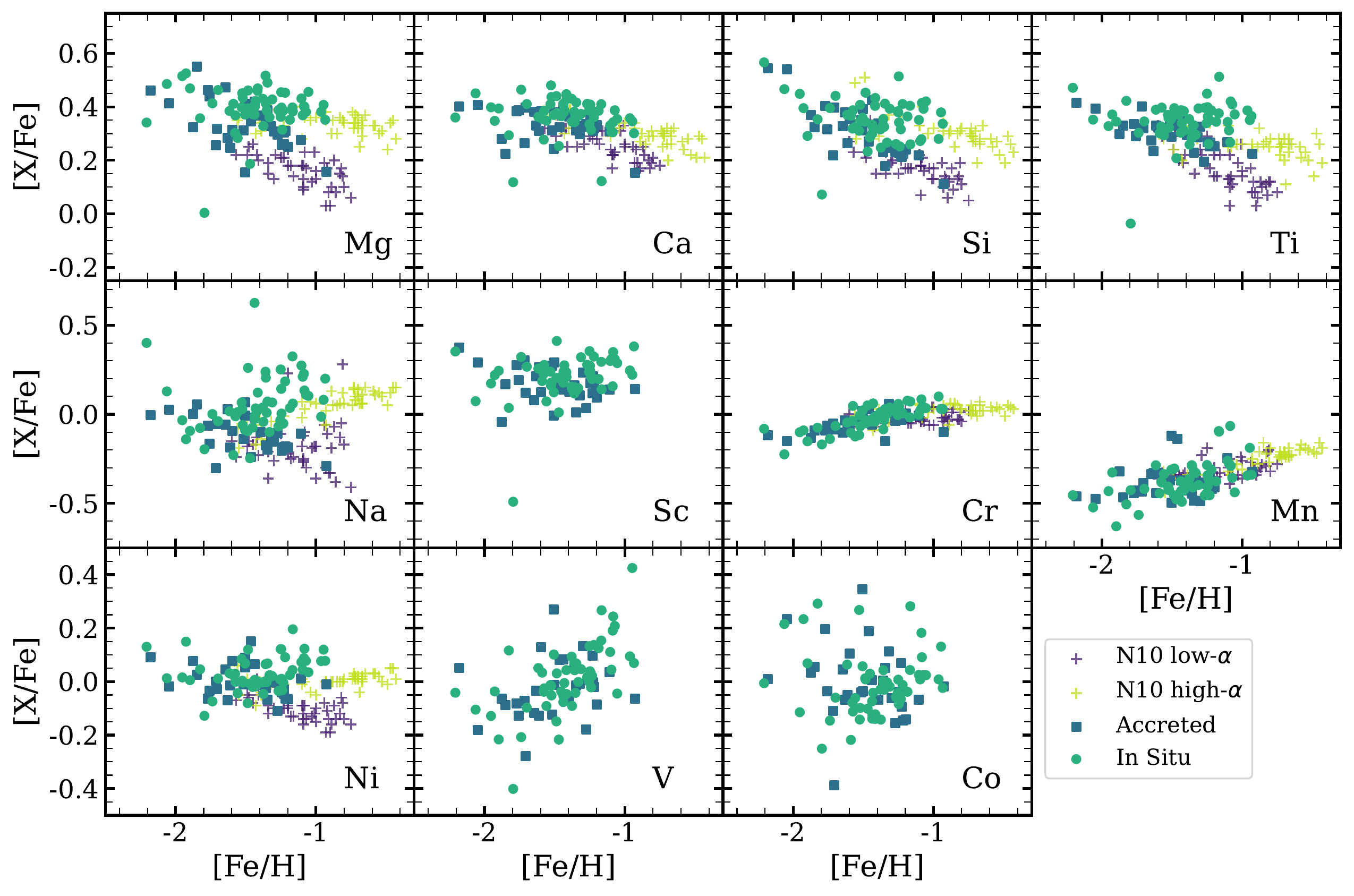}
    \caption{Same as Figure~\ref{fig:x_fe}, but with stars color coded by \textit{in situ} (teal circles) and accreted (blue squares) origin. We include the high-$\alpha$ (\textit{in situ}, lime green crosses) and low-$\alpha$ (accreted, purple crosses) from \citet{nissen2010, nissen2011} for comparison. Note that the y-axes ranges differ in each row.}
    \label{fig:xfe_kin}
\end{figure*}

In Figure~\ref{fig:xfe_kin}, we plot the [X/Fe] vs. [Fe/H] abundances of our sample, as in Figure~\ref{fig:x_fe}, but now color coded by the their \textit{in situ} vs. accreted origin. We include the abundances of subsolar metallicity dwarfs from \citet{nissen2010, nissen2011}, who identify two chemically distinct halo populations in the solar neighborhood \citep[likely corresponding to the \textit{in situ} thick disk and accreted GES stars;][]{hayes2018} with systematic differences in their abundances. Amongst the $\alpha$-elements, we find that the accreted stars have systematically lower [X/Fe] abundances than the \textit{in situ} stars at $\feh \sim -1$. This is most notable in Mg and Si, but identifiable in Ca and Ti as well. We find that the abundances of the \textit{in situ} stars overlap with the high-$\alpha$ population from \citet{nissen2010, nissen2011} and that the accreted stars overlap with the low-$\alpha$ population. At $\feh \sim -1.6$ and lower metallicities, the abundances of the two populations converge.

Conversely, we observe that the abundance trends of the \textit{in situ} and accreted stars overlap for most light odd-$Z$ and Fe-peak elements (Sc, Cr, Mn, V, and Co), in agreement with \citet{nissen2010, nissen2011} where data is available. Na and Ni more resemble the $\alpha$-elements, with the abundances of the \textit{in situ} and accreted stars diverging at $\feh \gtrsim 1.5$. The [Na/Fe] abundances show the largest systematic differences. Differences in star formation history, stellar IMF, CCSN black hole landscape, or the ratio of CCSN/SNIa enrichment, could cause the observed differences in abundances at fixed [Fe/H] between stars formed in the Milky Way and those formed in a massive satellite.

\begin{figure*}[!htb]
    \centering
    \includegraphics[width=\textwidth]{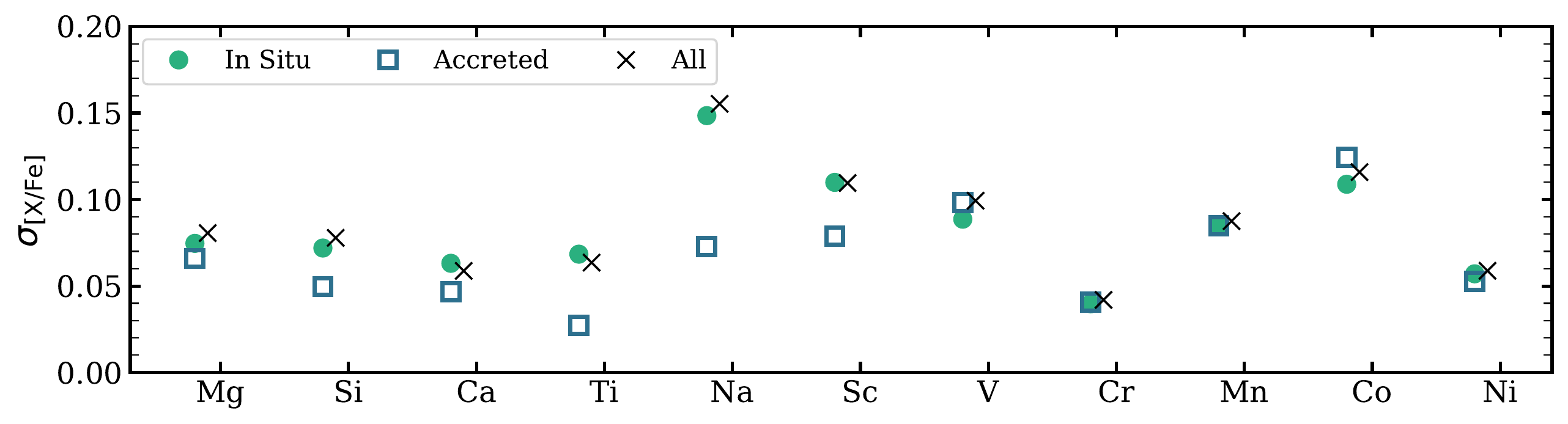}
    \caption{Intrinsic scatter in [X/Fe] for the \textit{in situ} subsample (teal circles), accreted subsample (blue empty squares) and full population (black X, relative to the two-parameter model).}
    \label{fig:sigma_kin}
\end{figure*}

We repeat the calculation of the intrinsic scatter in [X/Fe] relative to the two-parameter model (Section~\ref{subsec:intrin}) for the \textit{in situ} and accreted stars. In Figure~\ref{fig:sigma_kin}, we plot the resulting intrinsic scatter in [X/Fe] of the \textit{in situ} population, accreted population and full sample. We find only small differences between the scatter in these samples for all elements except Ti, Na, and Sc. For the other eight elements, the values of the intrinsic scatter about the two-parameter model in [X/Fe] for \textit{in situ} and accreted populations are within 0.03 dex, and within 0.01 dex for Mg, V, Cr, Mn, and Ni. The variations in the intrinsic scatter between the \textit{in situ} sample and full population are similarly small, with $\sigma_{\rm intrin}$ within 0.01 dex for all elements. If we remove outlier stars with a 3$\sigma$ cut, the intrinsic scatter of the \textit{in situ} and accreted samples is within 0.03 dex for all elements but Na, where the \textit{in situ} stars remain more scattered. The similarity of intrinsic scatter implies that the scatter in the full population is not driven primarily by the differences between \textit{in situ} and accreted populations with different enrichment histories. The populations' different enrichment histories cause diverging abundance trends, but they do not appear to impact the abundance scatter for most elements. 
     
\section{Discussion: The Origin of Intrinsic Scatter} \label{sec:discussion}

We find significant RMS intrinsic scatter, typically 0.05-0.13 dex ($\sim 10-30\%$), in [X/Fe] at fixed [Fe/H] and [X/Mg] at fixed [Mg/H], for each of the twelve elements examined in this paper.  For most elements the scatter is larger in the low metallicity half of our sample compared to the high metallicity half, by a factor 1.5-1.8 in [X/Mg].  The scatter is consistent between the kinematically identified {\it in situ} and accreted subsets of our sample, and removing the accreted stars does not noticeably decrease the scatter.  The dominant sources of these elements are expected to be CCSN and SNIa \citep{andrews2017,rybizki2017,johnson2019}.  The strongest candidate for a large AGB contribution is Na, as population studies at higher metallicity demonstrate a significant time-delayed contribution and SNIa production of Na is predicted to be small \citep{griffith2019,griffith2022,weinberg2019,weinberg2022}.

For each of these enrichment processes --- CCSN, SNIa, and AGB --- the ``population-averaged'' yield of a given element may depend on metallicity.  We use the term population-averaged to include averaging over the stellar IMF and other properties that can affect yields such as rotation or binary fraction.  CCSN arise from massive stars with short lifetimes, so for a population of stars formed at time $t$ the CCSN yield at $t+t'$ is time-independent for $t'>40$ Myr.  For SNIa the enrichment is extended in time, often modeled as a rate $R_{\rm Ia} \propto (t')^{-1.1}$ after a minimumum delay of 50-150 Myr \citep{maoz2017}.  The relative yield of individual elements is usually assumed to be independent of time-delay, but it could change systematically with $t'$ if the mass distribution of white dwarf progenitors changes (e.g., from sub-$M_{\rm ch}$ to $M_{\rm ch}$) between early and late SNIa.  AGB element ratios should have some time-dependence because the yields of different elements have different mass dependence, though in most cases the IMF-averaged yield is dominated by stars with lifetimes below 1 Gyr \citep[e.g.]{johnson2022}.

A necessary condition for intrinsic scatter is a mix of stars that have different enrichment histories.  These different histories may have arisen in distinct structures, such as disrupted dwarf galaxies, but they can also arise at different radii in a single progenitor, and in a chaotic proto-Galaxy they may simply arise in spatially distinct regions that do not efficiently mix their ISM with each other.  Given a mix of stellar population histories, there are a number of mechanisms that can produce scatter.  Our enumeration below is not exhaustive, but it encompasses the options that we think could most plausibly affect our sample at the observed level.  Section 4.5 of \cite{belokurov2022} provides an excellent discussion of abundance ratio scatter, complementing ours by focusing on the galactic scale events within which these mechanisms can operate.

First, scatter can arise from fluctuations in the AGB/SNIa/CCSN enrichment ratio at fixed metallicity.  We focus on SNIa/CCSN because AGB contributions are likely to be small for most of our elements.  At the metallicity range of the Galactic disk, variations in SNIa/CCSN likely dominate the scatter in $\afe$ ratios \citep{bertran2016,vincenzo2021}, and scatter in many abundance ratios decreases when this variation is controlled for \citep{ratcliffe2022,ting2022,weinberg2022}.  Differences in SNIa/CCSN at fixed [Fe/H], linked to differences in star formation efficiency, drive many of the abundance ratio differences among dwarf satellites and halo substructures \citep{hasselquist2021,horta2022}.  However, several arguments suggest that SNIa/CCSN variations are not the primary source of intrinsic scatter in our sample.  The high $\afe$ values of our stars (Si, Ca, and Ti as well as Mg) place them near the plateau that is usually interpreted as reflecting CCSN yield ratios with minimal SNIa contribution.  \cite{conroy2019} propose an alternative scenario in which the plateau in this metallicity range includes a significant SNIa contribution and the $\afe$ trend stays flat because of rapidly accelerating star formation; this scenario would leave some room for SNIa/CCSN fluctuations in our sample.  However, our sample exhibits scatter in [X/Fe] {\it within} the Fe-peak elements and scatter in [X/Mg] {\it within} the $\alpha$-elements, which is not expected if the scatter is dominated by SNIa/CCSN fluctuations.  Furthermore, while adding $\afe$ as a model parameter does reduce the intrinsic scatter, the effect is modest (see Fig.~\ref{fig:rms_2v3}).  If SNIa/CCSN variations were the dominant source of abundance ratio scatter, then controlling for a single ratio that is sensitive to SNIa/CCSN should reduce the remaining scatter to near zero.

Metallicity-dependent yields are a second potential source of scatter.  A star of a given metallicity is enriched by the yields from stars with a different metallicity.  For elements with metallicity-dependent yield, therefore, stars from regions with different metallicity history could exhibit different abundances of these elements at fixed [Fe/H] or [Mg/H].  For smooth star formation histories this effect is probably small (see Figure~17 of \citealt{weinberg2019}).  However, in a proto-Galaxy experiencing rapid bursts of star formation or chaotic accretion and outflow, abundance evolution can be complex and could be a source of scatter \citep{johnson2020,belokurov2022}.  Our sample exhibits substantial scatter for even-$Z$ elements that are not expected to have metallicity-dependent yields, though the relatively large scatter for Na, Sc, V, Mn, and Co could perhaps have a contribution from metallicity dependence.

Time-dependence of SNIa and AGB yields can in principle be a source of scatter because stars from regions with different star formation efficiencies and histories reach a given metallicity at different times.  However, this effect should be small for CCSN production because of the short massive star lifetimes, so it cannot explain the scatter we find in $\alpha$-elements.  The possible impact of time-dependent SNIa yields, on mean trends as well as scatter, is an interesting area for future work.

A fourth possible source of scatter is environmental variation of the IMF, so that stars formed in different regions would experience different population-averaged supernova yields.  For example, the IGIMF model \citep[integrated galaxial IMF,][]{weidner2005} posits that the high mass slope and cutoff of the IMF depends on the mass of star-forming gas available.  Counts of luminous stars in the 30 Doradus region of the Large Magellanic Cloud favor a top-heavy IMF relative to the \citet{kroupa2001} form \citep{schneider2017, bestenlehner2020}. Yields could also vary because of systematic variations in stellar rotation or binary properties.  We regard this class of explanations as exotic, but in principle it is a possible source of abundance scatter.

The fifth possible source of scatter, and one that does seem natural in the metallicity range of our sample, is stochastic sampling of the IMF, or more generally of the supernova population.  In this scenario, any given star does not sample the population-averaged yield but the particular realization of supernovae that enriched its patch of the ISM.  The key parameters controlling the scatter are the supernova-to-supernova variance of abundance ratios and the effective number of supernovae that contribute to a typical star's enrichment at a given metallicity.  We investigate the stochastic CCSN scenario quantitatively in Section~\ref{subsec:stoch} below.  Computations of stochastic SNIa yield variations require both yields as a function of progenitor properties \citep{lach2020} and a model of the distribution of these properties.  Because the SNIa rate is an order of magnitude lower than the CCSN rate, Poisson fluctuations are larger, and stochastic enrichment variations could be comparable even if the yield variation from one SNIa to another is smaller than that of CCSN.  However, stochastic SNIa sampling cannot explain the scatter for elements that have negligible SNIa contributions.

\subsection{Stochastic Sampling of the IMF}\label{subsec:stoch}

The case for which we can provide concrete calculations is stochastic sampling of the IMF producing CCSN yield fluctuations. To compare our observed trends with theoretical yields, we calculate the expected abundances and scatter in $\xfe$ and $\xmg$ from a stellar population with $N$ CCSN using theoretical yields. This requires CCSN yields from progenitors with a fine mass grid and a realistic explosion landscape. We choose to use the yields from the Z9.6+W18 explosion engine in \citet{sukhbold2016} because of their fine mass sampling (200 models from $9-120\,\Msun$) and physically motivated model for neutrino driven explosions. This yield set is at solar metallicity. No grid of CCSN yields with such fine progenitor mass sampling exists at $\feh\approx-1.5$. While we expect the absolute $\xfe$ and $\xmg$ abundance to vary between CCSN yield sets at solar and subsolar metallicity for elements with metallicity dependent yields, the scatter may be more robust to variations in $\feh$. We proceed with the Z9.6+W18 yields but note that this metallicity difference is a source of uncertainty in the predictions.

\begin{figure*}[!htb]
    \centering
    \includegraphics[width=\textwidth]{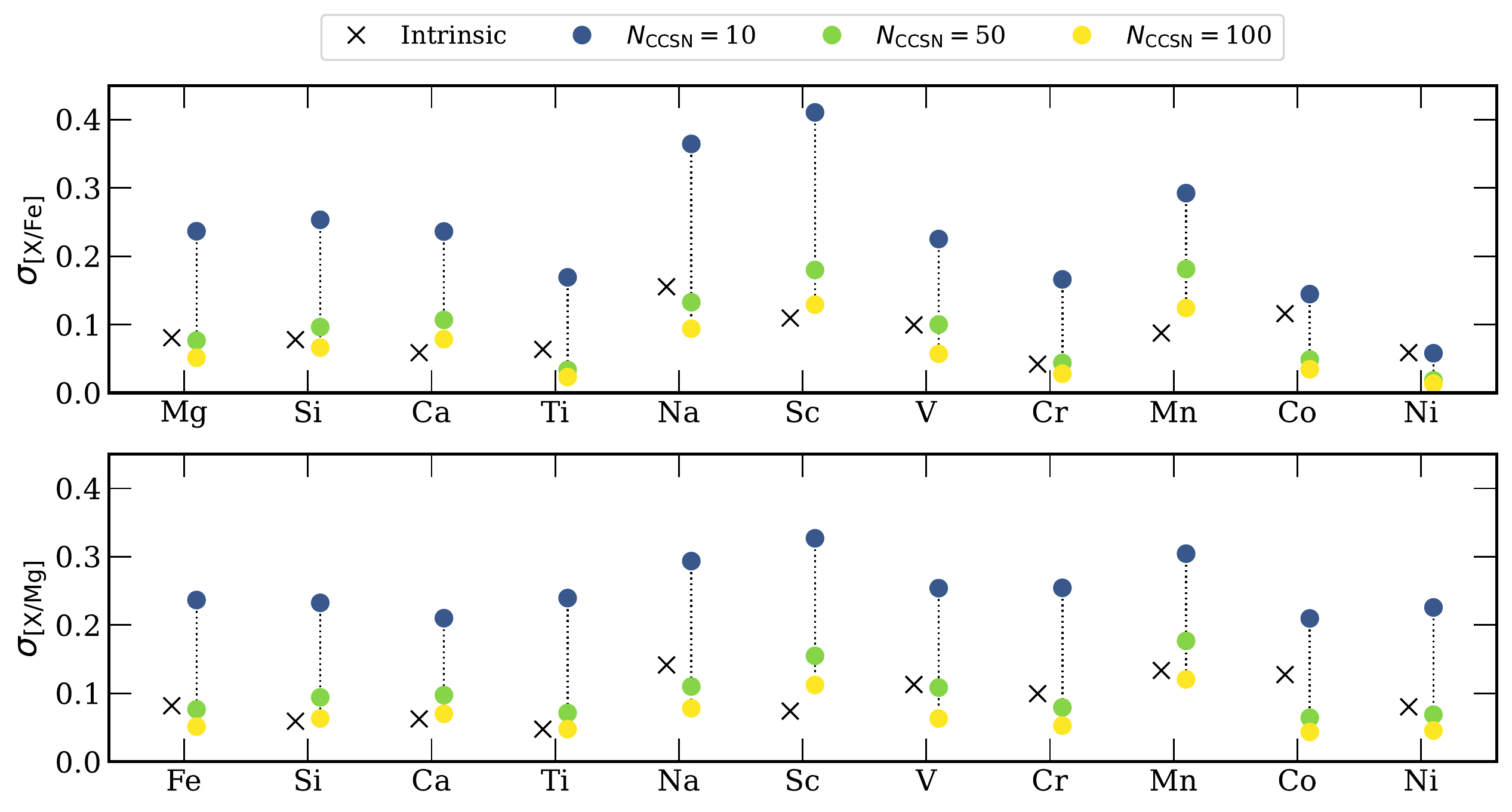}
    \caption{Top: Standard deviation of [X/Fe] from 1000 draws of 10 (dark blue), 50 (lime green), and 100 (yellow) CCSN using Z9.6+W18 yields from \citet{sukhbold2016}. We included the intrinsic scatter on [X/Fe] relative to the two-parameter model (black cross) for comparison. Bottom: Same as top panel, but for [X/Mg].}
    \label{fig:sigma_xmg}
\end{figure*}

As a measure of the predicted scatter from a stochastically sampled IMF, we calculate the standard deviation ($\sigma$) of $\xfe$ and $\xmg$ abundances from 1000 draws of stellar populations with $N= 10$, 50, and 100 CCSN. We randomly sample $N$ stars of progenitor masses $8-120\,\Msun$ from an IMF weighted distribution, using a \citet{kroupa2001} IMF. For each star, we interpolate its net yield from the Z9.6+W18 grid, employing the yield lookup tables from \texttt{VICE}\footnote{Versatile Integrator for Chemical Evolution; \url{https://pypi.org/project/vice/}} \citep{johnson2020}. The Z9.6+W18 explosion engine produces a complex CCSN landscape, with progenitor mass ranges of black hole formation mixed in with successful explosions. If the randomly drawn star explodes, it contributes net wind and explosive yields. If the star collapses to a black hole, it only contributes net wind yields. For each population size (10, 50, 100 CCSN), we draw 1000 random samples, sum the net yields, convert to $\xfe$ and $\xmg$ with \citet{asplund2009} solar abundances, and calculate the standard deviation of the abundance ratios. We plot $\sigma_{\rm\xfe}$ and $\sigma_{\rm\xmg}$ for the three choices of $N_{\rm CCSN}$ in the top and bottom panels of Figure~\ref{fig:sigma_xmg}, respectively, and report the values in Tables~\ref{tab:scat_xfe} and ~\ref{tab:scat_xmg}.

We find that the scatter in [X/Fe] and [X/Mg] roughly scales by $\sqrt{N_{\rm CCSN}}$, with fewer CCSN producing larger abundance scatter. Relative to Fe, the stochastically sampled Ni yields have the smallest scatter, with $\sigma_{\rm [Ni/Fe]}=0.058$ dex for 10 CCSN and $\sigma_{\rm [Ni/Fe]}=0.013$ dex for 100 CCSN. This behavior is not surprising, as the production of Fe is intimately linked to the production of Ni in CCSN \citep{sukhbold2016}. Ti, Cr, and Co also have small scatter, with $\sigma_{\rm [X/Fe]}<0.05$ dex for $N_{\rm CCSN}=100$. We predict the largest scatter for Na, Sc, and Mg, with $\sigma_{\rm [X/Fe]}>0.09$ dex for $N_{\rm CCSN}=100$. The mean $\sigma_{\rm [X/Fe]}$ across our eleven (non-Fe) elements for populations of 10, 50 and 100 CCSN are 0.23, 0.09, and 0.06 dex, respectively. The scatter in $\xmg$ tends to be slightly larger than that in [X/Fe], though the mean $\sigma_{\rm [X/Mg]}$ is only $\sim 0.01$ dex larger than the mean $\sigma_{\rm [X/Fe]}$. For scatter in [X/Mg] abundances, we again find that Ti, Cr, Co, and Ni have the smallest scatter while Na, Sc, and Mn have the largest.

\begin{figure*}[!htb]
    \centering
    \includegraphics[width=\textwidth]{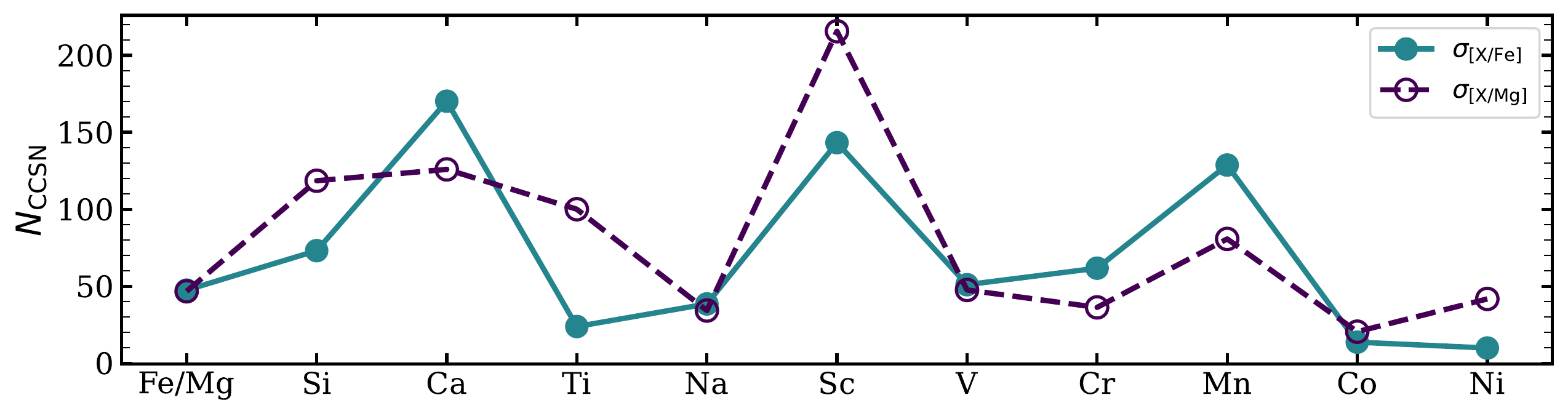}
    \caption{The number of stochastically sampled CCSN that would reproduce the observed intrinsic abundance scatter in [X/Fe] (blue, solid line) and [X/Mg] (dark purple, dashed line).}
    \label{fig:n_ccsn}
\end{figure*}

Figure~\ref{fig:sigma_xmg} plots the intrinsic scatter found in our observational analysis alongside the scatter from the theoretical CCSN yields. Relative to both Fe and Mg, the magnitude of the intrinsic scatter for most elements corresponds to stochastic enrichment from more than 10 but fewer than 100 CCSN. Encouragingly, most of the elements that show the largest observed scatter are also those with the largest predicted scatter, e.g., Na, Sc, V, Mn. The exception is Co, which has a high observed scatter but low predicted scatter, and Ni, which has a low observed scatter but still lower predicted scatter. If the model were perfect, it would explain all of the observed $\sigma_{\rm [X/Fe]}$ and $\sigma_{\rm [X/Mg]}$ values within a single value of the effective supernova number $N_{\rm CCSN}$. Figure~\ref{fig:n_ccsn} plots the value of $N_{\rm CCSN}$ that would reproduce each of the observed scatter values, based on interpolation from a grid of 17 values of $N_{\rm CCSN}$ between 5 and 200. 

As expected from Figure~\ref{fig:sigma_xmg}, most of the observed scatter values imply $N_{\rm CCSN}$ between 30 and 120. The values of $\sigma_{\rm [Ti/Fe]}$, $\sigma_{\rm [Co/Fe]}$, and $\sigma_{\rm [Ni/Fe]}$ imply low $N_{\rm CCSN}$ because they are high compared to the predicted scatter. The values of $\sigma_{\rm [Ca/Fe]}$, $\sigma_{\rm [Sc/Mg]}$, and $\sigma_{\rm [Mn/Fe]}$ imply high $N_{\rm CCSN}$ because they are low compared to the predicted scatter. Given the uncertainties in the model and the observationally inferred intrinsic scatter values, we regard the factor $\sim 3$ consistency in $N_{\rm CCSN}$ across most element ratios as a qualitative success of the stochastic IMF sampling scenarios. The median number of CCSN needed to reproduce the elemental scatter is $N_{\rm CCSN}\sim50$, though Ti, Co, and Ni would  need an additional source of intrinsic scatter if the model yields are correct.  If we repeat this analysis with the intrinsic scatter in [X/Fe] from the sample with clipped outliers, we find that the median number of CCSN increases to $\sim 60$.

We can further test the stochastic IMF sampling scenario by predicting the correlation of deviations in abundance ratios for pairs of elements. For the theoretical abundances, we define the abundance deviation as the difference between the predicted [X/Fe] value from one draw of $N$ CCSN and the mean [X/Fe] value of 1000 draws of $N$ CCSN ($\Delta \xfe = \xfe - \overline{\xfe}$). For the observed abundances, we define the abundance deviation as the difference between the measured [X/Fe] and the [X/Fe] predicted by the two-parameter model ($\Delta \xfe = \xfe_{\rm obs} - \xfe_{\rm pred}$). We calculate correlation coefficients $(r)$ between $\Delta \xfe$ and $\Delta \rm [Y/Fe]$, as well as $\Delta \xmg$ and $\Delta \rm [Y/Mg]$, for all pairs of elements for both the theoretical and observational abundances. Through visual inspection we find that pairs of deviations with $|r|>0.5$ show correlations that are readily discernible in a pairwise scatter plot.  We have checked that correlations that would be induced by photon-noise alone are small compared to the more prominent correlations we measure.

\begin{figure*}[!htb]
    \centering
    \includegraphics[width=\textwidth]{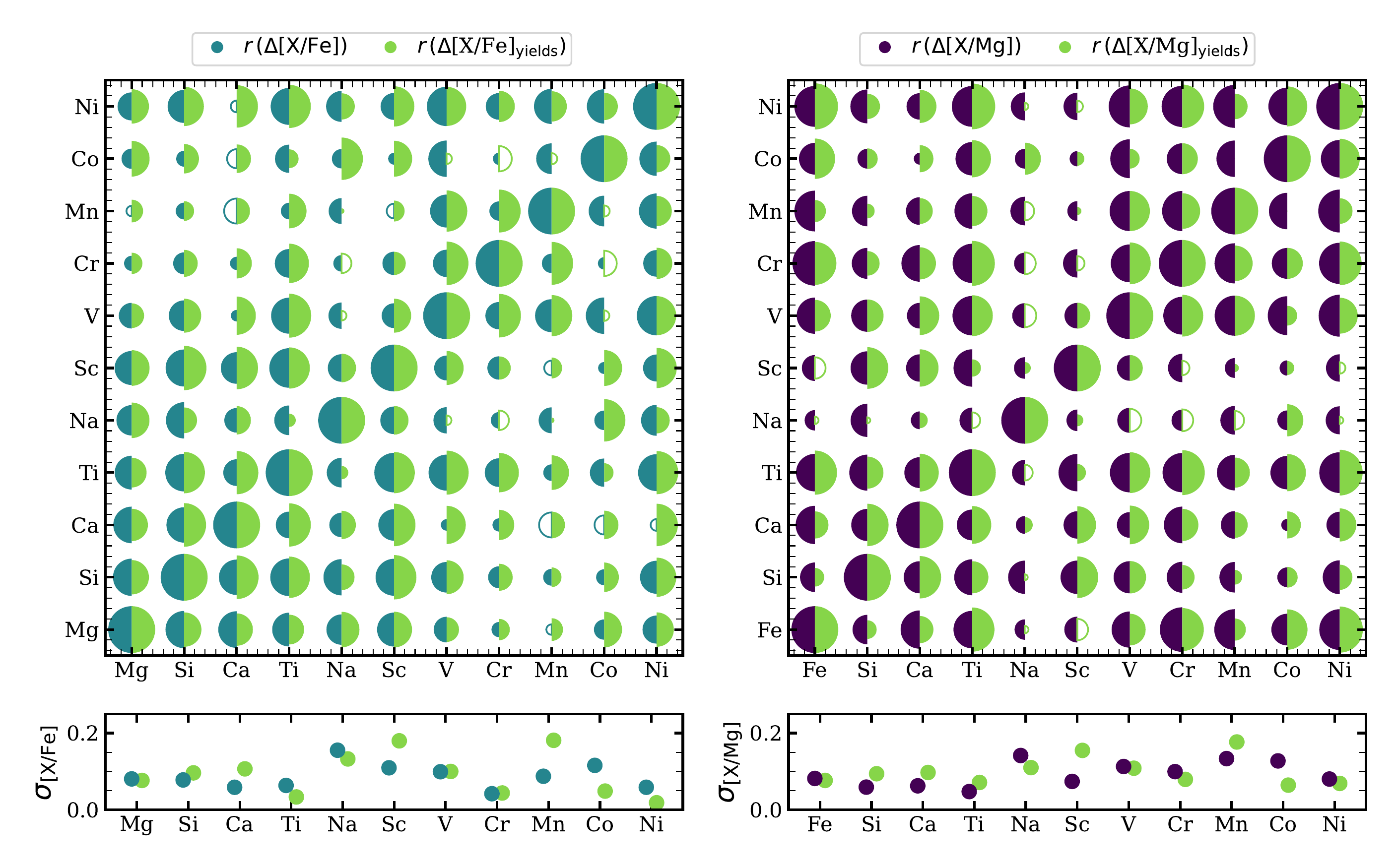}
    \caption{Top left: Correlation matrix of the observed deviations in [X/Fe] (blue) and the deviations predicted by the stochastic IMF sampling model (lime green). Positive correlation coefficients are shown as filled half circles and negative correlation coefficients as empty half circles. Agreement between the observed and predicted correlation coefficients can be assessed by comparing the left and right halves of each circle.  Correlations along the diagonal are 1.0 by definition, and for other coefficients the area of the half circle is proportional to the magnitude of the correlation. For scale, the correlation between the observed deviations (blue) of $\Delta$[Si/Fe] and $\Delta$[Ni/Fe] is 0.5. Top right: Same as the left, but for [X/Mg], with the observed correlation in dark purple. Bottom left: Same as top panel of Figure~\ref{fig:sigma_xmg}, but only including $N_{\rm CCSN}=50$ and with the intrinsic scatter in [X/Fe] about the two parameter model shown in blue. Bottom right: Same as bottom left but for Mg and with the intrinsic scatter about the two-parameter model in dark purple.}
    \label{fig:corrs}
\end{figure*}

The upper panels of Figure~\ref{fig:corrs} compare the observed and predicted correlations for $\Delta\xfe$ (left panel) and $\Delta \xmg$ (right panel).  For each element-pair, the areas of the left and right half circles are proportional to the magnitudes of the correlations, so similar areas indicate good agreement.  Unshaded half circles denote negative correlations.  One can gain physical intuition for the predicted correlations from Figure 12 of \cite{griffith2021a}, which plots element yields as a function of progenitor mass, with the caveat that these yields are then modulated by black hole formation and weighted by the IMF.  For example, the mass-dependence of Si yields is closer to that of Ca than Mg, so the Si-Ca correlation is larger than the Si-Mg correlation.  The mass-dependence of Na yields is different from that of most of the elements we measure, and the predicted correlation coefficients for Na are relatively small.  For convenient reference, the lower panels plot the intrinsic scatter about the two-parameter model and the expected scatter for $N_{\rm CCSN}=50$.  Although the magnitude of the RMS scatter scales as $N_{\rm CCSN}^{-1/2}$, the correlation coefficients plotted in the upper panel should be independent of $N_{\rm CCSN}$ provided that stochastic IMF sampling is indeed the dominant source of abundance scatter.

In the correlations of the deviations in [X/Fe] relative to the two-parameter model (left circle half in the left panel), we see large correlations within $\alpha$-elements and between Sc and the $\alpha$-elements, Ni and the Fe-peak elements, and V and the Fe-peak elements. We also observe large correlations of the deviations in [X/Mg] (left circle half in the right panel), especially amongst pairs of $\alpha$-elements and pairs of Fe-peak elements. Ti also shows strong correlations with the Fe-peak elements. If the size and shading of the left half of the circle matches that of the right half of the circle, this indicates that the correlation of the observed deviations agrees with the correlation predicted by our stochastic IMF sampling calculation.  For the most part we find good agreement in the sign and reasonable agreement in the magnitude of predicted and observed correlations, especially in [X/Mg] (e.g. (Ti, Fe), (V, Mn), and (Ni, Fe)), though with some exceptions (e.g. (Mn, Na)).

Although the results in Figures~\ref{fig:sigma_xmg}-\ref{fig:corrs} support the idea that much of the scatter in our sample arises from stochastic IMF sampling, it is somewhat difficult to separate this contribution from that of varying SNIa/CCSN ratio.  In particular, because variations in the compactness of the CCSN progenitor strongly affect the production of Fe-peak elements \citep{sukhbold2016,griffith2021b}, stochastic IMF sampling of the CCSN population produces fluctuations in Fe-peak/$\alpha$-element ratios that resemble the effects of fluctuating SNIa/CCSN enrichment.

\subsection{Implications of Inferred $N_{\rm CCSN}$}

In the stochastic IMF sampling scenario, a typical star forms from gas that is enriched by the ejecta of $N_{\rm CCSN}$ supernovae.  A different sample star of similar metallicity is enriched by a different $N_{\rm CCSN}$ supernovae and thus has a slightly different mix of elements.  Given the value of $N_{\rm CCSN}$ required to reproduce the intrinsic scatter in our sample, we can estimate the mixing mass $M_{\rm mix}$ of gas over which the ejecta of these supernovae must be mixed from the condition
\begin{equation}
    Z = \frac{y_{\rm Z}N_{\rm CCSN}}{M_{\rm mix}},
    \label{eq:Mmix}
\end{equation}
where $Z$ is the mass fraction of a given element and $y_{Z}$ is the average yield of that element per CCSN, in $\msun$.  The median [Mg/H] of our sample is $-1.1$.  For an illustrative calculation we take O as a representative element and assume that [O/H] = [Mg/H] $\approx -1$.  Using the IMF-averaged O yield $y_Z \approx 0.75 \msun$ predicted by the CCSN model used in Section~\ref{subsec:stoch}, $Z_{\rm O} = 0.1 Z_{\rm O\,\odot} = 6 \times 10^{-4}$, and $N_{\rm CCSN}=50$, Equation \ref{eq:Mmix} implies $M_{\rm mix} \approx 6 \times 10^4\msun$.  There are significant uncertainties in this estimate, notably the choice of the sample median as the representative metallicity and the choice of yield.  O production in the Z9.6+W18 model of \cite{sukhbold2016} is suppressed by black hole formation, and if all stars with $M=8-40\msun$ exploded the predicted yield would be a factor of 2.3 higher (\citealt{griffith2021a}, Figure 11).  Conversely, the Mg/O yield ratio of the \cite{sukhbold2016} models is sub-solar, and if we had used Mg as our representative element then our estimated $M_{\rm mix}$ would be only $2\times 10^4\msun$.

These uncertainties notwithstanding, the stochastic IMF sampling interpretation of the measured intrinsic scatter places the mixing mass in the $10^4-10^6\msun$ range characteristic of giant molecular clouds (GMCs).  In the present-day disk this would be a surprisingly low mixing mass, since a GMC forms only a few percent of its mass into stars and an average fluid element must cycle through the dense molecular phase 10-20 times before forming into stars \citep{semenov2017}.  We would expect the effective mass sampled by a forming star, and thus $N_{\rm CCSN}$, to be larger than the typical GMC mass by a similar factor. At a gas surface density of 10 $\msun \text{pc}^{-2}$, typical of the solar annulus, $10^5\msun$ corresponds to a region of only $(0.1\kpc)^2$, which also seems like a short lengthscale for mixing given $\sim$Gyr gas depletion times \citep{leroy2008} and turbulent velocities at the few km s$^{-1}$ level. However, $M_{\rm mix} \sim 10^5 \msun$ is perhaps a plausible value for the low metallicity phase of the proto-Galaxy, where star formation may occur in disconnected regions that do not efficiently share metals with each other.  The median metallicities of our low and high metallicity subsets differ by 0.4 dex.  If $M_{\rm mix}$ remained constant during this enrichment phase, we would expect the intrinsic scatter in the lower metallicity subset to be higher by a factor $\sqrt{2.5} \sim 1.6$.  This factor is in fact typical of what we find for many abundance ratios, which provides further circumstantial evidence for the stochastic IMF sampling interpretation.  

If $M_{\rm mix}$ were below $\sim 10^4 \msun$, the predicted scatter would exceed the level that we find in our sample.  There is a caveat to this statement, as Equation \ref{eq:Mmix} implicitly assumes that all of the metals produced by CCSN remain in the star-forming ISM.  If a large fraction of CCSN metals are ejected from the star-forming proto-Galaxy or locked in a hot phase with a very long cooling time, then the number of CCSN needed to enrich the remaining gas to a given metallicity is larger, implying larger $N_{\rm CCSN}$ and lower scatter for a given $M_{\rm mix}$.  At the opposite end, a value of $M_{\rm mix}$ larger than $\sim 10^6 \msun$ would imply a smaller intrinsic scatter than we find, so a mechanism other than stochastic CCSN sampling would have to dominate the scatter in abundance ratios.

The order-of-magnitude arguments presented here illustrate the power of abundance ratio scatter to test theoretical predictions of star formation, gas flows, and ISM mixing in the early Galaxy.  Neutron capture elements with larger AGB contributions, which we will investigate in future work, will provide further diagnostic power.

\section{Summary} \label{sec:summary}

In this work, we present the stellar abundances for Mg, Si, Ca, Ti, Na, Sc, V, Cr, Mn, Fe, Co, and Ni for 86 metal poor ($-2 \lesssim \feh \lesssim -1$) subgiants, derived from spectral synthesis of high-resolution PEPSI spectra. By targeting stars within a small range of $\logg$, we minimize the impact of systematic uncertainties with $\teff$ and $\logg$ on our abundances. We utilize the spectral analysis code iSpec and spectral synthesis code MOOG to calculate the LTE abundances of the 12 elements listed above (Table~\ref{tab:abunds}). We compare our abundance trends to those from GALAH DR3 \citep[][Figures~\ref{fig:x_fe} and~\ref{fig:x_mg}]{buder2021, griffith2022}, finding that most [X/Fe] vs. [Fe/H] and [X/Mg] vs. [Mg/H] trends are consistent with an extrapolation of the GALAH DR3 low-Ia medians. The initial [Mg/Fe] abundance falls below the observed plateau in the GALAH sample of at 0.3 dex at $\feh \sim -1$. We further find that the individual \ion{Mg}{1} lines are systematically offset above and below the median Mg abundance of all lines. We apply line-by-line offsets to the [Mg/H] values for all stars (Section~\ref{subsec:mg}) to alleviate the discrepancy in the abundances derived by individual lines and to align our [Mg/Fe] values with the plateau at 0.3 dex. 

With robust stellar abundances in hand, we present multiple diagnostics of the abundance scatter (Section~\ref{sec:scatter}). We estimate the photon-noise contributed scatter by varying the spectra about the observed flux by the noise and calculating the standard deviation of the abundances derived from 10 fluctuated spectra. We find that $\sigma_{\rm photon}$ is $<0.04$ dex for all elements in [X/Fe] and $<0.05$ dex in [X/Mg]. For most elements, the photon-noise is $\sim0.02-0.03$ dex. We then measure the RMS and intrinsic scatter in [X/Fe] and [X/Mg] about a two and three-parameter model. The magnitude of the RMS scatter is larger than the photon-noise in all cases, an indication that we are measuring intrinsic abundance variations. Relative to the two-parameter model, we find similar magnitudes of intrinsic scatter in [X/Fe] and [X/Mg] for the $\alpha$ and light odd-$Z$-elements. Conversely, the scatter in [X/Fe] for Fe-peak elements Cr, Mn, and Ni is smaller than the scatter in [X/Mg]. Adding a third model parameter ($Q_{\alpha}$) reduces the estimated intrinsic scatter for Mg, Ca, Si, and Ti by a factor of $\sim 1.5$, but it does not significantly impact the measured scatter in Fe-peak elements. We present measurements of scatter in Tables~\ref{tab:scat_xfe} and~\ref{tab:scat_xmg} and Figures~\ref{fig:rms_2v3} and~\ref{fig:rms_2v3_scatter}.

As we expect the level of scatter to vary with [Fe/H] and [Mg/H], we calculate the intrinsic scatter in the low and high-metallicity halves of our sample. We find that scatter below the median [Fe/H] and [Mg/H] tends to be larger than scatter above the medians (Figure~\ref{fig:intrin}). This effect is more prevalent in [X/Mg], where the scatter in the low-metallicity sample is larger than that of the high metallicity sample by a factor of 1.5 or more for all elements. We measure larger intrinsic scatter in the high-metallicity sample than the low-metallicity sample for [Na/Fe] and [Mn/Fe].

In Section~\ref{sec:kin} we investigate the kinematic groups present in our stellar sample, separating our stars into \textit{in situ} and accreted subsamples. We observe that the [X/Fe] vs. [Fe/H] abundance trends of the two subsamples diverge near $\feh$ of $-1$ for Mg, Ca, Si, Ti, Na, and Ni (Figure~\ref{fig:xfe_kin}). We find small variations in $\sigma_{\rm intrin}$ between the \textit{in situ} and accreted samples for all elements but Ti, Na, and Sc, and small variations between the \textit{in situ} sample and full population (Figure~\ref{fig:sigma_kin}). The similarity of intrinsic scatter between the populations indicates that contamination from accreted stars does not skew our measured intrinsic scatter or estimate of CCSN enrichment. While the chemical evolution history of the Milky Way disk and accreted satellites may differ, we find that the magnitude of intrinsic scatter in most [X/Fe] abundances does not. 

We considered a variety of possible origins for the intrinsic abundance ratio scatter in our sample.  The two most straightforward mechanisms are variation in the ratio of SNIa/CCSN enrichment and stochastic sampling of the supernova population.  For elements with metallicity-dependent yields, additional scatter could arise from complex enrichment histories shaped by star formation bursts, accretion events, and outflows.  Variation of SNIa/CCSN alone cannot explain all aspects of our data, as we find significant intrinsic scatter within the $\alpha$-elements and within the Fe-peak elements, and controlling for $\afe$ reduces but does not eliminate abundance ratio scatter.  Stochastic sampling causes the yields that enrich a given forming star to differ from the population-averaged yields, so it can produce scatter in individual abundance ratios as well as coherent groups of elements.  

We quantitatively investigated the hypothesis that the intrinsic scatter arises from stochastic sampling of CCSN progenitor masses, assuming a \cite{kroupa2001} IMF and the Z9.6+W18 model grid from \cite{sukhbold2016}.  Although these models are computed for solar metallicity progenitors, the grid has the fine mass resolution needed for this calculation, and it includes a realistic treatment of supernova-driven explosions and black hole formation.  This stochastic IMF sampling model is remarkably successful at explaining our observations.  The RMS intrinsic scatter of most abundance ratios is well reproduced if the effective number of supernovae contributing to a typical star's abundances is $N_{\rm CCSN} \sim 50$ (Figure~\ref{fig:sigma_xmg}).  The scatter of Ni and Co is high compared to the model predictions, implying a lower $N_{\rm CCSN} \sim 10$, while the scatter of Ca, Sc, and Mn is lower than predicted, implying $N_{\rm CCSN} \sim 150-200$ (Figure~\ref{fig:n_ccsn}).  The model predicts correlated abundance deviations --- e.g., stars high in [Si/Mg] are also likely to be high in [Ca/Mg], [Ti/Mg], [Sc/Mg], and [Ni/Mg] --- in a pattern that qualitatively matches our measurements (Figure~\ref{fig:corrs}).

Taking $N_{\rm CCSN} \sim 50$ as a representative value, we infer that a typical star near our sample's median metallicity ($\mgh \approx -1$, $\feh \approx -1.4$) is enriched by CCSN ejecta mixed over a gas mass $M_{\rm mix} \sim 6 \times 10^4 \msun$, roughly the mass scale of a present-day GMC.  If this mixing mass stays constant for stars in the $-2 < \feh < -1$ range of our sample then the scatter for our low metallicity and high metallicity subsets should differ by a factor $\sim 1.6$, in agreement with our measurements.  Although there are significant uncertainties in this $M_{\rm mix}$ estimate, it illustrates the ability of abundance scatter measurements to constrain the physics of star formation, gas flows, and ISM mixing in the early Galaxy.

Empirically, further insights should come from the abundance scatter of neutron capture elements, which have contributions from AGB enrichment and perhaps from rare exotic sources such as neutron star mergers or magnetar-driven supernovae.  We will undertake such analysis with our PEPSI spectra in future work.  Theoretically, our intrinsic scatter measurements provide a target for simulations of galaxy formation that have the resolution to track enrichment and feedback at the individual supernova level and ISM mixing processes at the $10^3-10^4 \msun$ mass scale.  Chemo-dynamical studies show that the metallicity range spanned by our sample is one over which the Milky Way transitioned from disordered kinematics to a rotationally supported system \citep{belokurov2022,conroy2022,rix2022}.  The detailed abundance patterns of stars across this transition, including the individual element scatter about average trends, will help to reveal how today's Galactic disk emerged from the chaotic environment of the proto-Galaxy.

\section{Acknowledgements} 

We thank Adam Leroy, Tuguldur Sukhbold, Yuan-Sen Ting, Todd Thompson, Fiorenzo Vincenzo, and the OSU Galaxy Group for helpful discussions of abundance scatter, stellar yields, and stochastic sampling of the IMF. We also thank the many LBT observers who collected data for this paper: S. Bose, N. Emami, P. Garnavich, R.T. Gatto, C. Howk, T. Jayasinghe, C. Kochanek, O. Kuhn, A. Pai Asnodkar, D. Reed, B. Rothberg, J. Sullivan, A. Tuli, P. Vallely, M. Whittle, J. Williams, and C. Wood.

This work was supported by NSF grant AST-1909841.

E.J.G. is supported by an NSF Astronomy and Astrophysics Postdoctoral Fellowship under award AST-2202135 and was supported during part of this work by an Ohio State University Presidential Graduate Fellowship.

The LBT is an international collaboration among institutions in the United States, Italy and Germany. LBT Corporation partners are: The University of Arizona on behalf of the Arizona Board of Regents; Istituto Nazionale di Astrofisica, Italy; LBT Beteiligungsgesellschaft, Germany, representing the Max-Planck Society, The Leibniz Institute for Astrophysics Potsdam, and Heidelberg University; The Ohio State University, representing OSU, University of Notre Dame, University of Minnesota and University of Virginia.

\software{Matplotlib \citep{hunter2007}, NumPy \citep{harris2020}, Pandas \citep{pandasa, pandasb}, Astropy \citep{astropy2013, astropy2018, astropy2022}, astroNN \citep{bovy2017}, EXOFASTv2 \citep{eastman2013, eastman2019}, iSpec \citep{blanco2014, blanco2019}, MOOG \citep{sneden1973}, galpy \citep{bovy2015, mackereth2018}}

\bibliography{sample631}{}

\begin{thebibliography}{}
\expandafter\ifx\csname natexlab\endcsname\relax\def\natexlab#1{#1}\fi
\providecommand{\url}[1]{\href{#1}{#1}}
\providecommand{\dodoi}[1]{doi:~\href{http://doi.org/#1}{\nolinkurl{#1}}}
\providecommand{\doeprint}[1]{\href{http://ascl.net/#1}{\nolinkurl{http://ascl.net/#1}}}
\providecommand{\doarXiv}[1]{\href{https://arxiv.org/abs/#1}{\nolinkurl{https://arxiv.org/abs/#1}}}

\bibitem[{{Abohalima} \& {Frebel}(2018)}]{abohalima2018}
{Abohalima}, A., \& {Frebel}, A. 2018, \apjs, 238, 36,
  \dodoi{10.3847/1538-4365/aadfe9}

\bibitem[{{Andrews} {et~al.}(2017){Andrews}, {Weinberg}, {Sch{\"o}nrich}, \&
  {Johnson}}]{andrews2017}
{Andrews}, B.~H., {Weinberg}, D.~H., {Sch{\"o}nrich}, R., \& {Johnson}, J.~A.
  2017, \apj, 835, 224, \dodoi{10.3847/1538-4357/835/2/224}

\bibitem[{{Asplund} {et~al.}(2009){Asplund}, {Grevesse}, {Sauval}, \&
  {Scott}}]{asplund2009}
{Asplund}, M., {Grevesse}, N., {Sauval}, A.~J., \& {Scott}, P. 2009, \araa, 47,
  481, \dodoi{10.1146/annurev.astro.46.060407.145222}

\bibitem[{{Astropy Collaboration} {et~al.}(2013){Astropy Collaboration},
  {Robitaille}, {Tollerud}, {Greenfield}, {Droettboom}, {Bray}, {Aldcroft},
  {Davis}, {Ginsburg}, {Price-Whelan}, {Kerzendorf}, {Conley}, {Crighton},
  {Barbary}, {Muna}, {Ferguson}, {Grollier}, {Parikh}, {Nair}, {Unther},
  {Deil}, {Woillez}, {Conseil}, {Kramer}, {Turner}, {Singer}, {Fox}, {Weaver},
  {Zabalza}, {Edwards}, {Azalee Bostroem}, {Burke}, {Casey}, {Crawford},
  {Dencheva}, {Ely}, {Jenness}, {Labrie}, {Lim}, {Pierfederici}, {Pontzen},
  {Ptak}, {Refsdal}, {Servillat}, \& {Streicher}}]{astropy2013}
{Astropy Collaboration}, {Robitaille}, T.~P., {Tollerud}, E.~J., {et~al.} 2013,
  \aap, 558, A33, \dodoi{10.1051/0004-6361/201322068}

\bibitem[{{Astropy Collaboration} {et~al.}(2018){Astropy Collaboration},
  {Price-Whelan}, {Sip{\H{o}}cz}, {G{\"u}nther}, {Lim}, {Crawford}, {Conseil},
  {Shupe}, {Craig}, {Dencheva}, {Ginsburg}, {Vand erPlas}, {Bradley},
  {P{\'e}rez-Su{\'a}rez}, {de Val-Borro}, {Aldcroft}, {Cruz}, {Robitaille},
  {Tollerud}, {Ardelean}, {Babej}, {Bach}, {Bachetti}, {Bakanov}, {Bamford},
  {Barentsen}, {Barmby}, {Baumbach}, {Berry}, {Biscani}, {Boquien}, {Bostroem},
  {Bouma}, {Brammer}, {Bray}, {Breytenbach}, {Buddelmeijer}, {Burke},
  {Calderone}, {Cano Rodr{\'\i}guez}, {Cara}, {Cardoso}, {Cheedella}, {Copin},
  {Corrales}, {Crichton}, {D'Avella}, {Deil}, {Depagne}, {Dietrich}, {Donath},
  {Droettboom}, {Earl}, {Erben}, {Fabbro}, {Ferreira}, {Finethy}, {Fox},
  {Garrison}, {Gibbons}, {Goldstein}, {Gommers}, {Greco}, {Greenfield},
  {Groener}, {Grollier}, {Hagen}, {Hirst}, {Homeier}, {Horton}, {Hosseinzadeh},
  {Hu}, {Hunkeler}, {Ivezi{\'c}}, {Jain}, {Jenness}, {Kanarek}, {Kendrew},
  {Kern}, {Kerzendorf}, {Khvalko}, {King}, {Kirkby}, {Kulkarni}, {Kumar},
  {Lee}, {Lenz}, {Littlefair}, {Ma}, {Macleod}, {Mastropietro}, {McCully},
  {Montagnac}, {Morris}, {Mueller}, {Mumford}, {Muna}, {Murphy}, {Nelson},
  {Nguyen}, {Ninan}, {N{\"o}the}, {Ogaz}, {Oh}, {Parejko}, {Parley}, {Pascual},
  {Patil}, {Patil}, {Plunkett}, {Prochaska}, {Rastogi}, {Reddy Janga},
  {Sabater}, {Sakurikar}, {Seifert}, {Sherbert}, {Sherwood-Taylor}, {Shih},
  {Sick}, {Silbiger}, {Singanamalla}, {Singer}, {Sladen}, {Sooley},
  {Sornarajah}, {Streicher}, {Teuben}, {Thomas}, {Tremblay}, {Turner},
  {Terr{\'o}n}, {van Kerkwijk}, {de la Vega}, {Watkins}, {Weaver}, {Whitmore},
  {Woillez}, {Zabalza}, \& {Astropy Contributors}}]{astropy2018}
{Astropy Collaboration}, {Price-Whelan}, A.~M., {Sip{\H{o}}cz}, B.~M., {et~al.}
  2018, \aj, 156, 123, \dodoi{10.3847/1538-3881/aabc4f}

\bibitem[{{Astropy Collaboration} {et~al.}(2022){Astropy Collaboration},
  {Price-Whelan}, {Lim}, {Earl}, {Starkman}, {Bradley}, {Shupe}, {Patil},
  {Corrales}, {Brasseur}, {N{"o}the}, {Donath}, {Tollerud}, {Morris},
  {Ginsburg}, {Vaher}, {Weaver}, {Tocknell}, {Jamieson}, {van Kerkwijk},
  {Robitaille}, {Merry}, {Bachetti}, {G{"u}nther}, {Aldcroft},
  {Alvarado-Montes}, {Archibald}, {B{'o}di}, {Bapat}, {Barentsen}, {Baz{'a}n},
  {Biswas}, {Boquien}, {Burke}, {Cara}, {Cara}, {Conroy}, {Conseil}, {Craig},
  {Cross}, {Cruz}, {D'Eugenio}, {Dencheva}, {Devillepoix}, {Dietrich},
  {Eigenbrot}, {Erben}, {Ferreira}, {Foreman-Mackey}, {Fox}, {Freij}, {Garg},
  {Geda}, {Glattly}, {Gondhalekar}, {Gordon}, {Grant}, {Greenfield}, {Groener},
  {Guest}, {Gurovich}, {Handberg}, {Hart}, {Hatfield-Dodds}, {Homeier},
  {Hosseinzadeh}, {Jenness}, {Jones}, {Joseph}, {Kalmbach}, {Karamehmetoglu},
  {Ka{l}uszy{'n}ski}, {Kelley}, {Kern}, {Kerzendorf}, {Koch}, {Kulumani},
  {Lee}, {Ly}, {Ma}, {MacBride}, {Maljaars}, {Muna}, {Murphy}, {Norman},
  {O'Steen}, {Oman}, {Pacifici}, {Pascual}, {Pascual-Granado}, {Patil},
  {Perren}, {Pickering}, {Rastogi}, {Roulston}, {Ryan}, {Rykoff}, {Sabater},
  {Sakurikar}, {Salgado}, {Sanghi}, {Saunders}, {Savchenko}, {Schwardt},
  {Seifert-Eckert}, {Shih}, {Jain}, {Shukla}, {Sick}, {Simpson},
  {Singanamalla}, {Singer}, {Singhal}, {Sinha}, {Sip{H{o}}cz}, {Spitler},
  {Stansby}, {Streicher}, {{{S}}umak}, {Swinbank}, {Taranu}, {Tewary},
  {Tremblay}, {Val-Borro}, {Van Kooten}, {Vasovi{'c}}, {Verma}, {de Miranda
  Cardoso}, {Williams}, {Wilson}, {Winkel}, {Wood-Vasey}, {Xue}, {Yoachim},
  {Zhang}, {Zonca}, \& {Astropy Project Contributors}}]{astropy2022}
{Astropy Collaboration}, {Price-Whelan}, A.~M., {Lim}, P.~L., {et~al.} 2022,
  ApJ, 935, 167, \dodoi{10.3847/1538-4357/ac7c74}

\bibitem[{{Belokurov} {et~al.}(2018){Belokurov}, {Erkal}, {Evans}, {Koposov},
  \& {Deason}}]{belokurov2018}
{Belokurov}, V., {Erkal}, D., {Evans}, N.~W., {Koposov}, S.~E., \& {Deason},
  A.~J. 2018, \mnras, 478, 611, \dodoi{10.1093/mnras/sty982}

\bibitem[{{Belokurov} \& {Kravtsov}(2022)}]{belokurov2022}
{Belokurov}, V., \& {Kravtsov}, A. 2022, \mnras, 514, 689,
  \dodoi{10.1093/mnras/stac1267}

\bibitem[{{Bensby} {et~al.}(2003){Bensby}, {Feltzing}, \&
  {Lundstr{\"o}m}}]{bensby2003}
{Bensby}, T., {Feltzing}, S., \& {Lundstr{\"o}m}, I. 2003, \aap, 410, 527,
  \dodoi{10.1051/0004-6361:20031213}

\bibitem[{{Bensby} {et~al.}(2014){Bensby}, {Feltzing}, \& {Oey}}]{bensby2014}
{Bensby}, T., {Feltzing}, S., \& {Oey}, M.~S. 2014, \aap, 562, A71,
  \dodoi{10.1051/0004-6361/201322631}

\bibitem[{{Bergemann}(2011)}]{bergemann2011}
{Bergemann}, M. 2011, \mnras, 413, 2184,
  \dodoi{10.1111/j.1365-2966.2011.18295.x}

\bibitem[{{Bergemann} \& {Cescutti}(2010)}]{bergemann2010a}
{Bergemann}, M., \& {Cescutti}, G. 2010, \aap, 522, A9,
  \dodoi{10.1051/0004-6361/201014250}

\bibitem[{{Bergemann} {et~al.}(2017){Bergemann}, {Collet}, {Amarsi}, {Kovalev},
  {Ruchti}, \& {Magic}}]{bergemann2017}
{Bergemann}, M., {Collet}, R., {Amarsi}, A.~M., {et~al.} 2017, \apj, 847, 15,
  \dodoi{10.3847/1538-4357/aa88cb}

\bibitem[{{Bergemann} \& {Gehren}(2008)}]{bergemann2008}
{Bergemann}, M., \& {Gehren}, T. 2008, \aap, 492, 823,
  \dodoi{10.1051/0004-6361:200810098}

\bibitem[{{Bergemann} {et~al.}(2013){Bergemann}, {Kudritzki}, {W{\"u}rl},
  {Plez}, {Davies}, \& {Gazak}}]{bergemann2013}
{Bergemann}, M., {Kudritzki}, R.-P., {W{\"u}rl}, M., {et~al.} 2013, \apj, 764,
  115, \dodoi{10.1088/0004-637X/764/2/115}

\bibitem[{{Bergemann} {et~al.}(2010){Bergemann}, {Pickering}, \&
  {Gehren}}]{bergemann2010b}
{Bergemann}, M., {Pickering}, J.~C., \& {Gehren}, T. 2010, \mnras, 401, 1334,
  \dodoi{10.1111/j.1365-2966.2009.15736.x}

\bibitem[{{Bertran de Lis} {et~al.}(2016){Bertran de Lis}, {Allende Prieto},
  {Majewski}, {Schiavon}, {Holtzman}, {Shetrone}, {Carrera}, {Garc{\'\i}a
  P{\'e}rez}, {M{\'e}sz{\'a}ros}, {Frinchaboy}, {Hearty}, {Nidever},
  {Zasowski}, \& {Ge}}]{bertran2016}
{Bertran de Lis}, S., {Allende Prieto}, C., {Majewski}, S.~R., {et~al.} 2016,
  \aap, 590, A74, \dodoi{10.1051/0004-6361/201527827}

\bibitem[{{Bestenlehner} {et~al.}(2020){Bestenlehner}, {Crowther},
  {Caballero-Nieves}, {Schneider}, {Sim{\'o}n-D{\'\i}az}, {Brands}, {de Koter},
  {Gr{\"a}fener}, {Herrero}, {Langer}, {Lennon}, {Ma{\'\i}z Apell{\'a}niz},
  {Puls}, \& {Vink}}]{bestenlehner2020}
{Bestenlehner}, J.~M., {Crowther}, P.~A., {Caballero-Nieves}, S.~M., {et~al.}
  2020, \mnras, 499, 1918, \dodoi{10.1093/mnras/staa2801}

\bibitem[{{Blanco-Cuaresma}(2019)}]{blanco2019}
{Blanco-Cuaresma}, S. 2019, \mnras, 486, 2075, \dodoi{10.1093/mnras/stz549}

\bibitem[{{Blanco-Cuaresma} {et~al.}(2014){Blanco-Cuaresma}, {Soubiran},
  {Heiter}, \& {Jofr{\'e}}}]{blanco2014}
{Blanco-Cuaresma}, S., {Soubiran}, C., {Heiter}, U., \& {Jofr{\'e}}, P. 2014,
  \aap, 569, A111, \dodoi{10.1051/0004-6361/201423945}

\bibitem[{{Bovy}(2015)}]{bovy2015}
{Bovy}, J. 2015, \apjs, 216, 29, \dodoi{10.1088/0067-0049/216/2/29}

\bibitem[{{Bovy}(2017)}]{bovy2017}
---. 2017, \mnras, 470, 1360, \dodoi{10.1093/mnras/stx1277}

\bibitem[{{Buder} {et~al.}(2021){Buder}, {Sharma}, {Kos}, {Amarsi},
  {Nordlander}, {Lind}, {Martell}, {Asplund}, {Bland-Hawthorn}, {Casey}, {de
  Silva}, {D'Orazi}, {Freeman}, {Hayden}, {Lewis}, {Lin}, {Schlesinger},
  {Simpson}, {Stello}, {Zucker}, {Zwitter}, {Beeson}, {Buck}, {Casagrande},
  {Clark}, {{\v{C}}otar}, {da Costa}, {de Grijs}, {Feuillet}, {Horner},
  {Kafle}, {Khanna}, {Kobayashi}, {Liu}, {Montet}, {Nandakumar}, {Nataf},
  {Ness}, {Spina}, {Tepper-Garc{\'\i}a}, {Ting}, {Traven},
  {Vogrin{\v{c}}i{\v{c}}}, {Wittenmyer}, {Wyse}, {{\v{Z}}erjal}, \& {GALAH
  Collaboration}}]{buder2021}
{Buder}, S., {Sharma}, S., {Kos}, J., {et~al.} 2021, \mnras, 506, 150,
  \dodoi{10.1093/mnras/stab1242}

\bibitem[{{Chaplin} {et~al.}(2020){Chaplin}, {Serenelli}, {Miglio}, {Morel},
  {Mackereth}, {Vincenzo}, {Kjeldsen}, {Basu}, {Ball}, {Stokholm}, {Verma},
  {Mosumgaard}, {Silva Aguirre}, {Mazumdar}, {Ranadive}, {Antia}, {Lebreton},
  {Ong}, {Appourchaux}, {Bedding}, {Christensen-Dalsgaard}, {Creevey},
  {Garc{\'\i}a}, {Handberg}, {Huber}, {Kawaler}, {Lund}, {Metcalfe}, {Stassun},
  {Bazot}, {Beck}, {Bell}, {Bergemann}, {Buzasi}, {Benomar}, {Bossini},
  {Bugnet}, {Campante}, {Orhan}, {Corsaro}, {Gonz{\'a}lez-Cuesta}, {Davies},
  {Di Mauro}, {Egeland}, {Elsworth}, {Gaulme}, {Ghasemi}, {Guo}, {Hall},
  {Hasanzadeh}, {Hekker}, {Howe}, {Jenkins}, {Jim{\'e}nez}, {Kiefer},
  {Kuszlewicz}, {Kallinger}, {Latham}, {Lundkvist}, {Mathur}, {Montalb{\'a}n},
  {Mosser}, {Bed{\'o}n}, {Nielsen}, {{\"O}rtel}, {Rendle}, {Ricker},
  {Rodrigues}, {Roxburgh}, {Safari}, {Schofield}, {Seager}, {Smalley},
  {Stello}, {Szab{\'o}}, {Tayar}, {Theme{\ss}l}, {Thomas}, {Vanderspek}, {van
  Rossem}, {Vrard}, {Weiss}, {White}, {Winn}, \& {Y{\i}ld{\i}z}}]{chaplin2020}
{Chaplin}, W.~J., {Serenelli}, A.~M., {Miglio}, A., {et~al.} 2020, Nature
  Astronomy, 4, 382, \dodoi{10.1038/s41550-019-0975-9}

\bibitem[{{Chen} {et~al.}(2022){Chen}, {Hayden}, {Sharma}, {Bland-Hawthorn},
  {Kobayashi}, \& {Karakas}}]{chen2022}
{Chen}, B., {Hayden}, M.~R., {Sharma}, S., {et~al.} 2022, arXiv e-prints,
  arXiv:2204.11413.
\newblock \doarXiv{2204.11413}

\bibitem[{{Chiappini} {et~al.}(1997){Chiappini}, {Matteucci}, \&
  {Gratton}}]{chiappini1997}
{Chiappini}, C., {Matteucci}, F., \& {Gratton}, R. 1997, \apj, 477, 765,
  \dodoi{10.1086/303726}

\bibitem[{{Clarke} {et~al.}(2019){Clarke}, {Debattista}, {Nidever}, {Loebman},
  {Simons}, {Kassin}, {Du}, {Ness}, {Fisher}, {Quinn}, {Wadsley}, {Freeman}, \&
  {Popescu}}]{clarke2019}
{Clarke}, A.~J., {Debattista}, V.~P., {Nidever}, D.~L., {et~al.} 2019, \mnras,
  484, 3476, \dodoi{10.1093/mnras/stz104}

\bibitem[{{Conroy} {et~al.}(2019){Conroy}, {Naidu}, {Zaritsky}, {Bonaca},
  {Cargile}, {Johnson}, \& {Caldwell}}]{conroy2019}
{Conroy}, C., {Naidu}, R.~P., {Zaritsky}, D., {et~al.} 2019, \apj, 887, 237,
  \dodoi{10.3847/1538-4357/ab5710}

\bibitem[{{Conroy} {et~al.}(2022){Conroy}, {Weinberg}, {Naidu}, {Buck},
  {Johnson}, {Cargile}, {Bonaca}, {Caldwell}, {Chandra}, {Han}, {Johnson},
  {Speagle}, {Ting}, {Woody}, \& {Zaritsky}}]{conroy2022}
{Conroy}, C., {Weinberg}, D.~H., {Naidu}, R.~P., {et~al.} 2022, arXiv e-prints,
  arXiv:2204.02989.
\newblock \doarXiv{2204.02989}

\bibitem[{{Cutri} {et~al.}(2003){Cutri}, {Skrutskie}, {van Dyk}, {Beichman},
  {Carpenter}, {Chester}, {Cambresy}, {Evans}, {Fowler}, {Gizis}, {Howard},
  {Huchra}, {Jarrett}, {Kopan}, {Kirkpatrick}, {Light}, {Marsh}, {McCallon},
  {Schneider}, {Stiening}, {Sykes}, {Weinberg}, {Wheaton}, {Wheelock}, \&
  {Zacarias}}]{cutri2003}
{Cutri}, R.~M., {Skrutskie}, M.~F., {van Dyk}, S., {et~al.} 2003, {2MASS All
  Sky Catalog of point sources.}

\bibitem[{{Cutri} {et~al.}(2021){Cutri}, {Wright}, {Conrow}, {Fowler},
  {Eisenhardt}, {Grillmair}, {Kirkpatrick}, {Masci}, {McCallon}, {Wheelock},
  {Fajardo-Acosta}, {Yan}, {Benford}, {Harbut}, {Jarrett}, {Lake}, {Leisawitz},
  {Ressler}, {Stanford}, {Tsai}, {Liu}, {Helou}, {Mainzer}, {Gettngs},
  {Gonzalez}, {Hoffman}, {Marsh}, {Padgett}, {Skrutskie}, {Beck}, {Papin}, \&
  {Wittman}}]{cutri2014}
{Cutri}, R.~M., {Wright}, E.~L., {Conrow}, T., {et~al.} 2021, VizieR Online
  Data Catalog, II/328

\bibitem[{{Das} {et~al.}(2020){Das}, {Hawkins}, \& {Jofr{\'e}}}]{das2020}
{Das}, P., {Hawkins}, K., \& {Jofr{\'e}}, P. 2020, \mnras, 493, 5195,
  \dodoi{10.1093/mnras/stz3537}

\bibitem[{{Eastman}(2017)}]{eastman2017}
{Eastman}, J. 2017, {EXOFASTv2: Generalized publication-quality exoplanet
  modeling code}, Astrophysics Source Code Library, record ascl:1710.003.
\newblock \doeprint{1710.003}

\bibitem[{{Eastman} {et~al.}(2013){Eastman}, {Gaudi}, \& {Agol}}]{eastman2013}
{Eastman}, J., {Gaudi}, B.~S., \& {Agol}, E. 2013, \pasp, 125, 83,
  \dodoi{10.1086/669497}

\bibitem[{{Eastman} {et~al.}(2019){Eastman}, {Rodriguez}, {Agol}, {Stassun},
  {Beatty}, {Vanderburg}, {Gaudi}, {Collins}, \& {Luger}}]{eastman2019}
{Eastman}, J.~D., {Rodriguez}, J.~E., {Agol}, E., {et~al.} 2019, arXiv
  e-prints, arXiv:1907.09480.
\newblock \doarXiv{1907.09480}

\bibitem[{{Ertl} {et~al.}(2016){Ertl}, {Janka}, {Woosley}, {Sukhbold}, \&
  {Ugliano}}]{ertl2016}
{Ertl}, T., {Janka}, H.~T., {Woosley}, S.~E., {Sukhbold}, T., \& {Ugliano}, M.
  2016, \apj, 818, 124, \dodoi{10.3847/0004-637X/818/2/124}

\bibitem[{{Fuhrmann}(1998)}]{fuhrmann1998}
{Fuhrmann}, K. 1998, \aap, 338, 161

\bibitem[{{Fulbright}(2000)}]{fulbright2000}
{Fulbright}, J.~P. 2000, \aj, 120, 1841, \dodoi{10.1086/301548}

\bibitem[{{Gaia Collaboration} {et~al.}(2018){Gaia Collaboration}, {Brown},
  {Vallenari}, {Prusti}, {de Bruijne}, {Babusiaux}, {Bailer-Jones}, {Biermann},
  {Evans}, {Eyer}, {Jansen}, {Jordi}, {Klioner}, {Lammers}, {Lindegren},
  {Luri}, {Mignard}, {Panem}, {Pourbaix}, {Randich}, {Sartoretti}, {Siddiqui},
  {Soubiran}, {van Leeuwen}, {Walton}, {Arenou}, {Bastian}, {Cropper},
  {Drimmel}, {Katz}, {Lattanzi}, {Bakker}, {Cacciari}, {Casta{\~n}eda},
  {Chaoul}, {Cheek}, {De Angeli}, {Fabricius}, {Guerra}, {Holl}, {Masana},
  {Messineo}, {Mowlavi}, {Nienartowicz}, {Panuzzo}, {Portell}, {Riello},
  {Seabroke}, {Tanga}, {Th{\'e}venin}, {Gracia-Abril}, {Comoretto},
  {Garcia-Reinaldos}, {Teyssier}, {Altmann}, {Andrae}, {Audard},
  {Bellas-Velidis}, {Benson}, {Berthier}, {Blomme}, {Burgess}, {Busso},
  {Carry}, {Cellino}, {Clementini}, {Clotet}, {Creevey}, {Davidson}, {De
  Ridder}, {Delchambre}, {Dell'Oro}, {Ducourant},
  {Fern{\'a}ndez-Hern{\'a}ndez}, {Fouesneau}, {Fr{\'e}mat}, {Galluccio},
  {Garc{\'\i}a-Torres}, {Gonz{\'a}lez-N{\'u}{\~n}ez}, {Gonz{\'a}lez-Vidal},
  {Gosset}, {Guy}, {Halbwachs}, {Hambly}, {Harrison}, {Hern{\'a}ndez},
  {Hestroffer}, {Hodgkin}, {Hutton}, {Jasniewicz}, {Jean-Antoine-Piccolo},
  {Jordan}, {Korn}, {Krone-Martins}, {Lanzafame}, {Lebzelter}, {L{\"o}ffler},
  {Manteiga}, {Marrese}, {Mart{\'\i}n-Fleitas}, {Moitinho}, {Mora}, {Muinonen},
  {Osinde}, {Pancino}, {Pauwels}, {Petit}, {Recio-Blanco}, {Richards},
  {Rimoldini}, {Robin}, {Sarro}, {Siopis}, {Smith}, {Sozzetti}, {S{\"u}veges},
  {Torra}, {van Reeven}, {Abbas}, {Abreu Aramburu}, {Accart}, {Aerts},
  {Altavilla}, {{\'A}lvarez}, {Alvarez}, {Alves}, {Anderson}, {Andrei},
  {Anglada Varela}, {Antiche}, {Antoja}, {Arcay}, {Astraatmadja}, {Bach},
  {Baker}, {Balaguer-N{\'u}{\~n}ez}, {Balm}, {Barache}, {Barata}, {Barbato},
  {Barblan}, {Barklem}, {Barrado}, {Barros}, {Barstow}, {Bartholom{\'e}
  Mu{\~n}oz}, {Bassilana}, {Becciani}, {Bellazzini}, {Berihuete}, {Bertone},
  {Bianchi}, {Bienaym{\'e}}, {Blanco-Cuaresma}, {Boch}, {Boeche}, {Bombrun},
  {Borrachero}, {Bossini}, {Bouquillon}, {Bourda}, {Bragaglia}, {Bramante},
  {Breddels}, {Bressan}, {Brouillet}, {Br{\"u}semeister}, {Brugaletta},
  {Bucciarelli}, {Burlacu}, {Busonero}, {Butkevich}, {Buzzi}, {Caffau},
  {Cancelliere}, {Cannizzaro}, {Cantat-Gaudin}, {Carballo}, {Carlucci},
  {Carrasco}, {Casamiquela}, {Castellani}, {Castro-Ginard}, {Charlot},
  {Chemin}, {Chiavassa}, {Cocozza}, {Costigan}, {Cowell}, {Crifo}, {Crosta},
  {Crowley}, {Cuypers}, {Dafonte}, {Damerdji}, {Dapergolas}, {David}, {David},
  {de Laverny}, {De Luise}, {De March}, {de Martino}, {de Souza}, {de Torres},
  {Debosscher}, {del Pozo}, {Delbo}, {Delgado}, {Delgado}, {Di Matteo},
  {Diakite}, {Diener}, {Distefano}, {Dolding}, {Drazinos}, {Dur{\'a}n},
  {Edvardsson}, {Enke}, {Eriksson}, {Esquej}, {Eynard Bontemps}, {Fabre},
  {Fabrizio}, {Faigler}, {Falc{\~a}o}, {Farr{\`a}s Casas}, {Federici},
  {Fedorets}, {Fernique}, {Figueras}, {Filippi}, {Findeisen}, {Fonti},
  {Fraile}, {Fraser}, {Fr{\'e}zouls}, {Gai}, {Galleti}, {Garabato},
  {Garc{\'\i}a-Sedano}, {Garofalo}, {Garralda}, {Gavel}, {Gavras}, {Gerssen},
  {Geyer}, {Giacobbe}, {Gilmore}, {Girona}, {Giuffrida}, {Glass}, {Gomes},
  {Granvik}, {Gueguen}, {Guerrier}, {Guiraud}, {Guti{\'e}rrez-S{\'a}nchez},
  {Haigron}, {Hatzidimitriou}, {Hauser}, {Haywood}, {Heiter}, {Helmi}, {Heu},
  {Hilger}, {Hobbs}, {Hofmann}, {Holland}, {Huckle}, {Hypki}, {Icardi},
  {Jan{\ss}en}, {Jevardat de Fombelle}, {Jonker}, {Juh{\'a}sz}, {Julbe},
  {Karampelas}, {Kewley}, {Klar}, {Kochoska}, {Kohley}, {Kolenberg},
  {Kontizas}, {Kontizas}, {Koposov}, {Kordopatis}, {Kostrzewa-Rutkowska},
  {Koubsky}, {Lambert}, {Lanza}, {Lasne}, {Lavigne}, {Le Fustec}, {Le
  Poncin-Lafitte}, {Lebreton}, {Leccia}, {Leclerc}, {Lecoeur-Taibi},
  {Lenhardt}, {Leroux}, {Liao}, {Licata}, {Lindstr{\o}m}, {Lister}, {Livanou},
  {Lobel}, {L{\'o}pez}, {Managau}, {Mann}, {Mantelet}, {Marchal}, {Marchant},
  {Marconi}, {Marinoni}, {Marschalk{\'o}}, {Marshall}, {Martino}, {Marton},
  {Mary}, {Massari}, {Matijevi{\v{c}}}, {Mazeh}, {McMillan}, {Messina},
  {Michalik}, {Millar}, {Molina}, {Molinaro}, {Moln{\'a}r}, {Montegriffo},
  {Mor}, {Morbidelli}, {Morel}, {Morris}, {Mulone}, {Muraveva}, {Musella},
  {Nelemans}, {Nicastro}, {Noval}, {O'Mullane}, {Ord{\'e}novic},
  {Ord{\'o}{\~n}ez-Blanco}, {Osborne}, {Pagani}, {Pagano}, {Pailler},
  {Palacin}, {Palaversa}, {Panahi}, {Pawlak}, {Piersimoni}, {Pineau}, {Plachy},
  {Plum}, {Poggio}, {Poujoulet}, {Pr{\v{s}}a}, {Pulone}, {Racero}, {Ragaini},
  {Rambaux}, {Ramos-Lerate}, {Regibo}, {Reyl{\'e}}, {Riclet}, {Ripepi}, {Riva},
  {Rivard}, {Rixon}, {Roegiers}, {Roelens}, {Romero-G{\'o}mez}, {Rowell},
  {Royer}, {Ruiz-Dern}, {Sadowski}, {Sagrist{\`a} Sell{\'e}s}, {Sahlmann},
  {Salgado}, {Salguero}, {Sanna}, {Santana-Ros}, {Sarasso}, {Savietto},
  {Schultheis}, {Sciacca}, {Segol}, {Segovia}, {S{\'e}gransan}, {Shih},
  {Siltala}, {Silva}, {Smart}, {Smith}, {Solano}, {Solitro}, {Sordo}, {Soria
  Nieto}, {Souchay}, {Spagna}, {Spoto}, {Stampa}, {Steele},
  {Steidelm{\"u}ller}, {Stephenson}, {Stoev}, {Suess}, {Surdej}, {Szabados},
  {Szegedi-Elek}, {Tapiador}, {Taris}, {Tauran}, {Taylor}, {Teixeira},
  {Terrett}, {Teyssandier}, {Thuillot}, {Titarenko}, {Torra Clotet}, {Turon},
  {Ulla}, {Utrilla}, {Uzzi}, {Vaillant}, {Valentini}, {Valette}, {van Elteren},
  {Van Hemelryck}, {van Leeuwen}, {Vaschetto}, {Vecchiato}, {Veljanoski},
  {Viala}, {Vicente}, {Vogt}, {von Essen}, {Voss}, {Votruba}, {Voutsinas},
  {Walmsley}, {Weiler}, {Wertz}, {Wevers}, {Wyrzykowski}, {Yoldas},
  {{\v{Z}}erjal}, {Ziaeepour}, {Zorec}, {Zschocke}, {Zucker}, {Zurbach}, \&
  {Zwitter}}]{gaia2018}
{Gaia Collaboration}, {Brown}, A.~G.~A., {Vallenari}, A., {et~al.} 2018, \aap,
  616, A1, \dodoi{10.1051/0004-6361/201833051}

\bibitem[{{Gaia Collaboration} {et~al.}(2021){Gaia Collaboration}, {Brown},
  {Vallenari}, {Prusti}, {de Bruijne}, {Babusiaux}, {Biermann}, {Creevey},
  {Evans}, {Eyer}, {Hutton}, {Jansen}, {Jordi}, {Klioner}, {Lammers},
  {Lindegren}, {Luri}, {Mignard}, {Panem}, {Pourbaix}, {Randich}, {Sartoretti},
  {Soubiran}, {Walton}, {Arenou}, {Bailer-Jones}, {Bastian}, {Cropper},
  {Drimmel}, {Katz}, {Lattanzi}, {van Leeuwen}, {Bakker}, {Cacciari},
  {Casta{\~n}eda}, {De Angeli}, {Ducourant}, {Fabricius}, {Fouesneau},
  {Fr{\'e}mat}, {Guerra}, {Guerrier}, {Guiraud}, {Jean-Antoine Piccolo},
  {Masana}, {Messineo}, {Mowlavi}, {Nicolas}, {Nienartowicz}, {Pailler},
  {Panuzzo}, {Riclet}, {Roux}, {Seabroke}, {Sordo}, {Tanga}, {Th{\'e}venin},
  {Gracia-Abril}, {Portell}, {Teyssier}, {Altmann}, {Andrae}, {Bellas-Velidis},
  {Benson}, {Berthier}, {Blomme}, {Brugaletta}, {Burgess}, {Busso}, {Carry},
  {Cellino}, {Cheek}, {Clementini}, {Damerdji}, {Davidson}, {Delchambre},
  {Dell'Oro}, {Fern{\'a}ndez-Hern{\'a}ndez}, {Galluccio}, {Garc{\'\i}a-Lario},
  {Garcia-Reinaldos}, {Gonz{\'a}lez-N{\'u}{\~n}ez}, {Gosset}, {Haigron},
  {Halbwachs}, {Hambly}, {Harrison}, {Hatzidimitriou}, {Heiter},
  {Hern{\'a}ndez}, {Hestroffer}, {Hodgkin}, {Holl}, {Jan{\ss}en}, {Jevardat de
  Fombelle}, {Jordan}, {Krone-Martins}, {Lanzafame}, {L{\"o}ffler}, {Lorca},
  {Manteiga}, {Marchal}, {Marrese}, {Moitinho}, {Mora}, {Muinonen}, {Osborne},
  {Pancino}, {Pauwels}, {Petit}, {Recio-Blanco}, {Richards}, {Riello},
  {Rimoldini}, {Robin}, {Roegiers}, {Rybizki}, {Sarro}, {Siopis}, {Smith},
  {Sozzetti}, {Ulla}, {Utrilla}, {van Leeuwen}, {van Reeven}, {Abbas}, {Abreu
  Aramburu}, {Accart}, {Aerts}, {Aguado}, {Ajaj}, {Altavilla}, {{\'A}lvarez},
  {{\'A}lvarez Cid-Fuentes}, {Alves}, {Anderson}, {Anglada Varela}, {Antoja},
  {Audard}, {Baines}, {Baker}, {Balaguer-N{\'u}{\~n}ez}, {Balbinot}, {Balog},
  {Barache}, {Barbato}, {Barros}, {Barstow}, {Bartolom{\'e}}, {Bassilana},
  {Bauchet}, {Baudesson-Stella}, {Becciani}, {Bellazzini}, {Bernet}, {Bertone},
  {Bianchi}, {Blanco-Cuaresma}, {Boch}, {Bombrun}, {Bossini}, {Bouquillon},
  {Bragaglia}, {Bramante}, {Breedt}, {Bressan}, {Brouillet}, {Bucciarelli},
  {Burlacu}, {Busonero}, {Butkevich}, {Buzzi}, {Caffau}, {Cancelliere},
  {C{\'a}novas}, {Cantat-Gaudin}, {Carballo}, {Carlucci}, {Carnerero},
  {Carrasco}, {Casamiquela}, {Castellani}, {Castro-Ginard}, {Castro Sampol},
  {Chaoul}, {Charlot}, {Chemin}, {Chiavassa}, {Cioni}, {Comoretto}, {Cooper},
  {Cornez}, {Cowell}, {Crifo}, {Crosta}, {Crowley}, {Dafonte}, {Dapergolas},
  {David}, {David}, {de Laverny}, {De Luise}, {De March}, {De Ridder}, {de
  Souza}, {de Teodoro}, {de Torres}, {del Peloso}, {del Pozo}, {Delbo},
  {Delgado}, {Delgado}, {Delisle}, {Di Matteo}, {Diakite}, {Diener},
  {Distefano}, {Dolding}, {Eappachen}, {Edvardsson}, {Enke}, {Esquej}, {Fabre},
  {Fabrizio}, {Faigler}, {Fedorets}, {Fernique}, {Fienga}, {Figueras},
  {Fouron}, {Fragkoudi}, {Fraile}, {Franke}, {Gai}, {Garabato},
  {Garcia-Gutierrez}, {Garc{\'\i}a-Torres}, {Garofalo}, {Gavras}, {Gerlach},
  {Geyer}, {Giacobbe}, {Gilmore}, {Girona}, {Giuffrida}, {Gomel}, {Gomez},
  {Gonzalez-Santamaria}, {Gonz{\'a}lez-Vidal}, {Granvik},
  {Guti{\'e}rrez-S{\'a}nchez}, {Guy}, {Hauser}, {Haywood}, {Helmi}, {Hidalgo},
  {Hilger}, {H{\l}adczuk}, {Hobbs}, {Holland}, {Huckle}, {Jasniewicz},
  {Jonker}, {Juaristi Campillo}, {Julbe}, {Karbevska}, {Kervella}, {Khanna},
  {Kochoska}, {Kontizas}, {Kordopatis}, {Korn}, {Kostrzewa-Rutkowska},
  {Kruszy{\'n}ska}, {Lambert}, {Lanza}, {Lasne}, {Le Campion}, {Le Fustec},
  {Lebreton}, {Lebzelter}, {Leccia}, {Leclerc}, {Lecoeur-Taibi}, {Liao},
  {Licata}, {Lindstr{\o}m}, {Lister}, {Livanou}, {Lobel}, {Madrero Pardo},
  {Managau}, {Mann}, {Marchant}, {Marconi}, {Marcos Santos}, {Marinoni},
  {Marocco}, {Marshall}, {Martin Polo}, {Mart{\'\i}n-Fleitas}, {Masip},
  {Massari}, {Mastrobuono-Battisti}, {Mazeh}, {McMillan}, {Messina},
  {Michalik}, {Millar}, {Mints}, {Molina}, {Molinaro}, {Moln{\'a}r},
  {Montegriffo}, {Mor}, {Morbidelli}, {Morel}, {Morris}, {Mulone}, {Munoz},
  {Muraveva}, {Murphy}, {Musella}, {Noval}, {Ord{\'e}novic}, {Orr{\`u}},
  {Osinde}, {Pagani}, {Pagano}, {Palaversa}, {Palicio}, {Panahi}, {Pawlak},
  {Pe{\~n}alosa Esteller}, {Penttil{\"a}}, {Piersimoni}, {Pineau}, {Plachy},
  {Plum}, {Poggio}, {Poretti}, {Poujoulet}, {Pr{\v{s}}a}, {Pulone}, {Racero},
  {Ragaini}, {Rainer}, {Raiteri}, {Rambaux}, {Ramos}, {Ramos-Lerate}, {Re
  Fiorentin}, {Regibo}, {Reyl{\'e}}, {Ripepi}, {Riva}, {Rixon}, {Robichon},
  {Robin}, {Roelens}, {Rohrbasser}, {Romero-G{\'o}mez}, {Rowell}, {Royer},
  {Rybicki}, {Sadowski}, {Sagrist{\`a} Sell{\'e}s}, {Sahlmann}, {Salgado},
  {Salguero}, {Samaras}, {Sanchez Gimenez}, {Sanna}, {Santove{\~n}a},
  {Sarasso}, {Schultheis}, {Sciacca}, {Segol}, {Segovia}, {S{\'e}gransan},
  {Semeux}, {Shahaf}, {Siddiqui}, {Siebert}, {Siltala}, {Slezak}, {Smart},
  {Solano}, {Solitro}, {Souami}, {Souchay}, {Spagna}, {Spoto}, {Steele},
  {Steidelm{\"u}ller}, {Stephenson}, {S{\"u}veges}, {Szabados}, {Szegedi-Elek},
  {Taris}, {Tauran}, {Taylor}, {Teixeira}, {Thuillot}, {Tonello}, {Torra},
  {Torra}, {Turon}, {Unger}, {Vaillant}, {van Dillen}, {Vanel}, {Vecchiato},
  {Viala}, {Vicente}, {Voutsinas}, {Weiler}, {Wevers}, {Wyrzykowski}, {Yoldas},
  {Yvard}, {Zhao}, {Zorec}, {Zucker}, {Zurbach}, \& {Zwitter}}]{gaia2021}
---. 2021, \aap, 649, A1, \dodoi{10.1051/0004-6361/202039657}

\bibitem[{{Gilmore} {et~al.}(2012){Gilmore}, {Randich}, {Asplund}, {Binney},
  {Bonifacio}, {Drew}, {Feltzing}, {Ferguson}, {Jeffries}, {Micela},
  {Negueruela}, {Prusti}, {Rix}, {Vallenari}, {Alfaro}, {Allende-Prieto},
  {Babusiaux}, {Bensby}, {Blomme}, {Bragaglia}, {Flaccomio}, {Fran{\c{c}}ois},
  {Irwin}, {Koposov}, {Korn}, {Lanzafame}, {Pancino}, {Paunzen},
  {Recio-Blanco}, {Sacco}, {Smiljanic}, {Van Eck}, {Walton}, {Aden}, {Aerts},
  {Affer}, {Alcala}, {Altavilla}, {Alves}, {Antoja}, {Arenou}, {Argiroffi},
  {Asensio Ramos}, {Bailer-Jones}, {Balaguer-Nunez}, {Bayo}, {Barbuy},
  {Barisevicius}, {Barrado y Navascues}, {Battistini}, {Bellas Velidis},
  {Bellazzini}, {Belokurov}, {Bergemann}, {Bertelli}, {Biazzo}, {Bienayme},
  {Bland-Hawthorn}, {Boeche}, {Bonito}, {Boudreault}, {Bouvier}, {Brandao},
  {Brown}, {de Bruijne}, {Burleigh}, {Caballero}, {Caffau}, {Calura},
  {Capuzzo-Dolcetta}, {Caramazza}, {Carraro}, {Casagrande}, {Casewell},
  {Chapman}, {Chiappini}, {Chorniy}, {Christlieb}, {Cignoni}, {Cocozza},
  {Colless}, {Collet}, {Collins}, {Correnti}, {Covino}, {Crnojevic}, {Cropper},
  {Cunha}, {Damiani}, {David}, {Delgado}, {Duffau}, {Edvardsson}, {Eldridge},
  {Enke}, {Eriksson}, {Evans}, {Eyer}, {Famaey}, {Fellhauer}, {Ferreras},
  {Figueras}, {Fiorentino}, {Flynn}, {Folha}, {Franciosini}, {Frasca},
  {Freeman}, {Fremat}, {Friel}, {Gaensicke}, {Gameiro}, {Garzon}, {Geier},
  {Geisler}, {Gerhard}, {Gibson}, {Gomboc}, {Gomez}, {Gonzalez-Fernandez},
  {Gonzalez Hernandez}, {Gosset}, {Grebel}, {Greimel}, {Groenewegen},
  {Grundahl}, {Guarcello}, {Gustafsson}, {Hadrava}, {Hatzidimitriou}, {Hambly},
  {Hammersley}, {Hansen}, {Haywood}, {Heber}, {Heiter}, {Held}, {Helmi},
  {Hensler}, {Herrero}, {Hill}, {Hodgkin}, {Huelamo}, {Huxor}, {Ibata},
  {Jackson}, {de Jong}, {Jonker}, {Jordan}, {Jordi}, {Jorissen}, {Katz},
  {Kawata}, {Keller}, {Kharchenko}, {Klement}, {Klutsch}, {Knude}, {Koch},
  {Kochukhov}, {Kontizas}, {Koubsky}, {Lallement}, {de Laverny}, {van Leeuwen},
  {Lemasle}, {Lewis}, {Lind}, {Lindstrom}, {Lobel}, {Lopez Santiago}, {Lucas},
  {Ludwig}, {Lueftinger}, {Magrini}, {Maiz Apellaniz}, {Maldonado}, {Marconi},
  {Marino}, {Martayan}, {Martinez-Valpuesta}, {Matijevic}, {McMahon},
  {Messina}, {Meyer}, {Miglio}, {Mikolaitis}, {Minchev}, {Minniti}, {Moitinho},
  {Momany}, {Monaco}, {Montalto}, {Monteiro}, {Monier}, {Montes}, {Mora},
  {Moraux}, {Morel}, {Mowlavi}, {Mucciarelli}, {Munari}, {Napiwotzki},
  {Nardetto}, {Naylor}, {Naze}, {Nelemans}, {Okamoto}, {Ortolani}, {Pace},
  {Palla}, {Palous}, {Parker}, {Penarrubia}, {Pillitteri}, {Piotto}, {Posbic},
  {Prisinzano}, {Puzeras}, {Quirrenbach}, {Ragaini}, {Read}, {Read}, {Reyle},
  {De Ridder}, {Robichon}, {Robin}, {Roeser}, {Romano}, {Royer}, {Ruchti},
  {Ruzicka}, {Ryan}, {Ryde}, {Santos}, {Sanz Forcada}, {Sarro Baro},
  {Sbordone}, {Schilbach}, {Schmeja}, {Schnurr}, {Schoenrich}, {Scholz},
  {Seabroke}, {Sharma}, {De Silva}, {Smith}, {Solano}, {Sordo}, {Soubiran},
  {Sousa}, {Spagna}, {Steffen}, {Steinmetz}, {Stelzer}, {Stempels},
  {Tabernero}, {Tautvaisiene}, {Thevenin}, {Torra}, {Tosi}, {Tolstoy}, {Turon},
  {Walker}, {Wambsganss}, {Worley}, {Venn}, {Vink}, {Wyse}, {Zaggia},
  {Zeilinger}, {Zoccali}, {Zorec}, {Zucker}, {Zwitter}, \& {Gaia-ESO Survey
  Team}}]{gilmore2012}
{Gilmore}, G., {Randich}, S., {Asplund}, M., {et~al.} 2012, The Messenger, 147,
  25

\bibitem[{{Gilmore} {et~al.}(2022){Gilmore}, {Randich}, {Worley}, {Hourihane},
  {Gonneau}, {Sacco}, {Lewis}, {Magrini}, {Francois}, {Jeffries}, {Koposov},
  {Bragaglia}, {Alfaro}, {Allende Prieto}, {Blomme}, {Korn}, {Lanzafame},
  {Pancino}, {Recio-Blanco}, {Smiljanic}, {Van Eck}, {Zwitter}, {Bensby},
  {Flaccomio}, {Irwin}, {Franciosini}, {Morbidelli}, {Damiani}, {Bonito},
  {Friel}, {Vink}, {Prisinzano}, {Abbas}, {Hatzidimitriou}, {Held}, {Jordi},
  {Paunzen}, {Spagna}, {Jackson}, {Maiz Apellaniz}, {Asplund}, {Bonifacio},
  {Feltzing}, {Binney}, {Drew}, {Ferguson}, {Micela}, {Negueruela}, {Prusti},
  {Rix}, {Vallenari}, {Bergemann}, {Casey}, {de Laverny}, {Frasca}, {Hill},
  {Lind}, {Sbordone}, {Sousa}, {Adibekyan}, {Caffau}, {Daflon}, {Feuillet},
  {Gebran}, {Gonzalez Hernandez}, {Guiglion}, {Herrero}, {Lobel}, {Merle},
  {Mikolaitis}, {Montes}, {Morel}, {Ruchti}, {Soubiran}, {Tabernero},
  {Tautvaisiene}, {Traven}, {Valentini}, {Van der Swaelmen}, {Villanova},
  {Viscasillas Vazquez}, {Bayo}, {Biazzo}, {Carraro}, {Edvardsson}, {Heiter},
  {Jofre}, {Marconi}, {Martayan}, {Masseron}, {Monaco}, {Walton}, {Zaggia},
  {Aguirre Borsen-Koch}, {Alves}, {Balaguer-Nunez}, {Barklem}, {Barrado},
  {Bellazzini}, {Berlanas}, {Binks}, {Bressan}, {Capuzzo-Dolcetta},
  {Casagrande}, {Casamiquela}, {Collins}, {D'Orazi}, {Dantas}, {Debattista},
  {Delgado-Mena}, {Di Marcantonio}, {Drazdauskas}, {Evans}, {Famaey},
  {Franchini}, {Fremat}, {Fu}, {Geisler}, {Gerhard}, {Gonzalez Solares},
  {Grebel}, {Gutierrez Albarran}, {Jimenez-Esteban}, {Jonsson},
  {Khachaturyants}, {Kordopatis}, {Kos}, {Lagarde}, {Ludwig}, {Mahy},
  {Mapelli}, {Marfil}, {Martell}, {Messina}, {Miglio}, {Minchev}, {Moitinho},
  {Montalban}, {Monteiro}, {Morossi}, {Mowlavi}, {Mucciarelli}, {Murphy},
  {Nardetto}, {Ortolani}, {Paletou}, {Palous}, {Pickering}, {Quirrenbach}, {Re
  Fiorentin}, {Read}, {Romano}, {Ryde}, {Sanna}, {Santos}, {Seabroke}, {Spina},
  {Steinmetz}, {Stonkute}, {Sutorius}, {Thevenin}, {Tosi}, {Tsantaki},
  {Wright}, {Wyse}, {Zoccali}, {Zorec}, \& {Zucker}}]{gilmore2022}
{Gilmore}, G., {Randich}, S., {Worley}, C.~C., {et~al.} 2022, arXiv e-prints,
  arXiv:2208.05432.
\newblock \doarXiv{2208.05432}

\bibitem[{{Griffith} {et~al.}(2019){Griffith}, {Johnson}, \&
  {Weinberg}}]{griffith2019}
{Griffith}, E., {Johnson}, J.~A., \& {Weinberg}, D.~H. 2019, \apj, 886, 84,
  \dodoi{10.3847/1538-4357/ab4b5d}

\bibitem[{{Griffith} {et~al.}(2021{\natexlab{a}}){Griffith}, {Weinberg},
  {Johnson}, {Beaton}, {Garc{\'\i}a-Hern{\'a}ndez}, {Hasselquist}, {Holtzman},
  {Johnson}, {J{\"o}nsson}, {Lane}, {Nataf}, \& {Roman-Lopes}}]{griffith2021a}
{Griffith}, E., {Weinberg}, D.~H., {Johnson}, J.~A., {et~al.}
  2021{\natexlab{a}}, \apj, 909, 77, \dodoi{10.3847/1538-4357/abd6be}

\bibitem[{{Griffith} {et~al.}(2021{\natexlab{b}}){Griffith}, {Sukhbold},
  {Weinberg}, {Johnson}, {Johnson}, \& {Vincenzo}}]{griffith2021b}
{Griffith}, E.~J., {Sukhbold}, T., {Weinberg}, D.~H., {et~al.}
  2021{\natexlab{b}}, \apj, 921, 73, \dodoi{10.3847/1538-4357/ac1bac}

\bibitem[{{Griffith} {et~al.}(2022){Griffith}, {Weinberg}, {Buder}, {Johnson},
  {Johnson}, \& {Vincenzo}}]{griffith2022}
{Griffith}, E.~J., {Weinberg}, D.~H., {Buder}, S., {et~al.} 2022, \apj, 931,
  23, \dodoi{10.3847/1538-4357/ac5826}

\bibitem[{{Grupp}(2004{\natexlab{a}})}]{grupp2004a}
{Grupp}, F. 2004{\natexlab{a}}, \aap, 420, 289,
  \dodoi{10.1051/0004-6361:20040971}

\bibitem[{{Grupp}(2004{\natexlab{b}})}]{grupp2004b}
---. 2004{\natexlab{b}}, \aap, 426, 309, \dodoi{10.1051/0004-6361:20040456}

\bibitem[{{Gustafsson} {et~al.}(2008){Gustafsson}, {Edvardsson}, {Eriksson},
  {J{\o}rgensen}, {Nordlund}, \& {Plez}}]{gustafsson2008}
{Gustafsson}, B., {Edvardsson}, B., {Eriksson}, K., {et~al.} 2008, \aap, 486,
  951, \dodoi{10.1051/0004-6361:200809724}

\bibitem[{{Harris} {et~al.}(2020){Harris}, {Millman}, {van der Walt},
  {Gommers}, {Virtanen}, {Cournapeau}, {Wieser}, {Taylor}, {Berg}, {Smith},
  {Kern}, {Picus}, {Hoyer}, {van Kerkwijk}, {Brett}, {Haldane}, {del R{\'\i}o},
  {Wiebe}, {Peterson}, {G{\'e}rard-Marchant}, {Sheppard}, {Reddy}, {Weckesser},
  {Abbasi}, {Gohlke}, \& {Oliphant}}]{harris2020}
{Harris}, C.~R., {Millman}, K.~J., {van der Walt}, S.~J., {et~al.} 2020, \nat,
  585, 357, \dodoi{10.1038/s41586-020-2649-2}

\bibitem[{{Hasselquist} {et~al.}(2021){Hasselquist}, {Hayes}, {Lian},
  {Weinberg}, {Zasowski}, {Horta}, {Beaton}, {Feuillet}, {Garro}, {Gallart},
  {Smith}, {Holtzman}, {Minniti}, {Lacerna}, {Shetrone}, {J{\"o}nsson},
  {Cioni}, {Fillingham}, {Cunha}, {O'Connell}, {Fern{\'a}ndez-Trincado},
  {Mu{\~n}oz}, {Schiavon}, {Almeida}, {Anguiano}, {Beers}, {Bizyaev},
  {Brownstein}, {Cohen}, {Frinchaboy}, {Garc{\'\i}a-Hern{\'a}ndez}, {Geisler},
  {Lane}, {Majewski}, {Nidever}, {Nitschelm}, {Povick}, {Price-Whelan},
  {Roman-Lopes}, {Rosado}, {Sobeck}, {Stringfellow}, {Valenzuela}, {Villanova},
  \& {Vincenzo}}]{hasselquist2021}
{Hasselquist}, S., {Hayes}, C.~R., {Lian}, J., {et~al.} 2021, \apj, 923, 172,
  \dodoi{10.3847/1538-4357/ac25f9}

\bibitem[{{Hawkins} {et~al.}(2015){Hawkins}, {Jofr{\'e}}, {Masseron}, \&
  {Gilmore}}]{hawkins2015}
{Hawkins}, K., {Jofr{\'e}}, P., {Masseron}, T., \& {Gilmore}, G. 2015, \mnras,
  453, 758, \dodoi{10.1093/mnras/stv1586}

\bibitem[{{Hayden} {et~al.}(2015){Hayden}, {Bovy}, {Holtzman}, {Nidever},
  {Bird}, {Weinberg}, {Andrews}, {Majewski}, {Allende Prieto}, {Anders},
  {Beers}, {Bizyaev}, {Chiappini}, {Cunha}, {Frinchaboy},
  {Garc{\'\i}a-Her{\'n}andez}, {Garc{\'\i}a P{\'e}rez}, {Girardi}, {Harding},
  {Hearty}, {Johnson}, {M{\'e}sz{\'a}ros}, {Minchev}, {O'Connell}, {Pan},
  {Robin}, {Schiavon}, {Schneider}, {Schultheis}, {Shetrone}, {Skrutskie},
  {Steinmetz}, {Smith}, {Wilson}, {Zamora}, \& {Zasowski}}]{hayden2015}
{Hayden}, M.~R., {Bovy}, J., {Holtzman}, J.~A., {et~al.} 2015, \apj, 808, 132,
  \dodoi{10.1088/0004-637X/808/2/132}

\bibitem[{{Hayes} {et~al.}(2018){Hayes}, {Majewski}, {Shetrone},
  {Fern{\'a}ndez-Alvar}, {Allende Prieto}, {Schuster}, {Carigi}, {Cunha},
  {Smith}, {Sobeck}, {Almeida}, {Beers}, {Carrera}, {Fern{\'a}ndez-Trincado},
  {Garc{\'\i}a-Hern{\'a}ndez}, {Geisler}, {Lane}, {Lucatello}, {Matthews},
  {Minniti}, {Nitschelm}, {Tang}, {Tissera}, \& {Zamora}}]{hayes2018}
{Hayes}, C.~R., {Majewski}, S.~R., {Shetrone}, M., {et~al.} 2018, \apj, 852,
  49, \dodoi{10.3847/1538-4357/aa9cec}

\bibitem[{{Hayes} {et~al.}(2022){Hayes}, {Masseron}, {Sobeck},
  {Garc{\'\i}a-Hern{\'a}ndez}, {Prieto}, {Beaton}, {Cunha}, {Hasselquist},
  {Holtzman}, {J{\"o}nsson}, {Majewski}, {Shetrone}, {Smith}, \&
  {Almeida}}]{hayes2022}
{Hayes}, C.~R., {Masseron}, T., {Sobeck}, J., {et~al.} 2022, \apjs, 262, 34,
  \dodoi{10.3847/1538-4365/ac839f}

\bibitem[{{Haywood} {et~al.}(2013){Haywood}, {Di Matteo}, {Lehnert}, {Katz}, \&
  {G{\'o}mez}}]{haywood2013}
{Haywood}, M., {Di Matteo}, P., {Lehnert}, M.~D., {Katz}, D., \& {G{\'o}mez},
  A. 2013, \aap, 560, A109, \dodoi{10.1051/0004-6361/201321397}

\bibitem[{{Heiter} {et~al.}(2021){Heiter}, {Lind}, {Bergemann}, {Asplund},
  {Mikolaitis}, {Barklem}, {Masseron}, {de Laverny}, {Magrini}, {Edvardsson},
  {J{\"o}nsson}, {Pickering}, {Ryde}, {Bayo Ar{\'a}n}, {Bensby}, {Casey},
  {Feltzing}, {Jofr{\'e}}, {Korn}, {Pancino}, {Damiani}, {Lanzafame}, {Lardo},
  {Monaco}, {Morbidelli}, {Smiljanic}, {Worley}, {Zaggia}, {Randich}, \&
  {Gilmore}}]{heiter2021}
{Heiter}, U., {Lind}, K., {Bergemann}, M., {et~al.} 2021, \aap, 645, A106,
  \dodoi{10.1051/0004-6361/201936291}

\bibitem[{{Helmi} {et~al.}(2018){Helmi}, {Babusiaux}, {Koppelman}, {Massari},
  {Veljanoski}, \& {Brown}}]{helmi2018}
{Helmi}, A., {Babusiaux}, C., {Koppelman}, H.~H., {et~al.} 2018, \nat, 563, 85,
  \dodoi{10.1038/s41586-018-0625-x}

\bibitem[{{Holtzman} {et~al.}(2018){Holtzman}, {Hasselquist}, {Shetrone},
  {Cunha}, {Allende Prieto}, {Anguiano}, {Bizyaev}, {Bovy}, {Casey},
  {Edvardsson}, {Johnson}, {J{\"o}nsson}, {Meszaros}, {Smith}, {Sobeck},
  {Zamora}, {Chojnowski}, {Fernandez-Trincado}, {Garcia-Hernandez}, {Majewski},
  {Pinsonneault}, {Souto}, {Stringfellow}, {Tayar}, {Troup}, \&
  {Zasowski}}]{holtzman2018}
{Holtzman}, J.~A., {Hasselquist}, S., {Shetrone}, M., {et~al.} 2018, \aj, 156,
  125, \dodoi{10.3847/1538-3881/aad4f9}

\bibitem[{{Horta} {et~al.}(2021){Horta}, {Schiavon}, {Mackereth}, {Pfeffer},
  {Mason}, {Kisku}, {Fragkoudi}, {Allende Prieto}, {Cunha}, {Hasselquist},
  {Holtzman}, {Majewski}, {Nataf}, {O'Connell}, {Schultheis}, \&
  {Smith}}]{horta2021}
{Horta}, D., {Schiavon}, R.~P., {Mackereth}, J.~T., {et~al.} 2021, \mnras, 500,
  1385, \dodoi{10.1093/mnras/staa2987}

\bibitem[{{Horta} {et~al.}(2022){Horta}, {Schiavon}, {Mackereth}, {Weinberg},
  {Hasselquist}, {Feuillet}, {O'Connell}, {Anguiano}, {Allende-Prieto},
  {Beaton}, {Bizyaev}, {Cunha}, {Geisler}, {Garc{\'\i}a-Hern{\'a}ndez},
  {Holtzman}, {J{\"o}nsson}, {Lane}, {Majewski}, {M{\'e}sz{\'a}ros}, {Minniti},
  {Nitschelm}, {Shetrone}, {Smith}, \& {Zasowski}}]{horta2022}
---. 2022, arXiv e-prints, arXiv:2204.04233.
\newblock \doarXiv{2204.04233}

\bibitem[{{Howes} {et~al.}(2016){Howes}, {Asplund}, {Keller}, {Casey}, {Yong},
  {Lind}, {Frebel}, {Hays}, {Alves-Brito}, {Bessell}, {Casagrande}, {Marino},
  {Nataf}, {Owen}, {Da Costa}, {Schmidt}, \& {Tisserand}}]{howes2016}
{Howes}, L.~M., {Asplund}, M., {Keller}, S.~C., {et~al.} 2016, \mnras, 460,
  884, \dodoi{10.1093/mnras/stw1004}

\bibitem[{{Hunter}(2007)}]{hunter2007}
{Hunter}, J.~D. 2007, Computing in Science and Engineering, 9, 90,
  \dodoi{10.1109/MCSE.2007.55}

\bibitem[{{Jacobson} {et~al.}(2015){Jacobson}, {Keller}, {Frebel}, {Casey},
  {Asplund}, {Bessell}, {Da Costa}, {Lind}, {Marino}, {Norris}, {Pe{\~n}a},
  {Schmidt}, {Tisserand}, {Walsh}, {Yong}, \& {Yu}}]{jacobson2015}
{Jacobson}, H.~R., {Keller}, S., {Frebel}, A., {et~al.} 2015, \apj, 807, 171,
  \dodoi{10.1088/0004-637X/807/2/171}

\bibitem[{{Jofr{\'e}} {et~al.}(2019){Jofr{\'e}}, {Heiter}, \&
  {Soubiran}}]{jofre2019}
{Jofr{\'e}}, P., {Heiter}, U., \& {Soubiran}, C. 2019, \araa, 57, 571,
  \dodoi{10.1146/annurev-astro-091918-104509}

\bibitem[{{Johnson}(2019)}]{johnson2019}
{Johnson}, J.~A. 2019, Science, 363, 474, \dodoi{10.1126/science.aau9540}

\bibitem[{{Johnson} \& {Weinberg}(2020)}]{johnson2020}
{Johnson}, J.~W., \& {Weinberg}, D.~H. 2020, \mnras, 498, 1364,
  \dodoi{10.1093/mnras/staa2431}

\bibitem[{{Johnson} {et~al.}(2022){Johnson}, {Weinberg}, {Vincenzo}, {Bird}, \&
  {Griffith}}]{johnson2022}
{Johnson}, J.~W., {Weinberg}, D.~H., {Vincenzo}, F., {Bird}, J.~C., \&
  {Griffith}, E.~J. 2022, arXiv e-prints, arXiv:2202.04666.
\newblock \doarXiv{2202.04666}

\bibitem[{{Johnson} {et~al.}(2021){Johnson}, {Weinberg}, {Vincenzo}, {Bird},
  {Loebman}, {Brooks}, {Quinn}, {Christensen}, \& {Griffith}}]{johnson2021}
{Johnson}, J.~W., {Weinberg}, D.~H., {Vincenzo}, F., {et~al.} 2021, \mnras,
  508, 4484, \dodoi{10.1093/mnras/stab2718}

\bibitem[{{Jonsell} {et~al.}(2005){Jonsell}, {Edvardsson}, {Gustafsson},
  {Magain}, {Nissen}, \& {Asplund}}]{jonsell2005}
{Jonsell}, K., {Edvardsson}, B., {Gustafsson}, B., {et~al.} 2005, \aap, 440,
  321, \dodoi{10.1051/0004-6361:20052797}

\bibitem[{{J{\"o}nsson} {et~al.}(2020){J{\"o}nsson}, {Holtzman}, {Allende
  Prieto}, {Cunha}, {Garc{\'\i}a-Hern{\'a}ndez}, {Hasselquist}, {Masseron},
  {Osorio}, {Shetrone}, {Smith}, {Stringfellow}, {Bizyaev}, {Edvardsson},
  {Majewski}, {M{\'e}sz{\'a}ros}, {Souto}, {Zamora}, {Beaton}, {Bovy}, {Donor},
  {Pinsonneault}, {Poovelil}, \& {Sobeck}}]{jonsson2020}
{J{\"o}nsson}, H., {Holtzman}, J.~A., {Allende Prieto}, C., {et~al.} 2020, \aj,
  160, 120, \dodoi{10.3847/1538-3881/aba592}

\bibitem[{{Kirby} {et~al.}(2011){Kirby}, {Cohen}, {Smith}, {Majewski}, {Sohn},
  \& {Guhathakurta}}]{kirby2011}
{Kirby}, E.~N., {Cohen}, J.~G., {Smith}, G.~H., {et~al.} 2011, \apj, 727, 79,
  \dodoi{10.1088/0004-637X/727/2/79}

\bibitem[{{Kroupa}(2001)}]{kroupa2001}
{Kroupa}, P. 2001, \mnras, 322, 231, \dodoi{10.1046/j.1365-8711.2001.04022.x}

\bibitem[{{Kurucz}(1992)}]{kurucz1992}
{Kurucz}, R.~L. 1992, in The Stellar Populations of Galaxies, ed. B.~{Barbuy}
  \& A.~{Renzini}, Vol. 149, 225

\bibitem[{{Lach} {et~al.}(2020){Lach}, {R{\"o}pke}, {Seitenzahl}, {Cot{\'e}},
  {Gronow}, \& {Ruiter}}]{lach2020}
{Lach}, F., {R{\"o}pke}, F.~K., {Seitenzahl}, I.~R., {et~al.} 2020, \aap, 644,
  A118, \dodoi{10.1051/0004-6361/202038721}

\bibitem[{{Leroy} {et~al.}(2008){Leroy}, {Walter}, {Brinks}, {Bigiel}, {de
  Blok}, {Madore}, \& {Thornley}}]{leroy2008}
{Leroy}, A.~K., {Walter}, F., {Brinks}, E., {et~al.} 2008, \aj, 136, 2782,
  \dodoi{10.1088/0004-6256/136/6/2782}

\bibitem[{{Li} {et~al.}(2022){Li}, {Aoki}, {Matsuno}, {Xing}, {Suda},
  {Tominaga}, {Chen}, {Honda}, {Ishigaki}, {Shi}, {Zhao}, \& {Zhao}}]{li2022}
{Li}, H., {Aoki}, W., {Matsuno}, T., {et~al.} 2022, \apj, 931, 147,
  \dodoi{10.3847/1538-4357/ac6514}

\bibitem[{{Lind} {et~al.}(2022){Lind}, {Nordlander}, {Wehrhahn}, {Montelius},
  {Osorio}, {Barklem}, {Af{\c{s}}ar}, {Sneden}, \& {Kobayashi}}]{lind2022}
{Lind}, K., {Nordlander}, T., {Wehrhahn}, A., {et~al.} 2022, \aap, 665, A33,
  \dodoi{10.1051/0004-6361/202142195}

\bibitem[{{Lombardo} {et~al.}(2022){Lombardo}, {Bonifacio}, {Fran{\c{c}}ois},
  {Hansen}, {Caffau}, {Hanke}, {Sk{\'u}lad{\'o}ttir}, {Arcones}, {Eichler},
  {Reichert}, {Psaltis}, {Koch Hansen}, \& {Sbordone}}]{lombardo2022}
{Lombardo}, L., {Bonifacio}, P., {Fran{\c{c}}ois}, P., {et~al.} 2022, \aap,
  665, A10, \dodoi{10.1051/0004-6361/202243932}

\bibitem[{{Lucey} {et~al.}(2019){Lucey}, {Hawkins}, {Ness}, {Asplund},
  {Bensby}, {Casagrande}, {Feltzing}, {Freeman}, {Kobayashi}, \&
  {Marino}}]{lucey2019}
{Lucey}, M., {Hawkins}, K., {Ness}, M., {et~al.} 2019, \mnras, 488, 2283,
  \dodoi{10.1093/mnras/stz1847}

\bibitem[{{Mackereth} \& {Bovy}(2018)}]{mackereth2018}
{Mackereth}, J.~T., \& {Bovy}, J. 2018, \pasp, 130, 114501,
  \dodoi{10.1088/1538-3873/aadcdd}

\bibitem[{{Mackereth} {et~al.}(2019){Mackereth}, {Schiavon}, {Pfeffer},
  {Hayes}, {Bovy}, {Anguiano}, {Allende Prieto}, {Hasselquist}, {Holtzman},
  {Johnson}, {Majewski}, {O'Connell}, {Shetrone}, {Tissera}, \&
  {Fern{\'a}ndez-Trincado}}]{mackereth2019}
{Mackereth}, J.~T., {Schiavon}, R.~P., {Pfeffer}, J., {et~al.} 2019, \mnras,
  482, 3426, \dodoi{10.1093/mnras/sty2955}

\bibitem[{{Majewski} {et~al.}(2017){Majewski}, {Schiavon}, {Frinchaboy},
  {Allende Prieto}, {Barkhouser}, {Bizyaev}, {Blank}, {Brunner}, {Burton},
  {Carrera}, {Chojnowski}, {Cunha}, {Epstein}, {Fitzgerald}, {Garc{\'\i}a
  P{\'e}rez}, {Hearty}, {Henderson}, {Holtzman}, {Johnson}, {Lam}, {Lawler},
  {Maseman}, {M{\'e}sz{\'a}ros}, {Nelson}, {Nguyen}, {Nidever}, {Pinsonneault},
  {Shetrone}, {Smee}, {Smith}, {Stolberg}, {Skrutskie}, {Walker}, {Wilson},
  {Zasowski}, {Anders}, {Basu}, {Beland}, {Blanton}, {Bovy}, {Brownstein},
  {Carlberg}, {Chaplin}, {Chiappini}, {Eisenstein}, {Elsworth}, {Feuillet},
  {Fleming}, {Galbraith-Frew}, {Garc{\'\i}a}, {Garc{\'\i}a-Hern{\'a}ndez},
  {Gillespie}, {Girardi}, {Gunn}, {Hasselquist}, {Hayden}, {Hekker}, {Ivans},
  {Kinemuchi}, {Klaene}, {Mahadevan}, {Mathur}, {Mosser}, {Muna}, {Munn},
  {Nichol}, {O'Connell}, {Parejko}, {Robin}, {Rocha-Pinto}, {Schultheis},
  {Serenelli}, {Shane}, {Silva Aguirre}, {Sobeck}, {Thompson}, {Troup},
  {Weinberg}, \& {Zamora}}]{majewski2017}
{Majewski}, S.~R., {Schiavon}, R.~P., {Frinchaboy}, P.~M., {et~al.} 2017, \aj,
  154, 94, \dodoi{10.3847/1538-3881/aa784d}

\bibitem[{{Maoz} \& {Graur}(2017)}]{maoz2017}
{Maoz}, D., \& {Graur}, O. 2017, \apj, 848, 25,
  \dodoi{10.3847/1538-4357/aa8b6e}

\bibitem[{{Mashonkina} {et~al.}(2007){Mashonkina}, {Korn}, \&
  {Przybilla}}]{mashonkina2007}
{Mashonkina}, L., {Korn}, A.~J., \& {Przybilla}, N. 2007, \aap, 461, 261,
  \dodoi{10.1051/0004-6361:20065999}

\bibitem[{{Matteucci} \& {Greggio}(1986)}]{matteucci1986}
{Matteucci}, F., \& {Greggio}, L. 1986, \aap, 154, 279

\bibitem[{{McMillan}(2017)}]{mcmillan2017}
{McMillan}, P.~J. 2017, \mnras, 465, 76, \dodoi{10.1093/mnras/stw2759}

\bibitem[{{McWilliam}(1997)}]{mcwilliam1997}
{McWilliam}, A. 1997, \araa, 35, 503, \dodoi{10.1146/annurev.astro.35.1.503}

\bibitem[{{Minchev} {et~al.}(2014){Minchev}, {Chiappini}, \&
  {Martig}}]{minchev2014}
{Minchev}, I., {Chiappini}, C., \& {Martig}, M. 2014, \aap, 572, A92,
  \dodoi{10.1051/0004-6361/201423487}

\bibitem[{{Moore} {et~al.}(1966){Moore}, {Minnaert}, \& {Houtgast}}]{moore1966}
{Moore}, C.~E., {Minnaert}, M.~G.~J., \& {Houtgast}, J. 1966, {The solar
  spectrum 2935 A to 8770 A}

\bibitem[{{Mucciarelli} {et~al.}(2013){Mucciarelli}, {Pancino}, {Lovisi},
  {Ferraro}, \& {Lapenna}}]{mucciarelli2013}
{Mucciarelli}, A., {Pancino}, E., {Lovisi}, L., {Ferraro}, F.~R., \& {Lapenna},
  E. 2013, \apj, 766, 78, \dodoi{10.1088/0004-637X/766/2/78}

\bibitem[{{Naidu} {et~al.}(2020){Naidu}, {Conroy}, {Bonaca}, {Johnson}, {Ting},
  {Caldwell}, {Zaritsky}, \& {Cargile}}]{naidu2020}
{Naidu}, R.~P., {Conroy}, C., {Bonaca}, A., {et~al.} 2020, \apj, 901, 48,
  \dodoi{10.3847/1538-4357/abaef4}

\bibitem[{{Nissen} \& {Schuster}(2010)}]{nissen2010}
{Nissen}, P.~E., \& {Schuster}, W.~J. 2010, \aap, 511, L10,
  \dodoi{10.1051/0004-6361/200913877}

\bibitem[{{Nissen} \& {Schuster}(2011)}]{nissen2011}
---. 2011, \aap, 530, A15, \dodoi{10.1051/0004-6361/201116619}

\bibitem[{pandas~development team(2020)}]{pandasa}
pandas~development team, T. 2020, pandas-dev/pandas: Pandas, latest,  Zenodo,
  \dodoi{10.5281/zenodo.3509134}

\bibitem[{{Pehlivan Rhodin} {et~al.}(2017){Pehlivan Rhodin}, {Hartman},
  {Nilsson}, \& {J{\"o}nsson}}]{pehlivan2017}
{Pehlivan Rhodin}, A., {Hartman}, H., {Nilsson}, H., \& {J{\"o}nsson}, P. 2017,
  \aap, 598, A102, \dodoi{10.1051/0004-6361/201629849}

\bibitem[{{Pejcha} \& {Thompson}(2015)}]{pejcha2015}
{Pejcha}, O., \& {Thompson}, T.~A. 2015, \apj, 801, 90,
  \dodoi{10.1088/0004-637X/801/2/90}

\bibitem[{{Ratcliffe} \& {Ness}(2022)}]{ratcliffe2022}
{Ratcliffe}, B., \& {Ness}, M. 2022, arXiv e-prints, arXiv:2206.02772.
\newblock \doarXiv{2206.02772}

\bibitem[{{Rix} {et~al.}(2022){Rix}, {Chandra}, {Andrae}, {Price-Whelan},
  {Weinberg}, {Conroy}, {Fouesneau}, {Hogg}, {De Angeli}, {Naidu}, {Xiang}, \&
  {Ruz-Mieres}}]{rix2022}
{Rix}, H.-W., {Chandra}, V., {Andrae}, R., {et~al.} 2022, arXiv e-prints,
  arXiv:2209.02722.
\newblock \doarXiv{2209.02722}

\bibitem[{{Roederer} {et~al.}(2014){Roederer}, {Preston}, {Thompson},
  {Shectman}, {Sneden}, {Burley}, \& {Kelson}}]{roederer2014}
{Roederer}, I.~U., {Preston}, G.~W., {Thompson}, I.~B., {et~al.} 2014, \aj,
  147, 136, \dodoi{10.1088/0004-6256/147/6/136}

\bibitem[{{Rybizki} {et~al.}(2017){Rybizki}, {Just}, \& {Rix}}]{rybizki2017}
{Rybizki}, J., {Just}, A., \& {Rix}, H.-W. 2017, \aap, 605, A59,
  \dodoi{10.1051/0004-6361/201730522}

\bibitem[{{Schlegel} {et~al.}(1998){Schlegel}, {Finkbeiner}, \&
  {Davis}}]{schlegel1998}
{Schlegel}, D.~J., {Finkbeiner}, D.~P., \& {Davis}, M. 1998, \apj, 500, 525,
  \dodoi{10.1086/305772}

\bibitem[{{Schneider} {et~al.}(2017){Schneider}, {Windsor}, {Cushing},
  {Kirkpatrick}, \& {Shkolnik}}]{schneider2017}
{Schneider}, A.~C., {Windsor}, J., {Cushing}, M.~C., {Kirkpatrick}, J.~D., \&
  {Shkolnik}, E.~L. 2017, \aj, 153, 196, \dodoi{10.3847/1538-3881/aa6624}

\bibitem[{{Sch{\"o}nrich} \& {Binney}(2009)}]{schonrich2009}
{Sch{\"o}nrich}, R., \& {Binney}, J. 2009, \mnras, 399, 1145,
  \dodoi{10.1111/j.1365-2966.2009.15365.x}

\bibitem[{{Semenov} {et~al.}(2017){Semenov}, {Kravtsov}, \&
  {Gnedin}}]{semenov2017}
{Semenov}, V.~A., {Kravtsov}, A.~V., \& {Gnedin}, N.~Y. 2017, \apj, 845, 133,
  \dodoi{10.3847/1538-4357/aa8096}

\bibitem[{{Skrutskie} {et~al.}(2006){Skrutskie}, {Cutri}, {Stiening},
  {Weinberg}, {Schneider}, {Carpenter}, {Beichman}, {Capps}, {Chester},
  {Elias}, {Huchra}, {Liebert}, {Lonsdale}, {Monet}, {Price}, {Seitzer},
  {Jarrett}, {Kirkpatrick}, {Gizis}, {Howard}, {Evans}, {Fowler}, {Fullmer},
  {Hurt}, {Light}, {Kopan}, {Marsh}, {McCallon}, {Tam}, {Van Dyk}, \&
  {Wheelock}}]{skrutskie2006}
{Skrutskie}, M.~F., {Cutri}, R.~M., {Stiening}, R., {et~al.} 2006, \aj, 131,
  1163, \dodoi{10.1086/498708}

\bibitem[{{Sneden}(1973)}]{sneden1973}
{Sneden}, C.~A. 1973, PhD thesis, University of Texas, Austin

\bibitem[{{Spitoni} {et~al.}(2019){Spitoni}, {Silva Aguirre}, {Matteucci},
  {Calura}, \& {Grisoni}}]{spitoni2019}
{Spitoni}, E., {Silva Aguirre}, V., {Matteucci}, F., {Calura}, F., \&
  {Grisoni}, V. 2019, \aap, 623, A60, \dodoi{10.1051/0004-6361/201834188}

\bibitem[{{Strassmeier} {et~al.}(2018){Strassmeier}, {Ilyin}, \&
  {Steffen}}]{strassmeier2018}
{Strassmeier}, K.~G., {Ilyin}, I., \& {Steffen}, M. 2018, \aap, 612, A44,
  \dodoi{10.1051/0004-6361/201731631}

\bibitem[{{Strassmeier} {et~al.}(2015){Strassmeier}, {Ilyin}, {J{\"a}rvinen},
  {Weber}, {Woche}, {Barnes}, {Bauer}, {Beckert}, {Bittner}, {Bredthauer},
  {Carroll}, {Denker}, {Dionies}, {DiVarano}, {D{\"o}scher}, {Fechner},
  {Feuerstein}, {Granzer}, {Hahn}, {Harnisch}, {Hofmann}, {Lesser}, {Paschke},
  {Pankratow}, {Plank}, {Pl{\"u}schke}, {Popow}, \&
  {Sablowski}}]{strassmeier2015}
{Strassmeier}, K.~G., {Ilyin}, I., {J{\"a}rvinen}, A., {et~al.} 2015,
  Astronomische Nachrichten, 336, 324, \dodoi{10.1002/asna.201512172}

\bibitem[{{Sukhbold} {et~al.}(2016){Sukhbold}, {Ertl}, {Woosley}, {Brown}, \&
  {Janka}}]{sukhbold2016}
{Sukhbold}, T., {Ertl}, T., {Woosley}, S.~E., {Brown}, J.~M., \& {Janka}, H.~T.
  2016, \apj, 821, 38, \dodoi{10.3847/0004-637X/821/1/38}

\bibitem[{{Ting} \& {Weinberg}(2022)}]{ting2022}
{Ting}, Y.-S., \& {Weinberg}, D.~H. 2022, \apj, 927, 209,
  \dodoi{10.3847/1538-4357/ac5023}

\bibitem[{{Tinsley}(1980)}]{tinsley1980}
{Tinsley}, B.~M. 1980, \fcp, 5, 287, \dodoi{10.48550/arXiv.2203.02041}

\bibitem[{{Vincenzo} {et~al.}(2021){Vincenzo}, {Weinberg}, {Miglio}, {Lane}, \&
  {Roman-Lopes}}]{vincenzo2021}
{Vincenzo}, F., {Weinberg}, D.~H., {Miglio}, A., {Lane}, R.~R., \&
  {Roman-Lopes}, A. 2021, \mnras, 508, 5903, \dodoi{10.1093/mnras/stab2899}

\bibitem[{{Weidner} \& {Kroupa}(2005)}]{weidner2005}
{Weidner}, C., \& {Kroupa}, P. 2005, \apj, 625, 754, \dodoi{10.1086/429867}

\bibitem[{{Weinberg} {et~al.}(2017){Weinberg}, {Andrews}, \&
  {Freudenburg}}]{weinberg2017}
{Weinberg}, D.~H., {Andrews}, B.~H., \& {Freudenburg}, J. 2017, \apj, 837, 183,
  \dodoi{10.3847/1538-4357/837/2/183}

\bibitem[{{Weinberg} {et~al.}(2019){Weinberg}, {Holtzman}, {Hasselquist},
  {Bird}, {Johnson}, {Shetrone}, {Sobeck}, {Allende Prieto}, {Bizyaev},
  {Carrera}, {Cohen}, {Cunha}, {Ebelke}, {Fernandez-Trincado},
  {Garc{\'\i}a-Hern{\'a}ndez}, {Hayes}, {J{\"o}nsson}, {Lane}, {Majewski},
  {Malanushenko}, {M{\'e}sz{\'a}ros}, {Nidever}, {Nitschelm}, {Pan}, {Rix},
  {Rybizki}, {Schiavon}, {Schneider}, {Wilson}, \& {Zamora}}]{weinberg2019}
{Weinberg}, D.~H., {Holtzman}, J.~A., {Hasselquist}, S., {et~al.} 2019, \apj,
  874, 102, \dodoi{10.3847/1538-4357/ab07c7}

\bibitem[{{Weinberg} {et~al.}(2022){Weinberg}, {Holtzman}, {Johnson}, {Hayes},
  {Hasselquist}, {Shetrone}, {Ting}, {Beaton}, {Beers}, {Bird}, {Bizyaev},
  {Blanton}, {Cunha}, {Fern{\'a}ndez-Trincado}, {Frinchaboy},
  {Garc{\'\i}a-Hern{\'a}ndez}, {Griffith}, {Johnson}, {J{\"o}nsson}, {Lane},
  {Leung}, {Mackereth}, {Majewski}, {M{\'e}sz{\'a}ros}, {Nitschelm}, {Pan},
  {Schiavon}, {Schneider}, {Schultheis}, {Smith}, {Sobeck}, {Stassun},
  {Stringfellow}, {Vincenzo}, {Wilson}, \& {Zasowski}}]{weinberg2022}
{Weinberg}, D.~H., {Holtzman}, J.~A., {Johnson}, J.~A., {et~al.} 2022, \apjs,
  260, 32, \dodoi{10.3847/1538-4365/ac6028}

\bibitem[{{W}es {M}c{K}inney(2010)}]{pandasb}
{W}es {M}c{K}inney. 2010, in {P}roceedings of the 9th {P}ython in {S}cience
  {C}onference, ed. {S}t\'efan van~der {W}alt \& {J}arrod {M}illman, 56 -- 61,
  \dodoi{10.25080/Majora-92bf1922-00a}

\bibitem[{{Xiang} {et~al.}(2019){Xiang}, {Ting}, {Rix}, {Sandford}, {Buder},
  {Lind}, {Liu}, {Shi}, \& {Zhang}}]{xiang2019}
{Xiang}, M., {Ting}, Y.-S., {Rix}, H.-W., {et~al.} 2019, \apjs, 245, 34,
  \dodoi{10.3847/1538-4365/ab5364}

\bibitem[{{Zacharias} {et~al.}(2004){Zacharias}, {Urban}, {Zacharias},
  {Wycoff}, {Hall}, {Monet}, \& {Rafferty}}]{zacharias2004}
{Zacharias}, N., {Urban}, S.~E., {Zacharias}, M.~I., {et~al.} 2004, \aj, 127,
  3043, \dodoi{10.1086/386353}

\end{thebibliography}
\bibliographystyle{aasjournal}

\end{document}